\documentclass{aa} 
\usepackage{graphicx}
\usepackage{txfonts}
\usepackage{natbib}
\usepackage[colorlinks]{hyperref}

\def\aap{A\&A}
\def\apj{ApJ}
\def\apjs{ApJS}
\def\aj{AJ}

\def \hi {\ion{H}{i}}

\def\kms{km\,s$^{-1}$}

\def\deg{\hbox{$^\circ$}}
\def\arcmin{\hbox{$^\prime$}}

\def\farcm{\hbox{$.\mkern-4mu^\prime$}}

\begin{document}

\title{Turbulent power distribution in the local interstellar medium }

   \subtitle{}

   \author{P.\ M.\ W.\ Kalberla \inst{1} \and U.\ Haud \inst{2} }

\institute{Argelander-Institut f\"ur Astronomie,
           Auf dem H\"ugel 71, 53121 Bonn, Germany \\
           \email{pkalberla@astro.uni-bonn.de}
           \and
           Tartu Observatory, University of Tartu,
           61602 T\~oravere, Tartumaa, Estonia }

   \authorrunning{P.\,M.\,W. Kalberla \& U.\ Haud } 

   \titlerunning{Turbulence in the ISM}

   \date{Received 30 October 2018 / Accepted  20 May 2019 }

% \abstract{}{}{}{}{} 
% 5 {} token are mandatory
  \abstract 
% context heading (optional) 
% {} leave it empty if necessary
{The interstellar medium (ISM)   on all scales is full of structures that can be used as
  tracers of processes that feed turbulence.  }
% aims heading (mandatory) 
{We used \hi\ survey data to derive global
  properties of the angular power distribution of the local ISM.  }
% methods heading (mandatory) 
{HI4PI observations on an nside = 1024 HEALPix grid and Gaussian
  components representing three phases, the cold, warm, and unstable
  lukewarm neutral medium (CNM, WNM, and LNM), were used for velocities
  $|v_{\mathrm{LSR}}| \leq 25$ \kms.  For high latitudes $|b| > 20\deg$ we
  generated apodized maps. After beam deconvolution we fitted angular power
  spectra.}
% results heading (mandatory)
{ Power spectra for observed column densities are exceptionally well
  defined and straight in log-log presentation with 3D power law indices
  $\gamma \geq -3$ for the local gas. For intermediate velocity
  clouds (IVCs) we derive $\gamma = -2.6$ and for high velocity clouds
  (HVCs) $\gamma = -2.0$. Single-phase power distributions for the CNM,
  LNM, and WNM are highly correlated and shallow with $ \gamma \sim
  -2.5$ for multipoles $l \leq 100$. Excess power from cold filamentary
  structures is observed at larger multipoles. The steepest single-channel  power spectra for the CNM are found at velocities with large
  CNM and low WNM phase fractions. }
% conclusions heading (optional), leave it empty if necessary 
{The phase space distribution in the local ISM is configured by phase
  transitions and needs to be described with three distinct different
  phases, being highly correlated but having distributions with
  different properties. Phase transitions cause locally hierarchical
  structures in phase space. The CNM is structured on small scales and
  is restricted in position-velocity space. The LNM as an interface to the WNM
  envelops the CNM. It extends to larger scales than the CNM and covers
   a wider range of velocities. Correlations between the phases are
  self-similar in velocity.  }

  % -------------------------------
  \keywords{Turbulence -- ISM: general -- ISM:  structure  }
% -------------------------------
  \maketitle
%
%________________________________________________________________
 
\section{Introduction}
\label{Intro}

Turbulence is ubiquitous in the interstellar medium (ISM). Neutral hydrogen is
  abundant and because of the easily observable \hi\ 21cm emission line
  it is one of their key diagnostics. However,  there are more bearings than
  just diagnostics. Already in the 1950s, soon after the detection of
  the 21cm line, Hendrik van de Hulst had ``the hope that you could do
  something by making a turbulence spectrum, a new trick to describe
  things.''\footnote[1]{Interview of Hendrik van de Hulst on 6 September
    1978 at Leiden by W. T. Sullivan III,
    \url{https://www.nrao.edu/archives/Sullivan/sullivan_interviewee_vandehulst.shtml}}
  van de Hulst was interested in measuring turbulence of the
  \hi\ distribution to gain insights into the nature of the
  ISM. Today we have a better background to formulate the question
of   how far turbulence is shaping the ISM and in particular the bistable
  \hi\ distribution. Turbulent motions determine the temporal and
  spatial structure of the gas pressure. If pressure fluctuations are
  sufficiently large, they  drive some of the gas into the cold
  phase. Thus, the \hi\ phase composition is affected by turbulence.
  We should also expect  that changes in the \hi\ phase composition
  have a noticeable imprint on the observable turbulent density and
  velocity distribution and focused on the \hi\ multiphase
  structure.

  For an introduction to topics addressed in the first paragraph we
  refer to the reviews about interstellar turbulence by
  \citet{Elmegreen2004} and \citet{Scalo200}. Our contemporary
  understanding of the multiphase structure of the ISM is based on
  seminal papers by \citet{MO77} and \citet{Wolfire2003}. Heating and
  cooling processes invoke thermal instabilities that tend to segregate
  the \hi\ into two distinct stable phases, the cold medium (CNM) and the warm medium (WNM). For  at least two decades there has been mounting evidence that a
  significant fraction of the \hi\ is in an intermediate unstable state,
  the lukewarm medium (LNM); for the most recent census of the phase fractions we
  refer to \citet{Murray2018b} and \citet{Kalberla2018}. Turbulence is
  considered as a driving mechanism that tends to produce strong
  nonlinear fluctuations in all the thermodynamic variables that are
  affecting thermal instabilities. \citet{Audit2005} find in this case
  large fractions of thermally unstable gas that would not exist without
  turbulent forcing. These fractions increase with the amplitude of the
  turbulent forcing and the cold and thermally unstable gas tends to be
  organized in filamentary structures. As a result, the standard
  two-phase model may need to be replaced by a phase continuum with at
  least three phases; we refer to the excellent review by
  \citet{Vazquez-Semadeni2012} and references therein.

  Analyzing Arecibo data, cold filamentary \hi\ structures that are
  aligned with the magnetic field have been found by \citet{Clark2014}
  at high Galactic latitudes. Similar large-scale structures all over
  the sky correlated with the magnetic field orientation implied by {\it
    Planck } 353 GHz polarized dust emission were reported by
  \citet{Kalberla2016}. Some of the observed anisotropic \hi\ structures
  show a clear association with magneto-ionic features
  (\citet{Kalberla2016b}, \citep{Kalberla2017}, and
    \citep{Jelic2018}). Derived power spectra for the \hi\ are in this
    case consistent with Kolmogorov turbulence, but anisotropies in
    narrow velocity intervals increase on average with spatial
    frequency, both as predicted by \citet{Goldreich1995} for
    incompressible magnetohydrodynamic (MHD) turbulence. These
    observations strongly support an MHD origin of the turbulence
    similar to the conception brought forward by \citet{Heiles2005} and
    \citet{HeilesCrutcher2005} who argued for an equipartition between
    turbulent and magnetic field energy.

  \hi\ observations are typically organized as data cubes in
  position-position-velocity (PPV). Power spectra of emission from
  narrow velocity channels are affected by velocity fluctuations that
  cause shallower slopes than spectra derived from broad channels. This
  is the result of an excess of small features from unrelated structures
  that blend by velocity crowding on the line of sight. The basic
  recipes used to analyze these data and disentangle velocity and density
  fields, called velocity channel analysis (VCA), were given by
  \citet{Lazarian2000}. Velocity fluctuation can in principle mimic
  density structures, an effect described as velocity crowding by
  \citet{Burton1972} and \citet{Lazarian2000}. Recently, a vivid debate
 has  been raised about structures seen in \hi\ channel
  maps. \citet{Lazarian2018} interpret cold filamentary structures
  observed by \citet{Clark2014} as being caused by velocity
  caustics. \citet{Clark2019} object and interpret these filaments as
  genuine \hi\ density structures that are associated with dust.
  \citet{Yuen2019} reinforce arguments given by \citet{Lazarian2018}.

  In our analysis we make use of a Gaussian decomposition of the HI4PI
  survey \citep{Kalberla2018}. We determine discrete spatial power
  spectra for CNM, LNM, and WNM at multipoles $l \la 1023$. When
  separating \hi\ phases we discover limitations on VCA, constraints
  specified previously by \citet{Kolmogorov1941} as restrictions on
  locally homogeneous and isotropic structures. The filamentary features
  debated by \citet{Lazarian2018}, \citet{Clark2019}, and
  \citet{Yuen2019} are identified as coherent CNM structures, resulting
  from phase transitions.

  The observed cold filamentary \hi\ column density structures are
  likely fibers with cylindrical geometry \citep{Clark2014} or
  projections from edge-on sheets and have velocity gradients
  perpendicular to the sheets.  \citet{Kalberla2016b} and
  \citet{Kalberla2017} report in addition a steepening of the spectral
  indices at velocities that are dominated by cold gas, indicating phase
  transitions caused by thermal instabilities. This steepening is
  opposite to predictions from numerical simulations by
  \citet{Saury2014}, but agrees well with more recent investigations by
  \citet{Wareing2019}. The relation between spectral indices and phase
  transitions is one of our topics. We also consider correlations in
  position and velocity space, relating the more extended LNM as an
  envelope to the CNM.
  
The local \hi\ needs to be distinguished from intermediate and high
velocity clouds (IVCs and HVCs, respectively). These features are
thought to be located in the Galactic halo, IVCs at kpc distances and
probably originating from a Galactic fountain and HVCs at larger
distances with an external origin
(\citet{Wakker1997} and \citet{vanWoerden2004}). Both cloud types have a two-phase
structure, but are distinct from Galactic \hi\ because of 
their morphology and their distribution in center velocities and
velocity widths \citep{Haud2008}. We consider the question of  how much the
IVC and HVC power spectra differ from the local distribution. 
  
%\subsection{Previously derived spectral indices}
%\label{Previous}
  
Several turbulence studies are available in the literature
(\citet{Crovisier1983}, \citet{Kalberla1983}, \citet{Green1993},
\citet{Stanimirovic1999}, \citet{Deshpande2000} ,\citet{Dickey2001},
\citet{Elmegreen2001}, \citet{Stanimirovic2001}, \citet{Miville2003},
\citet{Khalil2006}, \citet{Miville2007}, \citet{Chepurnov2010b},
\citet{Roy2010}, \citet{Dedes2012}, \citet{Roy2012}, \citet{Pingel2013},
\citet{Martin2015}, \citet{Kalberla2016b}, \citet{Kalberla2017},
\citet{Blagrave2017}, \citet{Pingel2018}, and \citet{Choudhuri2018}),
yet a major problem is that the scatter of derived power indices is
appreciable. The published range for 3D power law indices is $-2.2 >
\gamma > -4$.

We  study how closely spectral indices depend on the composition of
the \hi\ in different phases, but we must also consider  that the large
scatter in $\gamma$ may be caused by the derivation of the results from
small patches on the sky under very different physical conditions: 
absorption data from \hi\ with considerable optical depth up to emission
from diffuse high latitude gas, distant \hi\ close to the Galactic plane
up to local gas. Different distances imply variable linear scales
perpendicular to but also along the line of sight. Last but not
least, instrumental issues such as beam effects, instrumental noise, and
apodization may cause uncertainties for the derived parameters. In some
of the publications these instrumental issues are not discussed, others
mention these effects and argue that beam smoothing and noise do not
affect their analysis. Only a few publications consider instrumental
biases in detail when fitting spectral
indices. \citet[][Fig. 22]{Kalberla2016b} demonstrated that a 3D
spectral index of $\gamma = -2.97$ can steepen to $\gamma = -3.4$ if the
beam correction is disregarded.  Apodization and and instrumental
noise can also lead to biases \citep[][Appendix A]{Kalberla2017}.

Biases caused by neglecting instrumental issues may even be larger than
noted in the previous example.  \citet{Blagrave2017} rectify a
  previously determined value of $\gamma = -3.6 \pm 0.2$
  \citep{Miville2003}    toward  Ursa Major to $\gamma = -2.68 \pm
  0.14$ and explain biases in their Appendix E. The independent
confirmation with $\gamma = -2.68 \pm 0.07$ by \citet{Kalberla2016b}
demonstrates that a proper data reduction can deliver reliable results
that are telescope independent. Particularly problematic is the slope
of $-3.6$ that was considered in the review by \citet[][Sect. 4.22,
  Fig. 10]{Hennebelle2012} as characteristic for high latitude
\hi\ emission. This incorrect value has in turn propagated to other
publications and is still cited in the most recent papers without
mentioning the revised value of $\gamma = -2.68$ for the Ursa Major region.

%\subsection{Layout}
%\label{Layout}

This paper is organized as follows. In Sect. \ref{power_spectra} we
briefly describe the data reduction. Details are given in Appendix
\ref{Obs}, and throughout the paper we discuss possible instrumental
biases. In Sect. \ref{GaussPower} we derive the spatial power
distribution for different \hi\ phases and correlations between WNM,
LNM, and CNM power spectra. Narrowband (thin slices in velocity)
spectral indices are derived in Sect. \ref{veldependent}, dependences 
of spectral indices on the velocity channel width (thick slices) are
discussed in Sect. \ref{velwidth}. In Sect. \ref{Constraints} we discuss
restrictions on the analysis of the turbulent flow in relation to the
\citet{Kolmogorov1941} theory for turbulence of locally homogeneous and
isotropic structures. Spectral indices for the density and velocity
correlation functions are discussed in Sect. \ref{VCArevisited}. In
Sect. \ref{IVC_HVC} we derive power spectra for IVCs and HVCs.  In
Sect. \ref{outer} we discuss the power at low multipoles and the outer
scale of turbulence.  We summarize and discuss our results in
Sect. \ref{Summary}.

\section{Derivation of power spectra }
\label{power_spectra}

We use high resolution 21 cm line data from the HI4PI survey
\citep{Winkel2016c}.  To calculate the spatial power spectra we use
ANAFAST  provided by version 3.40 of the HEALPix software
package\footnote[2]{https://sourceforge.net/projects/healpix/}
\citep{Gorski2005}. For multipoles $l$ this routine calculates the power
spectrum $P(l)$ of a HEALPix map\footnote[3]{For calculations
  of the CMB, power spectra of the kind $l (l+1) C_{\mathrm l} $ are
usually  considered. Here we do not follow this convention, but use simply $
  C_{\mathrm l} $. This way the power exponent $\gamma$ increases by
  two.}
\begin{equation}  % Eq. 1
P(l) = C_{\mathrm l}(l) \propto l^\gamma,
\label{eq:gamma}
\end{equation}
where $ C_{\mathrm l}(l) $ are the correlation coefficients; we also define
 a power law index $\gamma$.  Since our signal is bandwidth limited,
we use in general $ l < l_{\mathrm{max}} =1023 $.

The parameter $P(l)$, however, is not the observed power spectrum. The observations
are affected by instrumental issues, data processing, and even by
  selection effects from windowing particular regions on the sky. First,  the intensity distribution on the sky is smoothed by the
effective beam function $B_{\mathrm{data}}$ of the telescope. This
includes beam smoothing as well as smoothing caused by the gridding
process, both causing an artificial steepening of the observed
  power distribution. In addition, some noise power
$N_{\mathrm{oise}}$, depending on instrument and observing method, is
added by the receiving system
\begin{equation} % Eq. 2
 P_{\mathrm {obs}}(l) = P(l) \cdot B_{\mathrm{data}}^2(l) + N_{\mathrm{oise}}(l).
\label{eq:Power_obs_1}
\end{equation}
The beam-corrected noise contribution $
  N_{\mathrm{oise}}(l)/B_{\mathrm{data}}^2(l)$ can be critical since
  beam effects suffer from a vanishing beam response
  $B_{\mathrm{data}}(l)$ at high multipoles.

The observed power distribution is further degraded by the window
function, causing a convolution of the observable power distribution
with the Fourier transform of the window function. These effects can be
mitigated by the use of a proper apodization \citep{Harris1978}. Finally, fitting the power spectra includes statistical and
systematical errors.

All these issues are discussed in detail in Appendix \ref{Obs}. There
we specify in particular the dependence of the noise term
$N_{\mathrm{oise}}(l)$ on gridding and data processing and
demonstrate that the noise term is unimportant for our application, using
HI4PI data with an excellent signal-to-noise  ratio. For $l < 1023 $
  we obtain the
clean and transparent relation 
\begin{equation} % Eq. 3
P(l) = P_{\mathrm{obs}}(l) / B^2_{\mathrm{data}}(l)
\label{eq:Power_cor2_1}.
\end{equation}

%=========================================================================
\begin{figure*}[th] %%  1
   \centering
   \includegraphics[width=9cm]{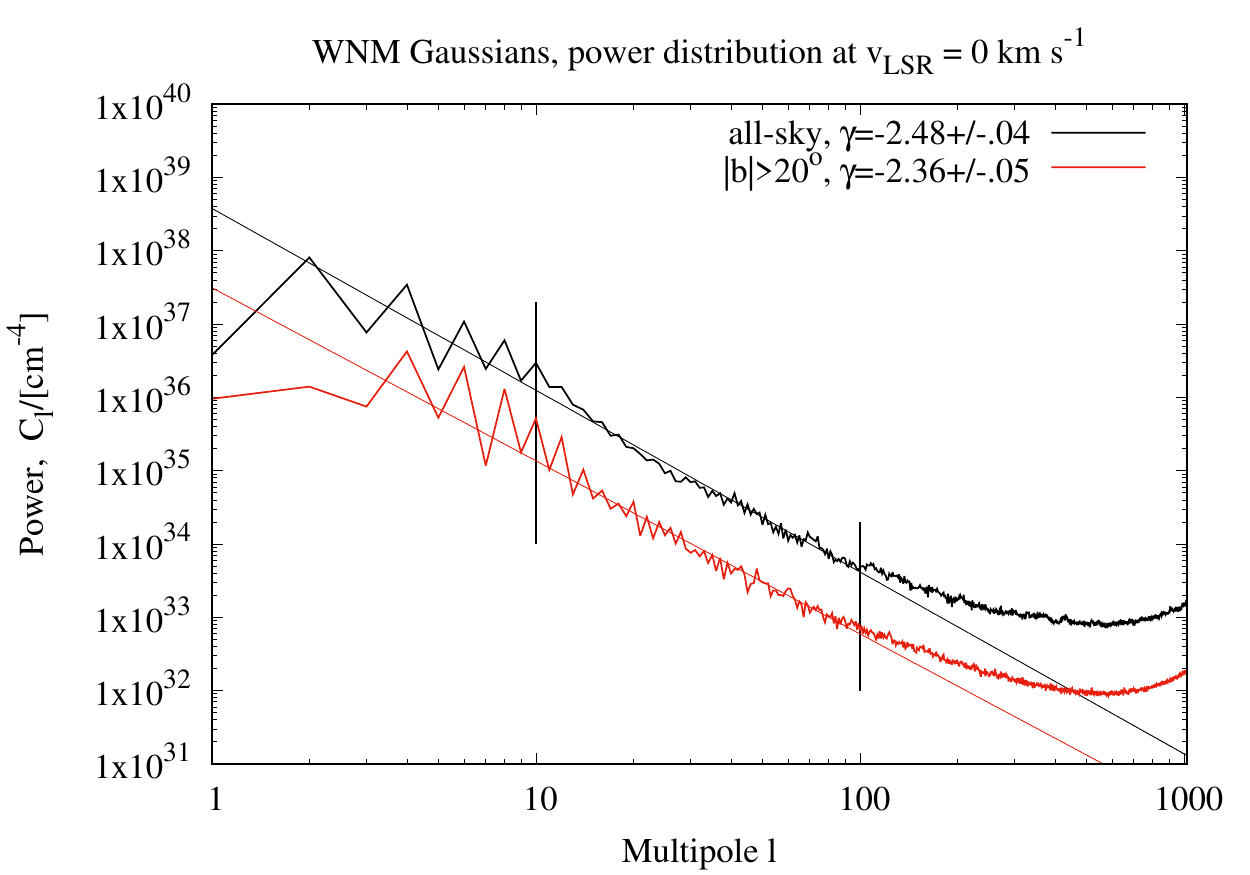}
   \includegraphics[width=9cm]{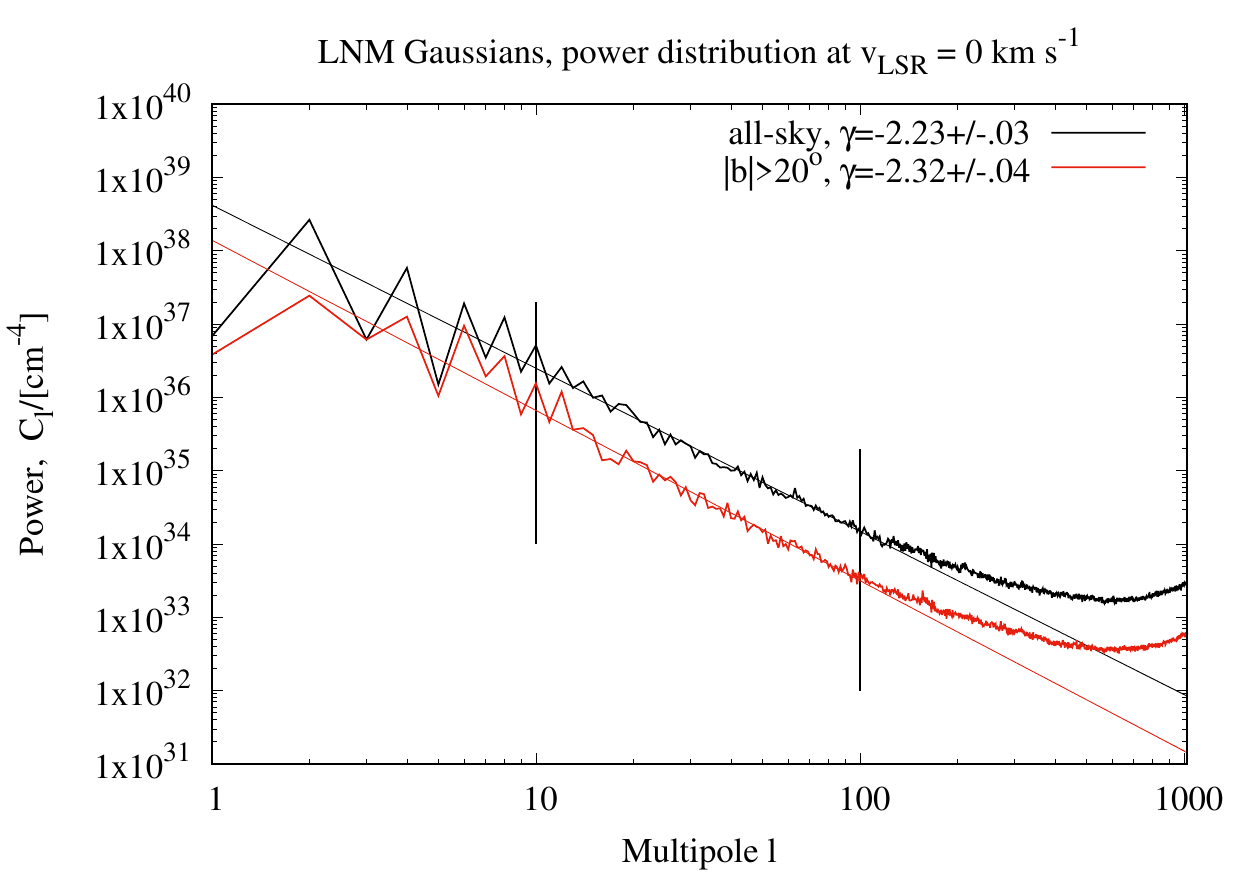}
   \includegraphics[width=9cm]{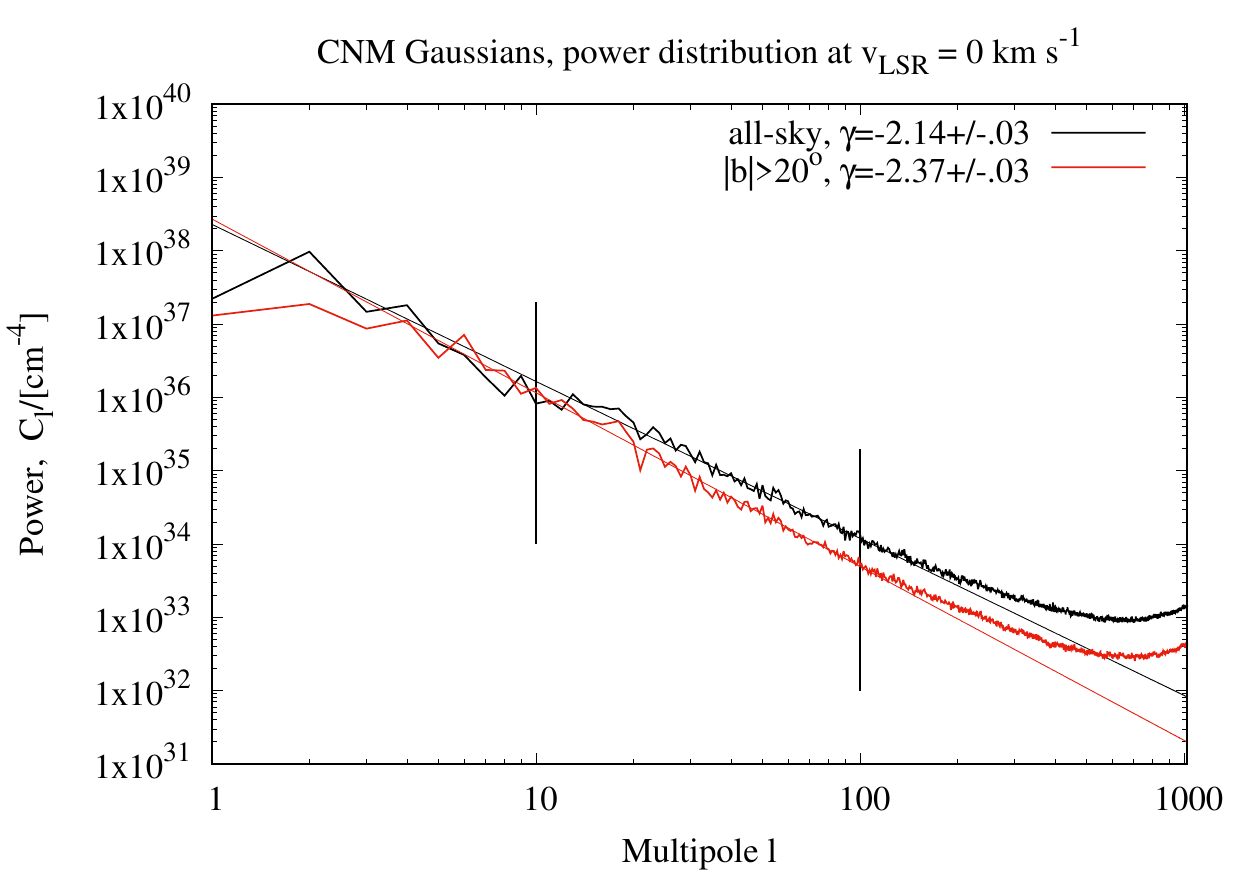}
   \includegraphics[width=9cm]{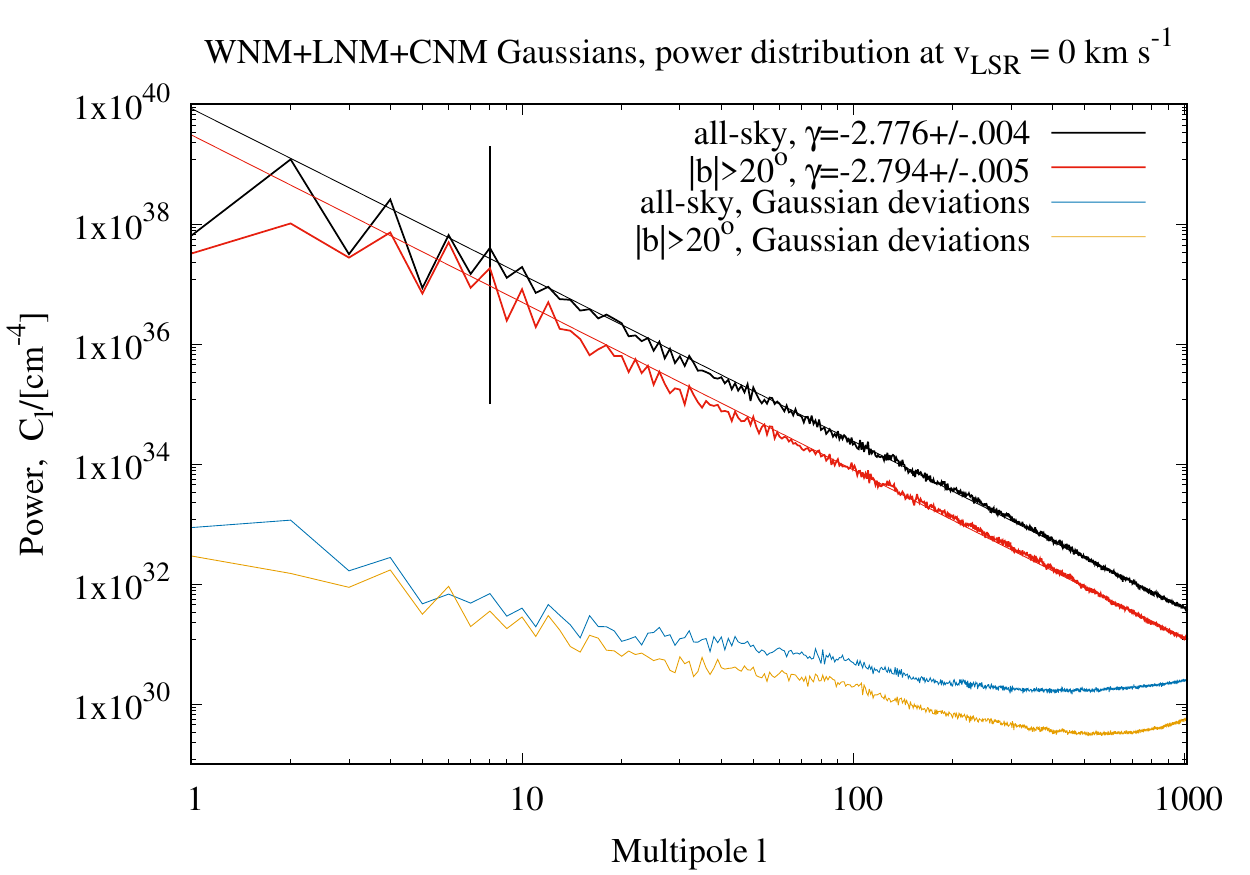}
   \caption{Power distributions for different \hi\ phases at
     $v_{\mathrm{LSR}} = 0 $ \kms.  Top left: WNM; top right: LNM;
     bottom left: CNM; and bottom right: Sum of all phases with
     uncertainties from the Gaussian decomposition (cyan and
     orange). Black lines show all-sky data, red lines are for $|b| > 20
     \deg$. Spectral indices $\gamma$ for CNM, LNM, and CNM are derived
     at $10 < l < 100$ and for the sum of all phases at $l > 8$ as
     indicated by the vertical lines. }
   \label{Fig_Gauss_0}
\end{figure*}
%=========================================================================

\section{Spatial power distribution for different \hi\ phases}
\label{GaussPower}

For $l \ga l_{\mathrm {crit}} \sim 8$ the beam corrected power law
  spectra $P(l)$ for the observed column density distributions are in
  most cases close to a straight line in log-log presentation and can be
  fit well with a constant spectral index (see, e.g., Fig.
  \ref{Fig_cor_noise}). The spectral index for $l \la l_{\mathrm {crit}}$
  is rather flat and not well defined. The critical multipole
  $l_{\mathrm {crit}}$ depends on the outer scale of turbulence $L$ and
  the scale height $H$ of the turbulent medium, $l_{\mathrm {crit}} \sim
  2 \pi H/L$. We postpone the detailed discussion of the low multipoles
  to Sect. \ref{outer}.

  Turbulence in a two-phase \hi\ medium was previously  considered
  by \citet{Martin2015}. In their study the CNM was defined by Gaussian
  components with Doppler temperatures $T_{\mathrm D} < 443$ K. These
  authors found in the case of the north ecliptic pole that the spectral
  index of $-2.86 \pm 0.04 $ for the total \hi\ changes to $-1.9 \pm 0.1
  $ for the CNM and $-2.7 \pm 0.1 $ for the WNM. In addition to changes
  in the spectral index they found ``an additional uncharacterized noise
  component in the $N_{\hi}$ maps near the pixel scale, reflected at
  high spatial frequencies in the power spectrum.''

Our analysis is based on a three-phase all-sky Gaussian decomposition
of the local \hi\ gas  \citep{Kalberla2018}. The different phases were
shown to have different spatial distributions. The CNM is clumpy and
embedded in the spatially more extended LNM which covers in addition a
larger velocity spread around the CNM with narrow lines. Observed column
densities are anti-correlated. The WNM with broad lines is embedding
both  LNM and CNM. These phase space relations should have
correspondences in the associated power distributions.

\subsection{Power spectra for CNM, LNM, and WNM column densities}
\label{Gauss_NH}

 The power spectra in Fig. \ref{Fig_Gauss_0} were calculated for the
 column density distributions shown by \citet{Kalberla2018} in their
 Fig. 9 on top. Figure \ref{Fig_Gauss_0} shows the spatial power
 distributions for column densities from Gaussian components assigned to
 WNM (top left), LNM (top right), and CNM (bottom left) for a single channel at $ v_{\mathrm{LSR}} = 0 $ \kms.

Using the sum of all phases to calculate the power spectra for all
Gaussian components results in the distributions shown in the bottom
panel of Fig. \ref{Fig_Gauss_0} on the right-hand side. These are
essentially the power spectra shown in Fig. \ref{Fig_cor_noise} with
minor deviations. Brightness temperatures restored from Gaussian
components may deviate within the noise from observations, furthermore
components that most probably are caused by radio frequency interference
or instrumental problems have been rejected. Such a Gaussian based
cleaning process has been proven  a very efficient tool to eliminate
residual instrumental problems \citep{Kalberla2015}. The power spectra
of the resulting deviations from the observed \hi\ distribution are
plotted in the lower right panel of Fig. \ref{Fig_Gauss_0} to
demonstrate that these deviations, including the noise floor, are far
below the individual power spectra (see Appendix \ref{Obs}). The shape of
the power distribution from the expected scatter caused by the Gaussian
decomposition (Fig. \ref{Fig_Gauss_0}, bottom right) does not reflect
the derived strong increase in  the power for the individual phases. We
conclude that the enhanced power at high multipoles $l \ga 100$ must be
significant and caused by CNM structures on small scales. These
structures were  described  by \citet{Clark2014} and
\citet{Kalberla2016} as filamentary. They are unresolved by HI4PI and
even by the narrow Arecibo beam, indicating a linear scale of $\la 0.1$
pc.

%=========================================================================

\begin{figure}[th] %%  2
    \centering
    \includegraphics[width=9cm]{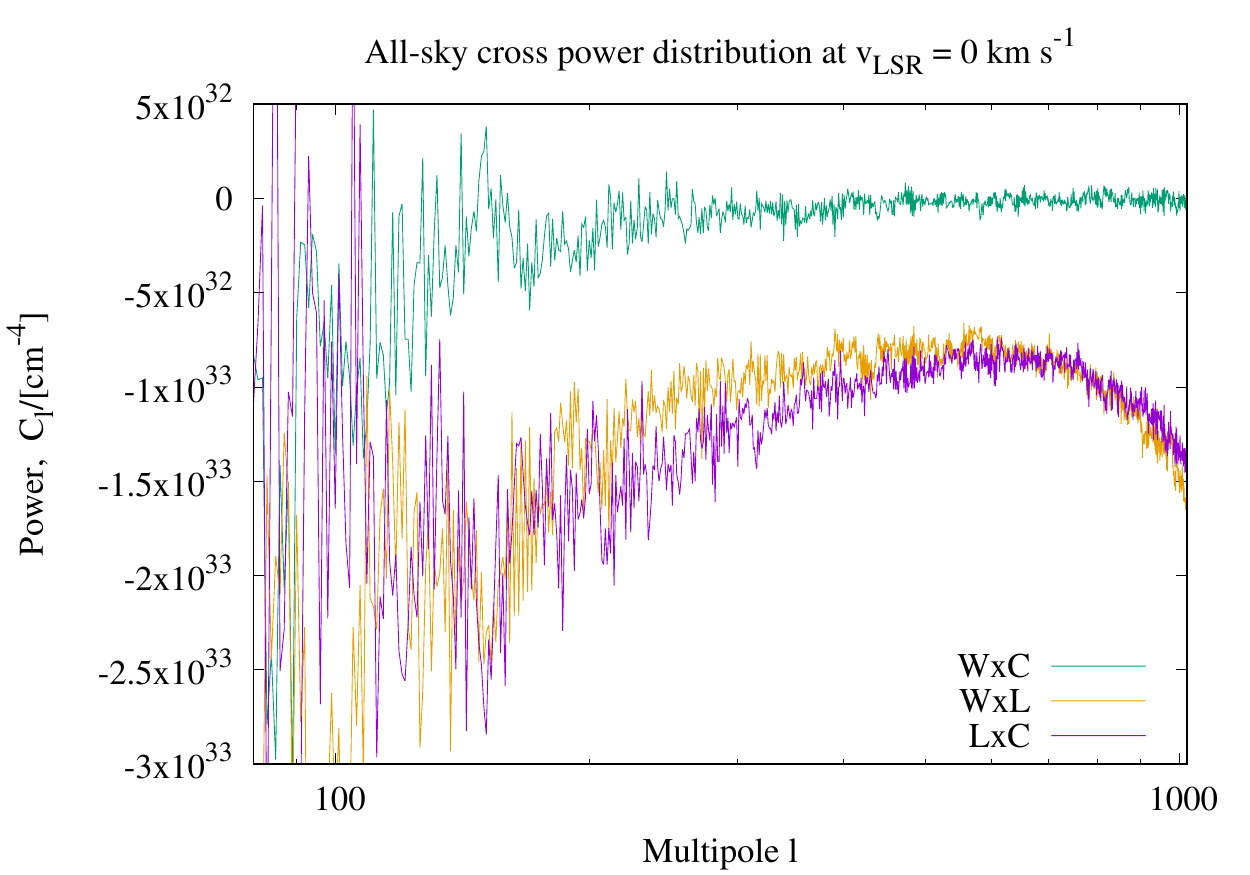}
    \caption{Cross power spectra between WNM and CNM ($P_{\mathrm
        {WxC}}$), WNM and LNM ($P_{\mathrm {WxL}}$), and LNM and CNM
      ($P_{\mathrm {LxC}}$) at $v_{\mathrm{LSR}} = 0 $ \kms.  }
   \label{Fig_Cross_Power}
\end{figure}
%=========================================================================

\begin{figure}[th] %%  3
    \centering
    \includegraphics[width=9cm]{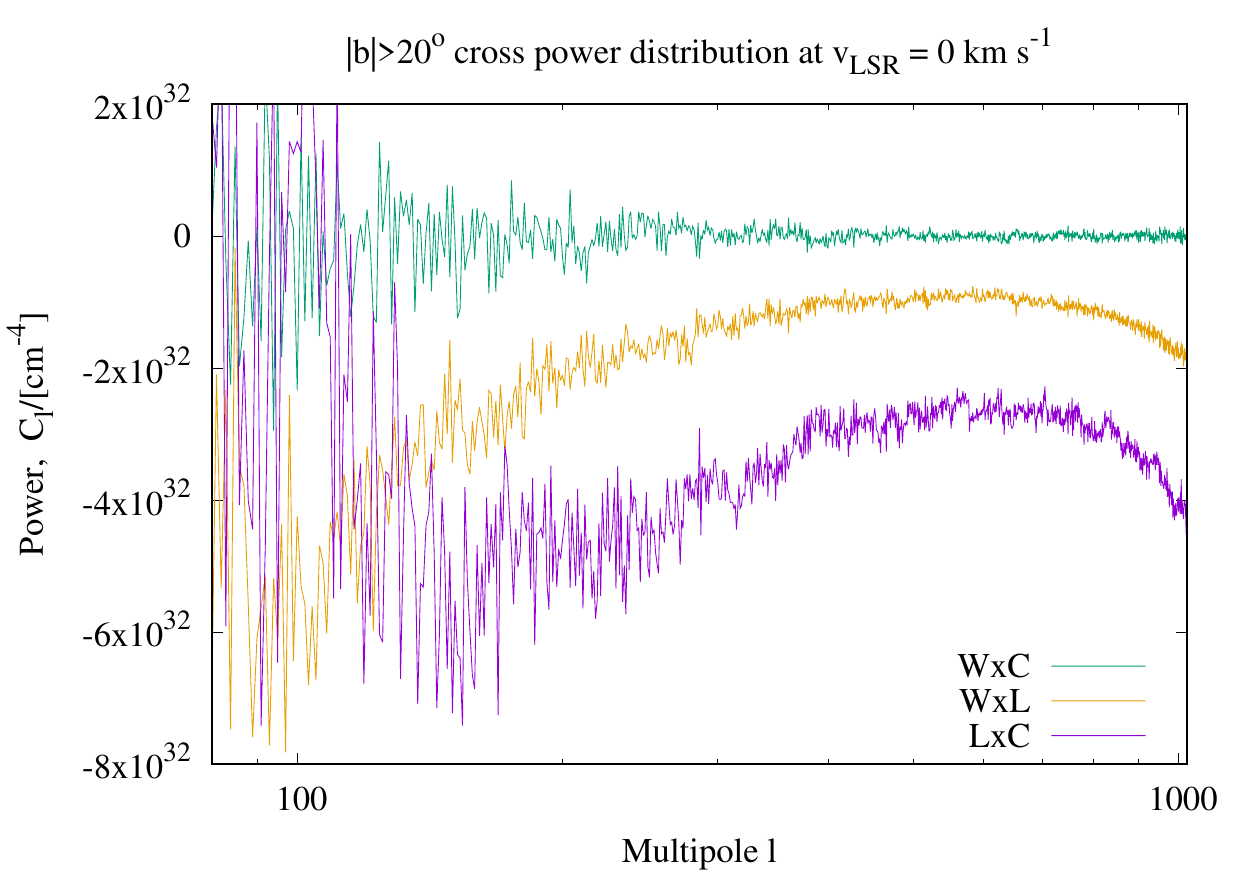}
    \caption{Cross power spectra between WNM and CNM ($P_{\mathrm
        {WxC}}$), WNM and LNM ($P_{\mathrm {WxL}}$), and LNM and CNM
      ($P_{\mathrm {LxC}}$) at $v_{\mathrm{LSR}} = 0 $ \kms. }
   \label{Fig_Cross_Power_high}
\end{figure}
%=========================================================================

It is highly unexpected to find small-scale structures in the WNM power
spectra at high multipoles; however, this is just a consequence of the
correlations between the \hi\ phases. Phase transitions generate CNM
structures on small scales with some unstable LNM around the CNM
\citep{Kalberla2018}. It is easy to understand that such structures are
also present in the more extended LNM, but they are necessarily also
reflected in the WNM power distribution with enhancements at high
multipoles. The WNM distribution is not smooth at high spatial
frequencies. 

The power distribution $P_{\mathrm {TOT}}$ from all Gaussian components
in the lower right panel of Fig. \ref{Fig_Gauss_0} can also be
calculated by summing up all auto power spectra for the individual
phases and all associated cross power spectra $P_{\mathrm
  {WxC}}$, $P_{\mathrm {WxL}}$, and $P_{\mathrm {LxC}}$ between the
phases that are needed for a complete description of a three phase
medium:
\begin{equation} % Eq. 6 
   P_{\mathrm {TOT}} = P_{\mathrm {WNM}} + P_{\mathrm {LNM}} +
   P_{\mathrm {CNM}} + 2 [ P_{\mathrm {WxC}} + P_{\mathrm {WxL}} +
     P_{\mathrm {LxC}} ].
 \label{eq:crossPower}
\end{equation}
 The cross terms are shown in Figs.  \ref{Fig_Cross_Power} and
 \ref{Fig_Cross_Power_high}. In particular, multipoles $l \ga 100$ are of
 interest. The auto power enhancements for $P_{\mathrm {WNM}}$,
 $P_{\mathrm {LNM}}$, and $P_{\mathrm {CNM}}$ at high multipoles are
 caused by pronounced systematic anti-correlations in the cross power
 between WNM and LNM ($P_{\mathrm {WxL}}$), and LNM and CNM
 ($P_{\mathrm {LxC}}$). There is only a weak anti-correlation between
 WNM and CNM ($P_{\mathrm {WxC}}$), an effect that was noted  by
 \citet[][Sect. 4.2]{Kalberla2018}. We note that in
 Figs. \ref{Fig_Cross_Power} and \ref{Fig_Cross_Power_high}, unlike
  all the other power law plots, we do not use a logarithmic
 scale. The negative cross power on low multipoles is significant.

The power distributions shown in Fig. \ref{Fig_Gauss_0} for column
densities in individual phases deviate significantly from the power
spectra of the total \hi, combining all phases. All single-phase slopes,
derived at $10 < l < 100$, are shallow, with $\gamma \sim -2.5$ in
comparison to $\gamma \sim -3$ for $P_{\mathrm {TOT}}$.  In all cases we
find enhanced power for $l \ga 100$, strongly increasing to high
multipoles. These issues can only be explained by the fact that the
column density distributions in different phases are highly correlated.
The cross terms in Eq. \ref{eq:crossPower} are important and describe
the interplay between different phases.

It is important to make this point clear. As an example we consider a
layered structure, consisting of several completely independent and
uncorrelated sheets of different phases along the line of sight. In this
case the cross terms in Eq. \ref{eq:crossPower} would vanish, resulting
in $P_{\mathrm {TOT}} = P_{\mathrm {WNM}} + P_{\mathrm {LNM}} +
P_{\mathrm {CNM}} $. This, obviously, is not observed. Another example
for vanishing cross terms is the case of disjunct spatial
distributions. Separate power spectra from different hemispheres, as
shown in Fig \ref{Fig_EBHIS_GASS}, may be added since the cross terms
for disjunct regions are zero. The established assumption, in the case of a
two-phase  medium, that ``the density fluctuations are spatially
separated in two media and therefore their correlation is likely to be
negligible'' \citep[][Sect. 4.3]{Lazarian2000} is not
justified\footnote[4]{In Eq. 51 of \citet{Lazarian2000} the
  cross term between CNM and WNM in the case of a two-component medium is
  defined without the factor two in our Eq. \ref{eq:crossPower}.}. The
\hi\ with its different phases needs to be considered  a composite. It
is not possible to describe the \hi\ with several independent phases.

%Correlations between the phases were already shown by \citet[][Sect. 4.2]{Kalberla2018}

%=========================================================================

\begin{figure*}[th] %%  4
  \centering
   \includegraphics[width=9cm]{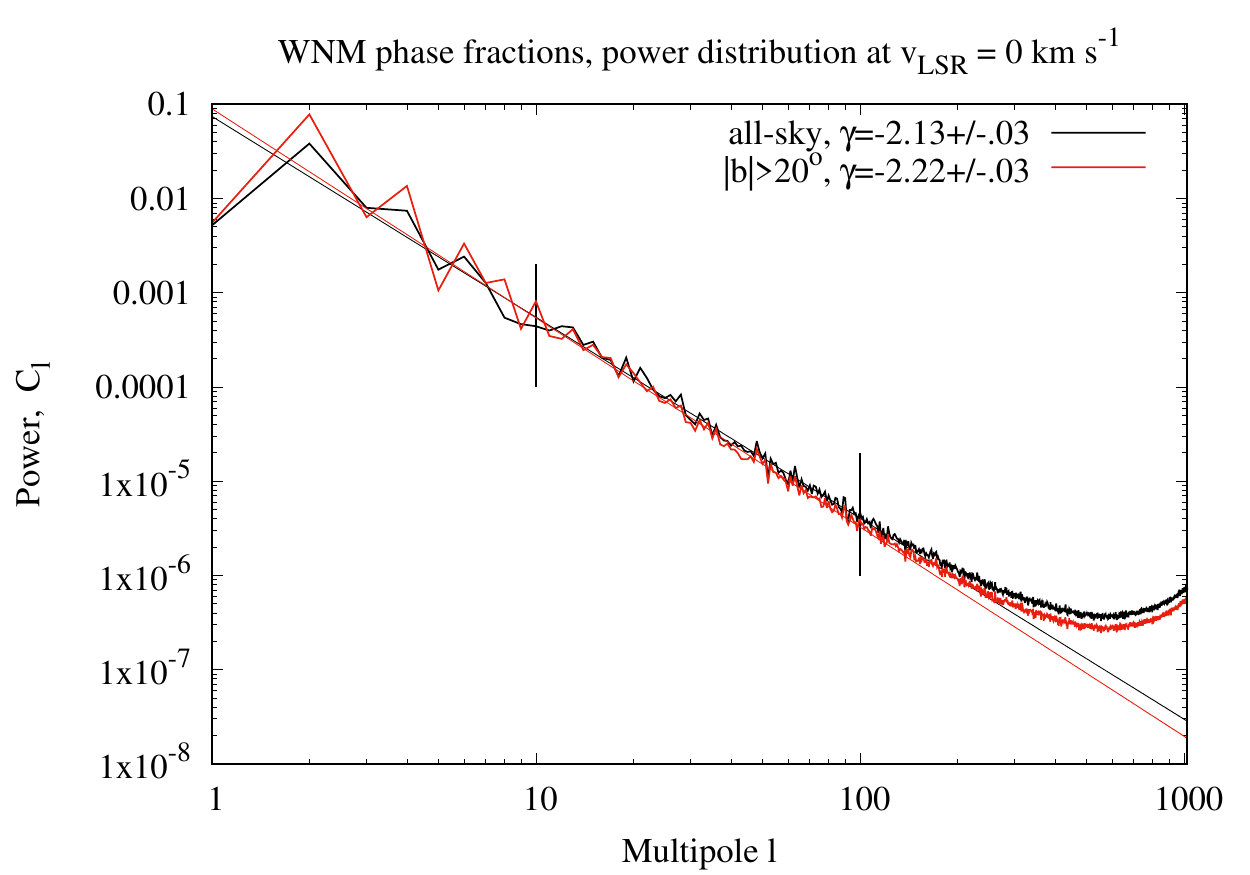}
   \includegraphics[width=9cm]{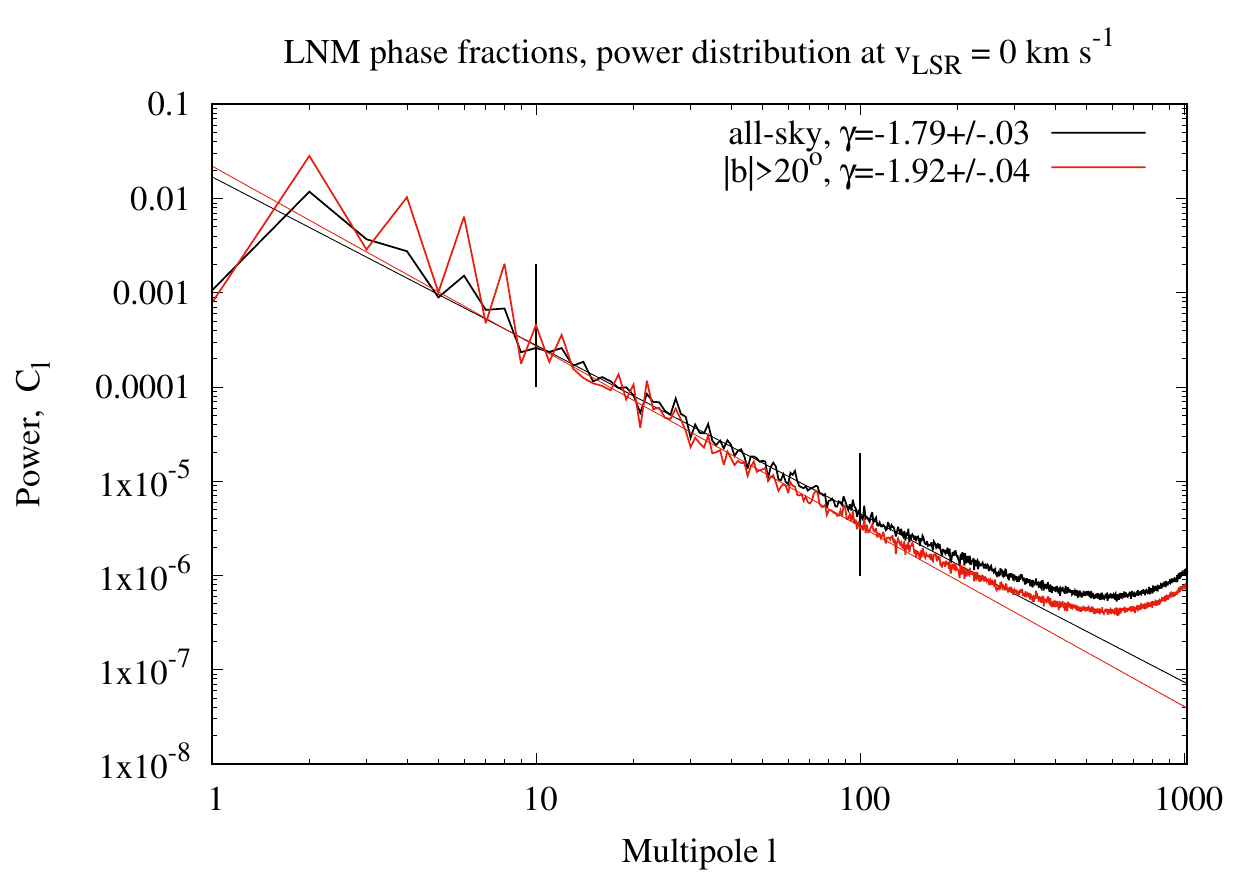}
   \includegraphics[width=9cm]{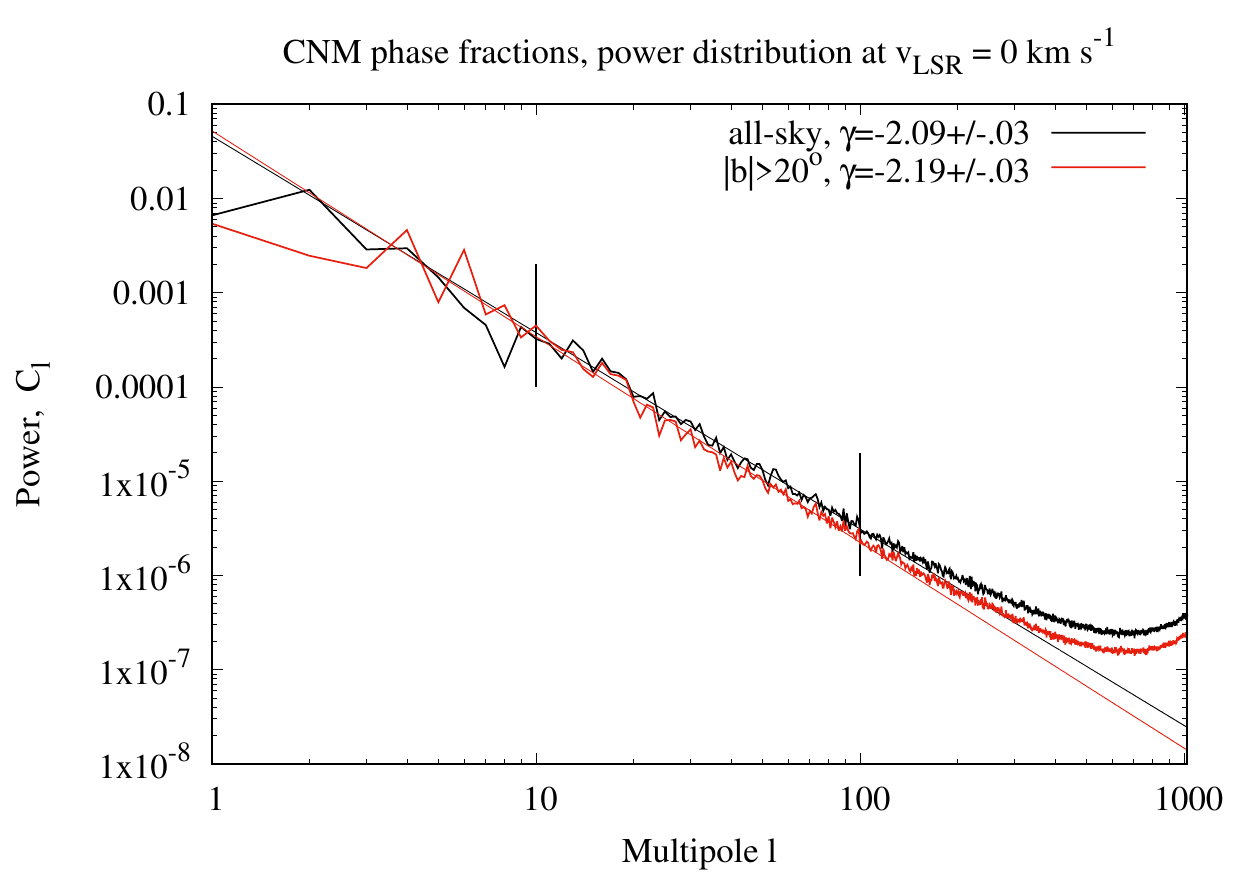}
   \includegraphics[width=9cm]{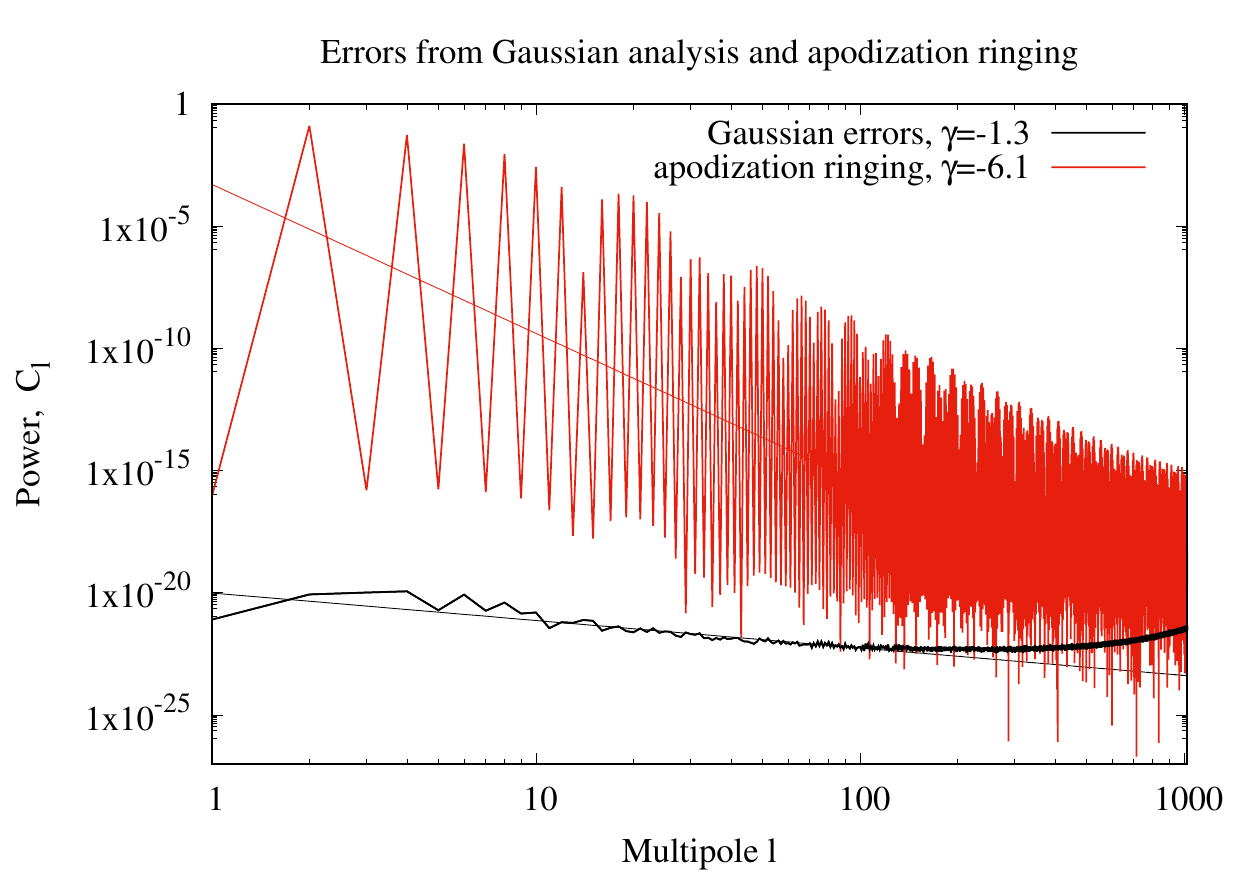}
   \caption{Power distributions for phase fractions in different
     \hi\ phases at $v_{\mathrm{LSR}} = 0 $ \kms.  Top left: WNM; top
     right: LNM; bottom left: CNM. Black lines show all-sky data, red
     lines are for $|b| > 20 \deg$. Spectral indices are derived at 
     $10 < l < 100$ as indicated by the vertical lines. Bottom right:
     Uncertainties from the Gaussian decomposition (black) and ringing
     caused by apodization (red). }
   \label{Fig_f_Gauss_0}
\end{figure*}

%=========================================================================

\subsection{Optical depth effects}
\label{tau}

The \hi\ survey data we analyzed  may suffer from optical depth
effects. For the CNM   the optical depth is expected to increase
with decreasing spin temperature, though a general physically significant
relationship cannot be established \citep{Heiles2003}. An increase in
the optical depth for the clumpy CNM can lead to a systematical
underestimation of the observed column densities and this implies that
the power at high multipoles is underestimated. In turn, observed power
spectra may be steepened artificially.

We use an empirical correction derived by \citet{Lee2015} from Arecibo
observations in direction to the Perseus molecular cloud which is
consistent with data from \citet{Heiles2003} and which was confirmed
later by \citet{Murray2018a} and \citet{Murray2018b}. Observed column densities
$N_\mathrm{Hobs}$ are accordingly biased by a factor
\begin{equation}
f = \mathrm{log_{10}}( N_\mathrm{Hobs}/ 10^{20})\ (0.25 \pm 0.02) + (0.87 \pm 0.02)
\label{eq:Tau}
,\end{equation}
and we apply a correction $N_\mathrm{Hcorr} = f \cdot N_\mathrm{Hobs}$
for $f > 1$, this concerns CNM with $N_\mathrm{Hobs} \ga 0.6~
10^{20}\ {\rm cm^{-2}}$. Using this correction we recalculate the power
spectra for $v_{\mathrm{LSR}} = 0 $ \kms, as shown in
Fig. \ref{Fig_Gauss_0}. We find in all cases that power spectra for the
CNM and the total \hi\ get flatter. However, the resulting changes of the
spectral indices are typically less than half of the rather low formal
uncertainties of the fit. The spectral indices are within the errors
unaffected by optical depth effects; only in one case do we find a slight
change. In the case of the total all-sky \hi\ column densities the index
changes from $\gamma = -2.776 \pm 0.004 $ to $\gamma = -2.768 \pm 0.003
$. We conclude that optical depth effects cannot explain the observed
steepening of the single-channel power spectra close to zero
velocities. Optical depth effects are noticeable at high column
densities in the Galactic plane. High latitude power spectra are
essentially unaffected and we draw our conclusions predominantly from
these data.

\subsection{Power spectra for CNM, LNM, and WNM phase fractions}
\label{Gauss_f}

The power spectra discussed in Sect. \ref{Gauss_NH} are caused by
  column density fluctuations of individual \hi\ phases. Hence there are
  two effects competing with each other, fluctuations in column density
  and in phase composition. We want to determine power spectra merely
  related to individual \hi\ phases, and therefore consider  average phase
  fractions along the line of sight. 

Phase fractions $f_{\mathrm{P}}(_{v1}^{v2})$ depend in general on the
velocity range and, following \citet{Kalberla2018}, for $ v1 <
v_{\mathrm{LSR}} < v2$ are defined as
\begin{equation}
f_{\mathrm{P}}(_{v1}^{v2}) = \frac { \int_{v1}^{v2} T_\mathrm{bP}(v_{\mathrm{LSR}})
  \delta v_{\mathrm{LSR}} } {\int_{v1}^{v2} T_\mathrm{b}(v_{\mathrm{LSR}})\delta
  v_{\mathrm{LSR}} }
\label{EQ_f}
,\end{equation}
where $T_\mathrm{b}$ is the observed brightness temperature, while
$T_\mathrm{bP}$ stands for the brightness temperature contribution from
phase P, P = CNM, LNM, or WNM. Since the dependences on column
  densities cancel in this expression, the weights for the decomposition
  into different phases are identical for all lines of
  sight. Derived multiphase power spectra are essentially deconvolved
  for column density effects.  For a visualization of the spatial column
  density distributions for different phases in comparison to associated
  phase fractions, we refer to \citet[][Figs. 9 and 10]{Kalberla2018}.

In Fig. \ref{Fig_f_Gauss_0} we show
power spectra from single-channel phase fractions at the velocity
$v_{\mathrm{LSR}} = 0$ \kms.
Comparing the power spectra from Fig. \ref{Fig_f_Gauss_0} with those from 
Fig. \ref{Fig_Gauss_0} shows that now the spectra for all-sky and high
latitudes are in  far better agreement than before, but spectra for $|b|
> 20 \deg$ are slightly steeper than the all-sky spectra. We observe
significantly increased power at multipoles $l \la l_{\mathrm
  {crit}}$. In the case of phase fractions the cosecant latitude effect of
the Galactic disk is removed and possible biases (see Sect. \ref{outer})
for low multipoles are minimized.  The spectral indices for the phase
fractions are $-1.8 > \gamma > -2.2$ and are significantly flatter than
$-2.14 > \gamma > -2.48$ for the column density distributions discussed
in Sect. \ref{Gauss_NH}. This flattening is most pronounced for the LNM
and can be explained by the absence of column density fluctuations.

Phase fractions need to sum up to one, the bottom panel on the right 
side of Fig. \ref{Fig_f_Gauss_0} shows in the all-sky case the
completely negligible power spectrum for deviations caused by the
Gaussian decomposition (black). We note that this power spectrum now reflects
 the increased uncertainties at low multipoles from baseline
uncertainties at the field boundaries discussed in Sect. \ref{Noise}.
The red curve shows apodization errors at high latitudes. They cause
a considerable ringing, but fortunately such unavoidable errors are also
not significant since they drop off very fast at high multipoles.

Phase fractions, defined by Eq. \ref{EQ_f} are useful to study the gas
composition independent from its total column density. These data do not
represent any reasonable model for the multiphase \hi\ distribution, but
are useful to consider  how strictly power spectra depend on the
phase composition. Phase transitions cause shallow power spectra for $l
\la 100$ and enhanced power at higher multipoles.

\section{Velocity dependent narrowband spectral indices}
\label{veldependent}

%=========================================================================

\begin{figure}[th] %%  5
   \centering
   \includegraphics[width=9cm]{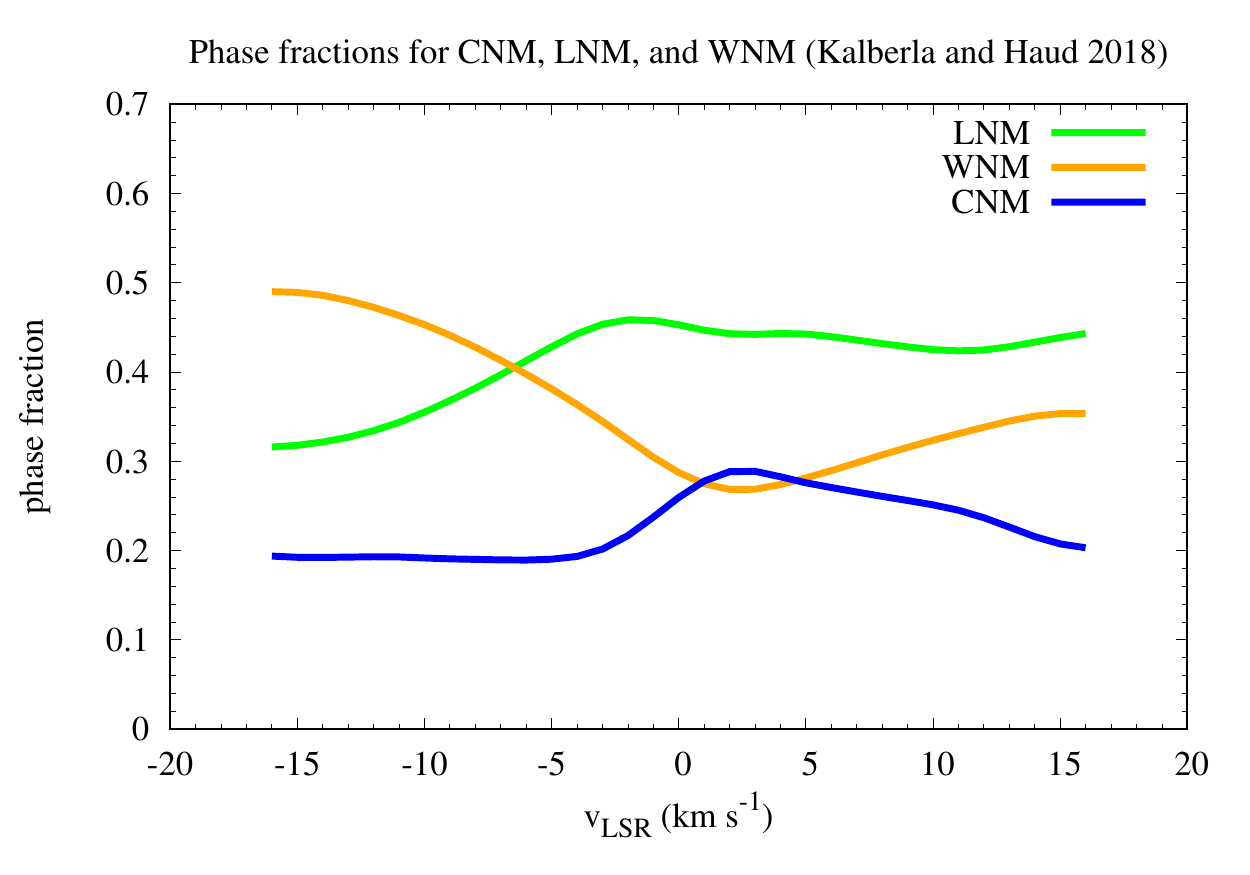}
   \caption{Velocity distribution of average \hi\ phase fractions for
     CNM, LNM, and WNM using unpublished data from \citet{Kalberla2018}.  }
   \label{Fig_VelPhase}
\end{figure}

%=========================================================================

We consider now velocity dependences in the spectral indices.  Figure
\ref{Fig_VelPhase} shows the average distribution of phase fractions for
the local ISM at high latitudes in the velocity range $-16 <
v_{\mathrm{LSR}} < 16 $ \kms\ \citep[][Sect. 3]{Kalberla2018}. There are
considerable fluctuations in  the phase fractions and an imprint on the
power distribution for individual phases at these velocities may be
possible.

\begin{figure}[th] %%  6
    \centering
    \includegraphics[width=9cm]{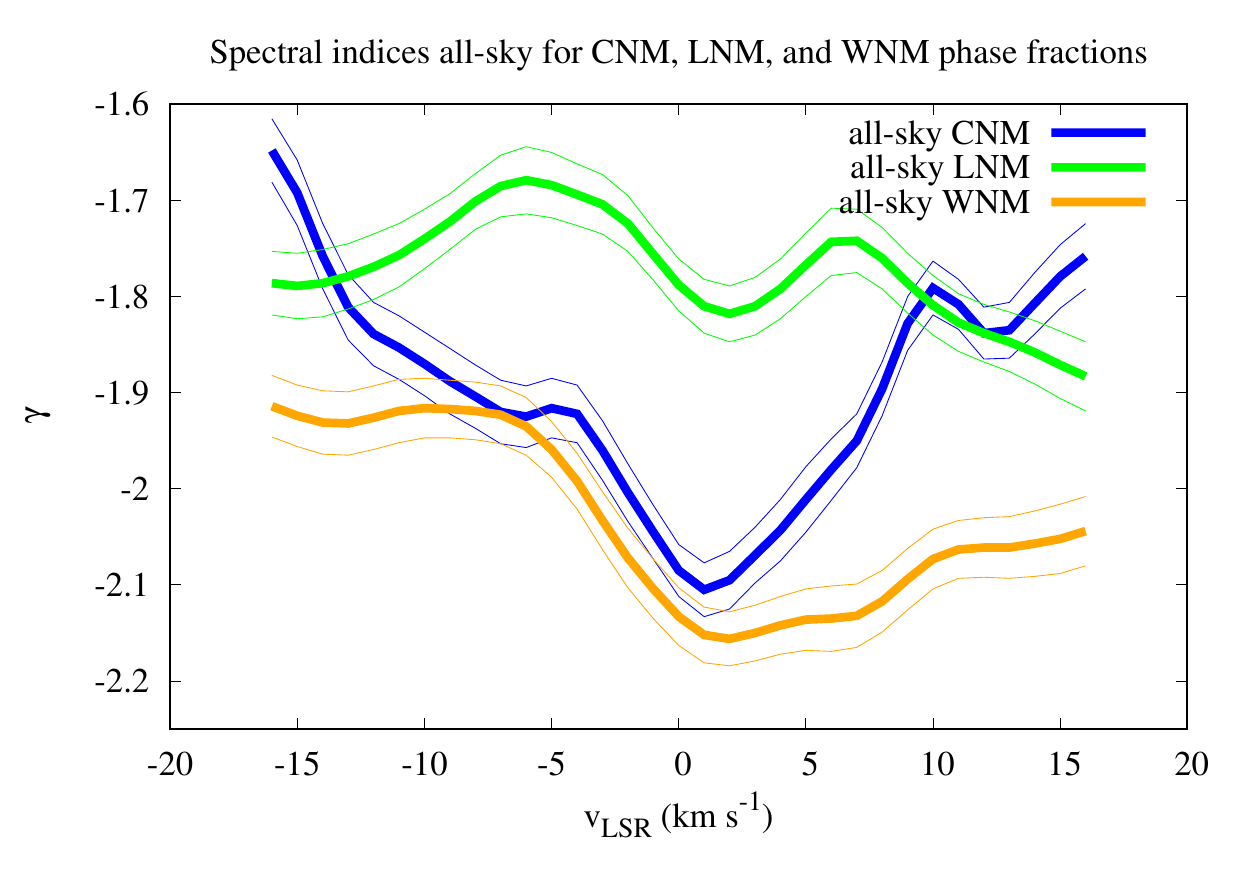}
    \caption{All-sky velocity dependences of spectral indices for
      \hi\ phase fractions.  The thin lines represent the scatter for
      one-sigma uncertainties. }
   \label{Fig_X_f_gamma}
\end{figure}

\begin{figure}[th] %%  7
    \centering
    \includegraphics[width=9cm]{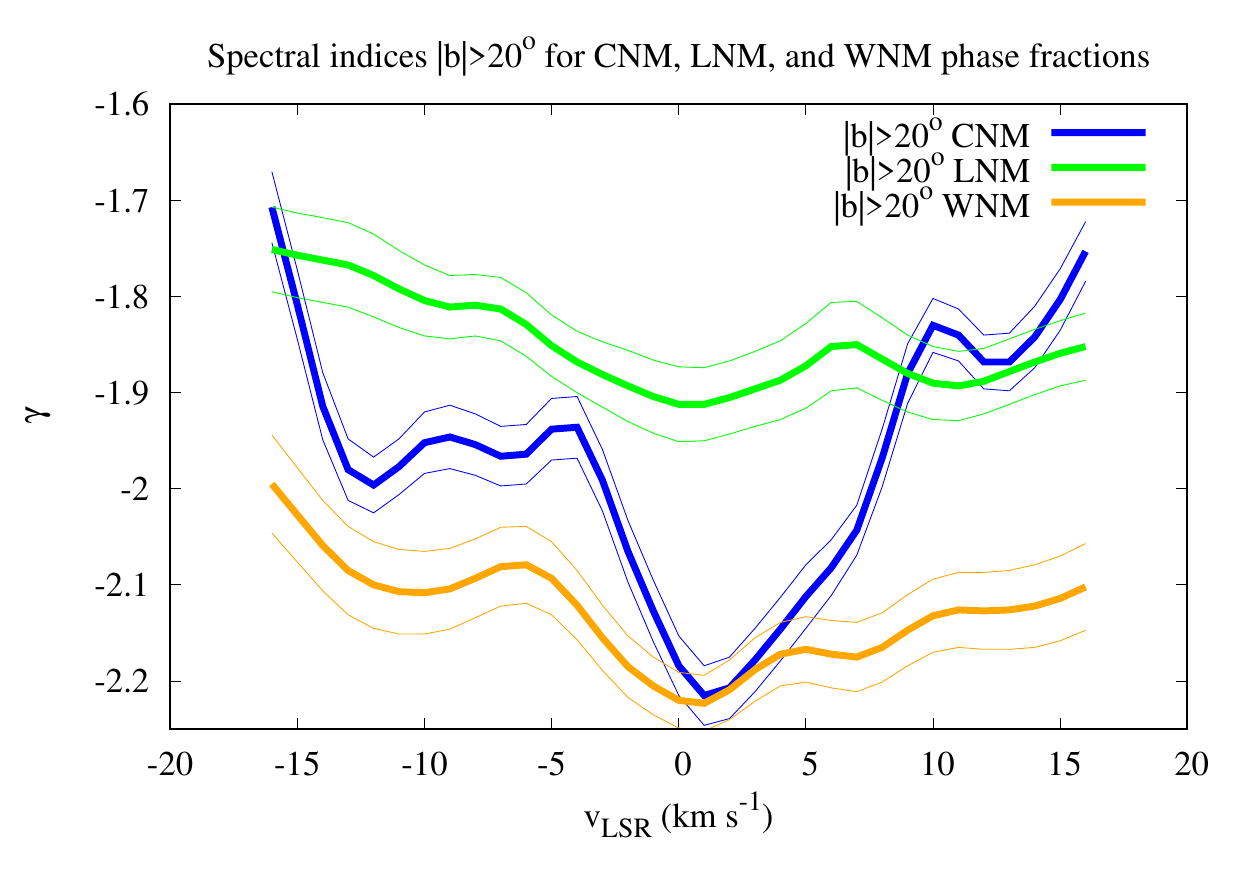}
    \caption{High latitude velocity dependences of spectral indices for
      \hi\ phase fractions.  The thin lines represent the scatter for
      one-sigma uncertainties.  }
   \label{Fig_X_f_gamma_high}
\end{figure}

We consider first the model distributions of phase fractions that are
unaffected by column density fluctuations. The velocity dependences of
narrowband spectral indices for multipoles $10 < l < 100$ are shown in
Fig. \ref{Fig_X_f_gamma} for all-sky and in
Fig. \ref{Fig_X_f_gamma_high} for $|b| > 20 \deg$. We observe a general
steepening of the spectral indices close to $ v_{\mathrm{LSR}} \sim 1$
\kms. The strongest effect is found for the CNM. The enhancement of the
average CNM phase fraction at this velocity (Fig. \ref{Fig_VelPhase})
clearly leads  to a pronounced steepening of the CNM power
spectra.  The steepening of the spectral indices of the other phases is
noteworthy, but may be a consequence of the coupling of the power spectra
between different phases according to Eq. \ref{eq:crossPower}.

\begin{figure}[th] %%  8
    \centering
    \includegraphics[width=9cm]{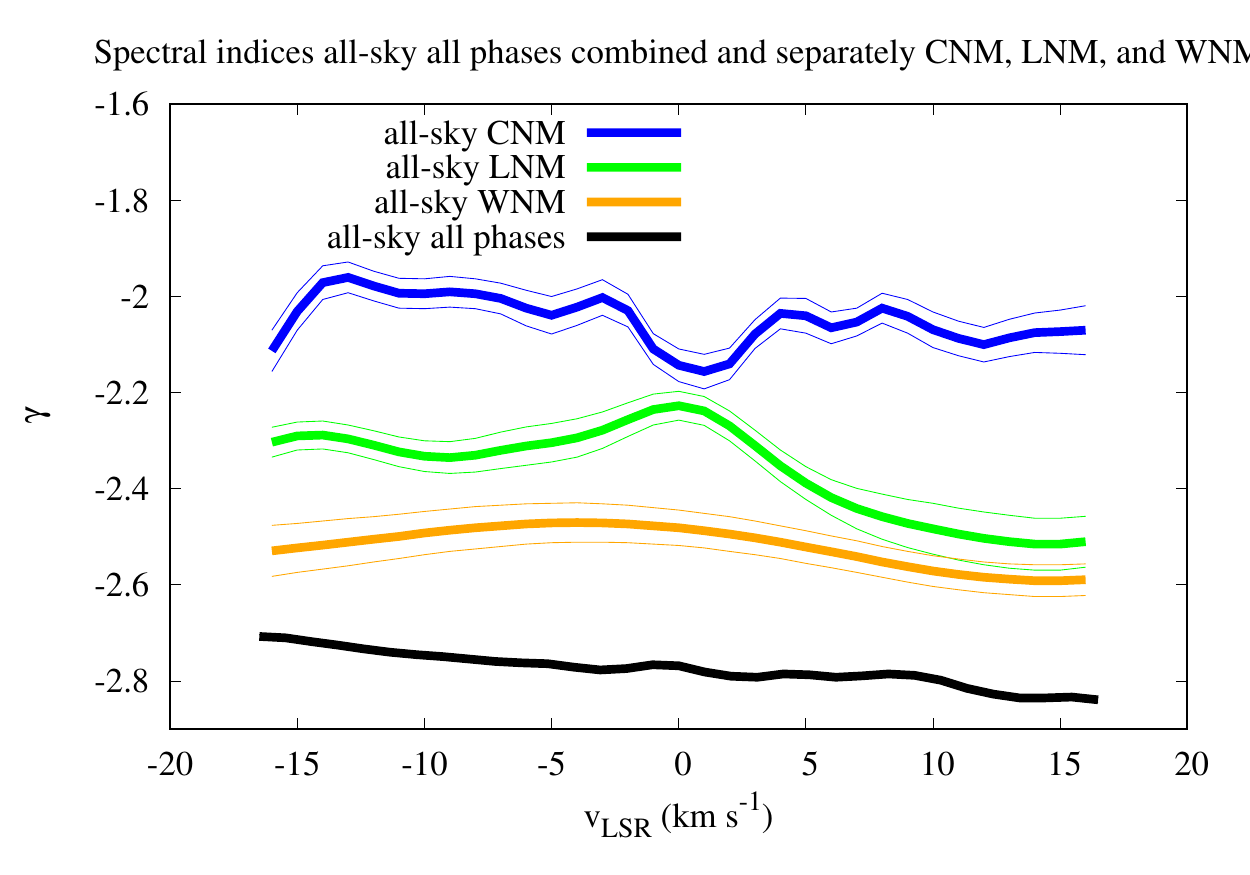}
    \caption{All-sky velocity dependences of spectral indices for
      individual phases, and the total \hi\ distribution (black).
      The thin lines represent the scatter for one-sigma uncertainties when the total \hi\ distribution (black) uncertainties are
      within the thickness of the line. }
   \label{Fig_X_gamma}
\end{figure}

\begin{figure}[th] %%  9
    \centering
    \includegraphics[width=9cm]{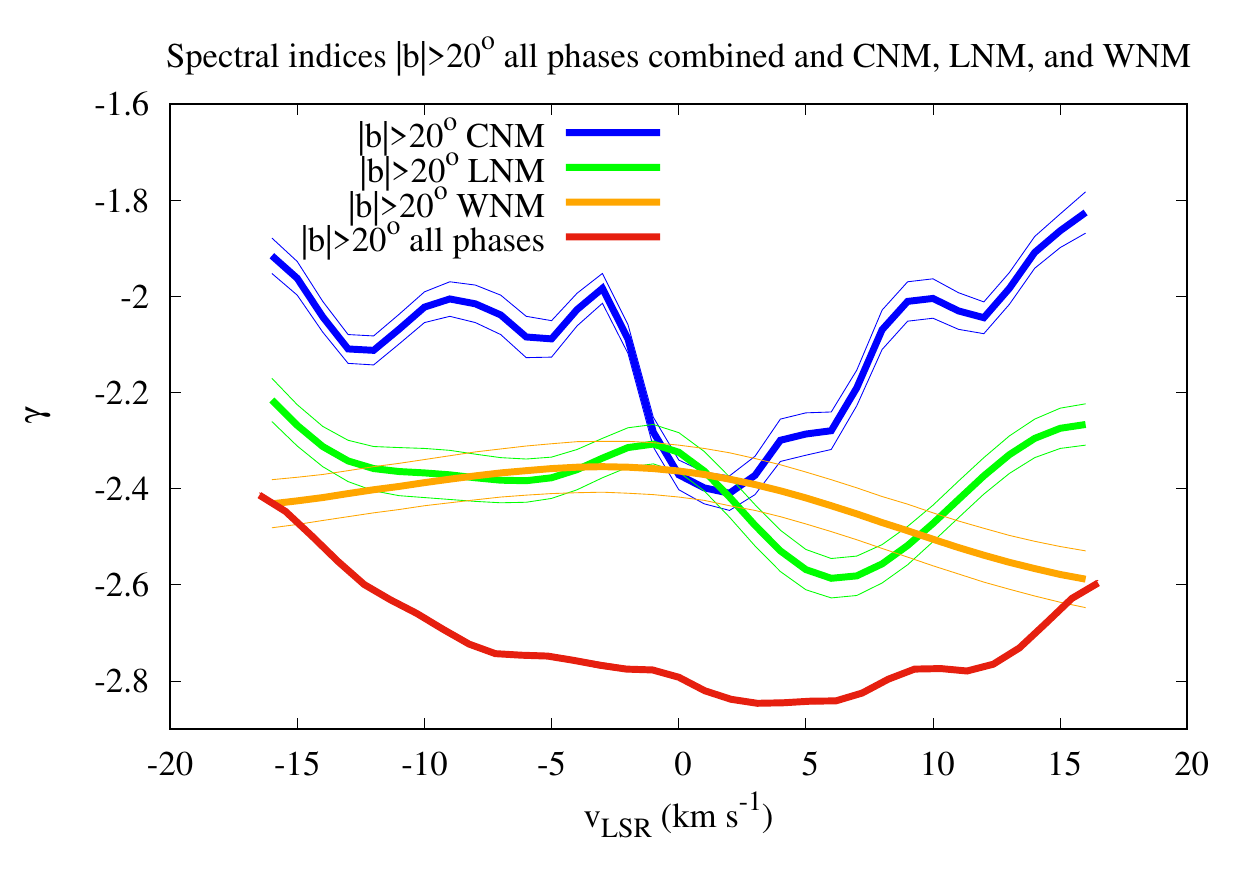}
    \caption{High latitude velocity dependences of spectral indices for
      individual phases, and the total \hi\ distribution (red).
      The thin lines represent the scatter for one-sigma uncertainties when the total \hi\ distribution (red) uncertainties are
      within the thickness of the line.  }
   \label{Fig_X_gamma_high}
\end{figure}

Next we consider the power spectra derived from the Gaussian components
of different \hi\ phases. Figures \ref{Fig_X_gamma} and \ref{Fig_X_gamma_high}
show still local minima for the CNM spectral indices, however the
spectral indices for the other phases are now mostly anti-correlated;
the WNM has a particularly smooth distribution. Spectral indices derived
from power spectra of the total observed \hi\ also show  smooth changes.
For high latitudes (Fig. \ref{Fig_X_gamma_high}, red) we find a
steepening in anti-correlation to the WNM, but with local minima that are
affected by the CNM and LNM. In the all-sky case
(Fig. \ref{Fig_X_gamma}, black) we observe a smooth trend, but influences
from the CNM and LNM are weak.

%=====================================================================

%=====================================================================

%=====================================================================

%========================================================================

%========================================================================
%=========================================================================

Velocity dependences for the spectral indices have been observed before
(\citet{Kalberla2016b} and \citet{Kalberla2017}).  Recently, 
\citet{Choudhuri2018} have  presented data that show a steepening of the
spectral indices at velocities that are considered to be representative
for the CNM. Spectral indices for $^{13}$CO 2--1 and $^{12}$CO 3--2
channel maps in the Perseus cloud were observed by \citet{Sun2006} with
a steepening of the indices that can be described as an average
structure of the index spectrum similar to the line profile.

Numerical studies of the condensation of WNM into CNM structures caused
by turbulence and thermal instabilities were conducted by
\citet{Saury2014}. They found that turbulence plays a key role in
the structure of the cold medium; it can induce the formation of CNM
when the WNM is pressurized and put it in a thermally unstable state.  They
found evidence for subsonic turbulence with a shallow power index
$\gamma \sim -2.4$. Recent high resolution simulations by
\citet{Wareing2019} have shown that the thermal instability dynamically
can form sheets and filaments on typical scales of 0.1 to 0.3 pc, explaining that as the structure grows in the simulation, the density
power spectrum rapidly rises and steepens.
\citet[][Fig. 10]{Wareing2019} derive $\gamma \sim -3,$ but the
steepening is most prominent on scales below 1 pc.

The preferred width of the filaments in the range 0.1 to 0.3 pc is
consistent with unresolved cold small-scale structures observed by
\citet{Clark2014} and \citet{Kalberla2016}. Furthermore, these
cold filaments have  characteristic radial velocities close
to zero \kms.  The single-phase power spectra show a strong increase at
the corresponding multipoles $l \sim 1023$.

%=========================================================================
\begin{figure}[th] %%  10
    \centering
    \includegraphics[width=9cm]{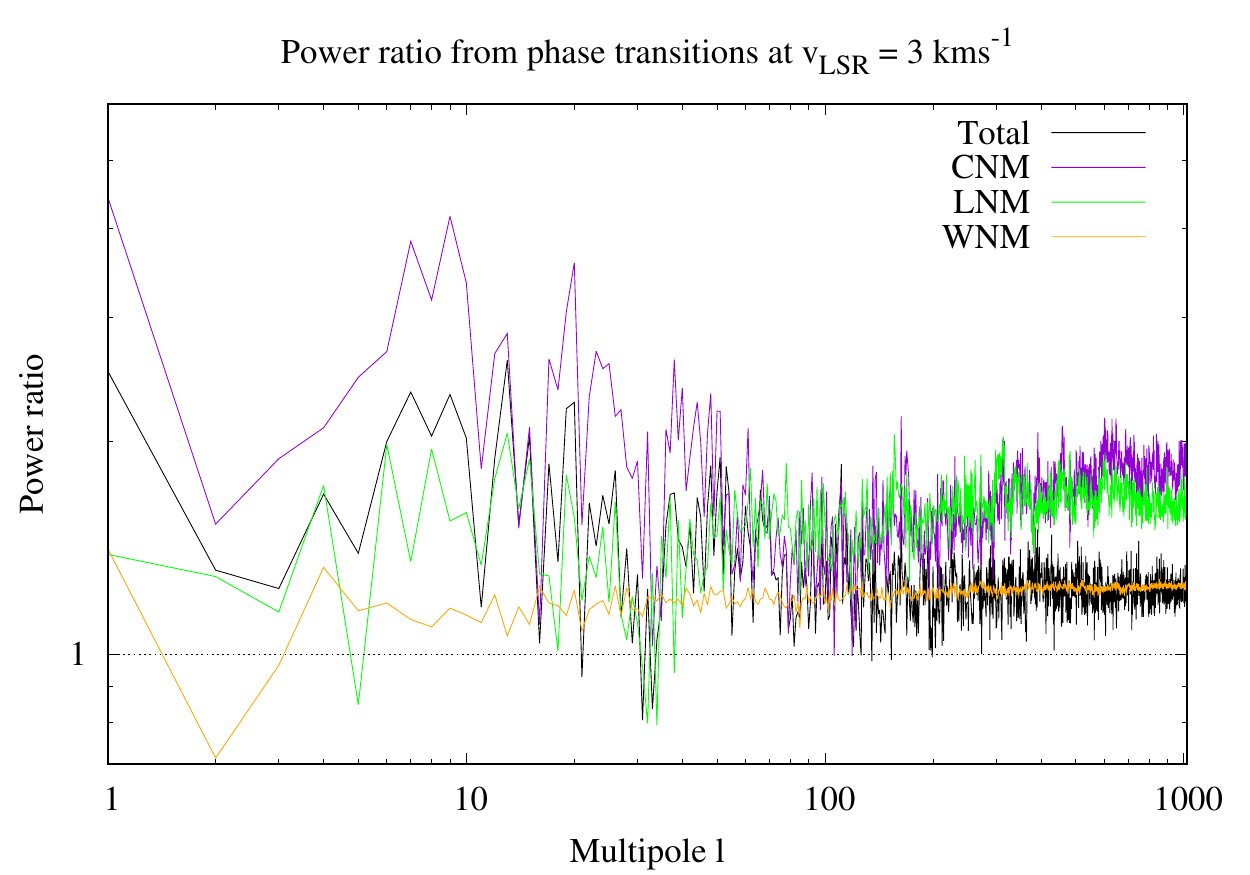}
    \caption{Power ratio $\mathfrak{R}(l)$ according to Eq. \ref{eq:R}.
      The ON velocity is $ v_{\mathrm{LSR}} = 3$ \kms; OFF is calculated
      as the average of single-channel slices at $ v_{\mathrm{LSR}} = -3$
      and 7 \kms. The dotted line indicates $\mathfrak{R}(l) = 1$. }
   \label{Fig_TI_plot_1}
\end{figure}
%=========================================================================

\begin{figure}[th] %%  11
    \centering
    \includegraphics[width=9cm]{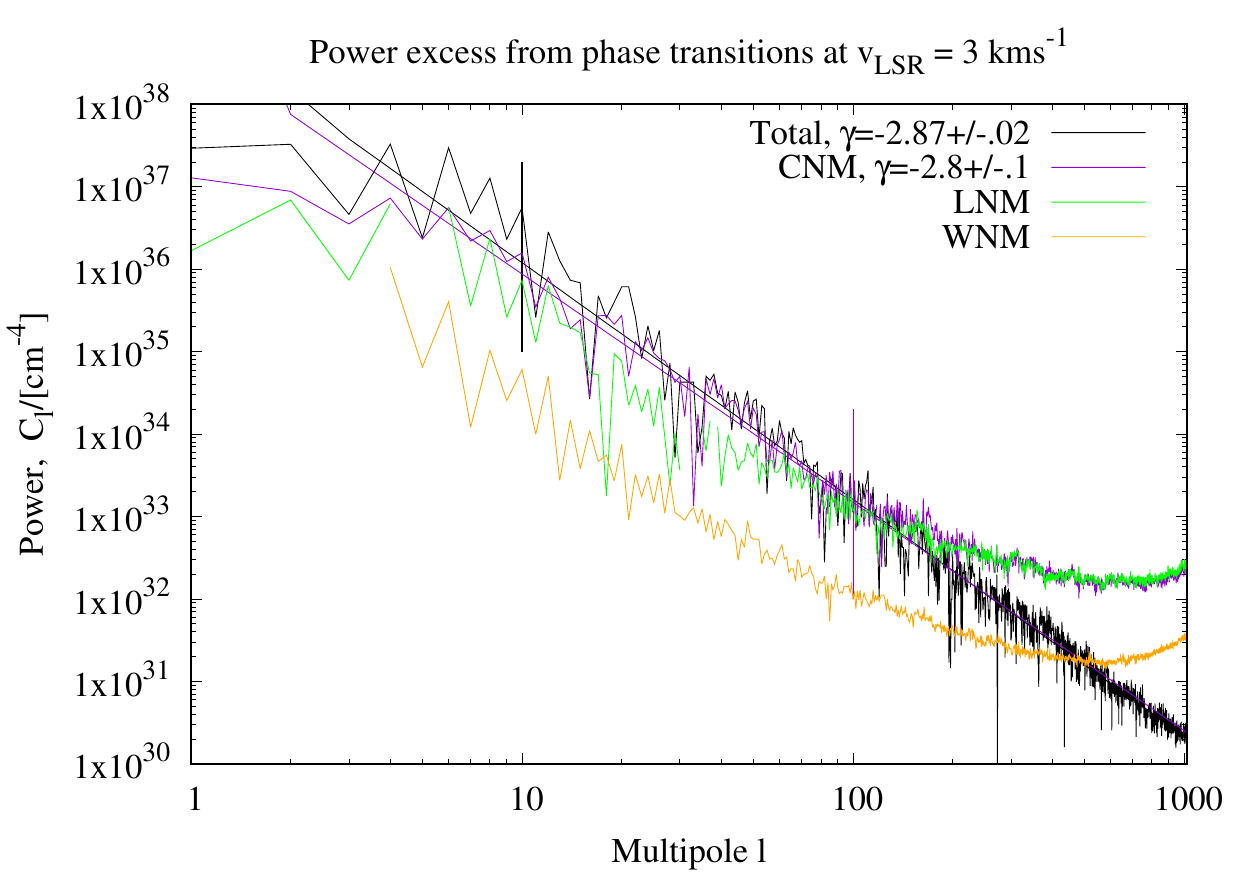}
    \caption{Power ratio $\mathfrak{E}(l)$ according to Eq. \ref{eq:E}.
      The ON velocity is $ v_{\mathrm{LSR}} = 3$ \kms; OFF is calculated
      as the average of single-channel slices at $ v_{\mathrm{LSR}} = -3$
      and 7 \kms. The spectral index for the total HI is calculated for
      $l > 10$; in the case of the CNM for $10 < l < 100$, the horizontal
      lines indicate these limits. }
   \label{Fig_TI_plot_2}
\end{figure}
%=========================================================================

%\subsection{Power distribution affected by phase transitions}
%\label{TI}

A steepening of thin velocity slice \hi\ power spectra in a narrow
velocity range, associated with a decrease in the WNM fraction and the
coexistence of cold anisotropic CNM filamentary structures was reported 
before for three fields at intermediate latitudes by
\citet[][Sect. 6]{Kalberla2017}. The interpretation was that phase
transitions, indicated by changes in \hi\ phase fractions, modify the
narrowband power distribution. For a determination of dependences
between power spectra and phase transitions these authors proposed  
isolating particular cold \hi\ components from the rest of the
\hi\ distribution in order to differentiate between ON velocities at the
line centers and appropriate OFF velocities in the line
wings. Accordingly we define here
\begin{equation}
\mathfrak{R}(l) = P_{\rm ON}(l) / P_{\rm OFF}(l)
\label{eq:R}
\end{equation}
as the power ratio and 
\begin{equation}
\mathfrak{E}(l) = P_{\rm ON}(l) - P_{\rm OFF}(l)
\label{eq:E}
\end{equation}
as the power excess, relating single-channel ON and OFF power
distributions; $P_{\rm OFF}(l)$ is defined as the average single-channel
power at two OFF positions. 

The choice of the ON velocity is obvious, we observe at $
v_{\mathrm{LSR}} = 3$ \kms\ the highest CNM and at the same time the
lowest WNM ratio (Fig. \ref{Fig_VelPhase}). The high latitude data also show
at this velocity  a well-defined minimum for the CNM spectral index
(Fig. \ref{Fig_X_gamma_high}). From this plot  the definition for
the OFF velocities at $ v_{\mathrm{LSR}} = 3 \pm 5$ \kms\ is also well
defined, the CNM power law steepening is limited to the range $\Delta
v_{\mathrm{LSR}} \sim 10 $ \kms; we refer to Sects. \ref{coherence} to
\ref{linewidth} for further discussion of the velocity spread.

Figure \ref{Fig_TI_plot_1} shows the power ratios $\mathfrak{R}(l)$.
We observe $\mathfrak{R}(l) \ga 1$ for all multipoles except $l \sim
30$. For the WNM $\mathfrak{R}(l)$ is flat but LNM and CNM show
enhancements at high multipoles. Strong fluctuations are also observable
at multipoles $10 \la l \la 50$.

Figure \ref{Fig_TI_plot_2} shows the power excess $\mathfrak{E}(l)$.
In comparison to Fig. \ref{Fig_Gauss_0} ($\gamma = -2.776 \pm 0.004 $
for the total \hi\ and $\gamma = -2.37 \pm 0.03 $ for the CNM) we
observe steep power spectra with $\gamma = -2.87 \pm 0.02$ for the total
\hi\ for $l \ga 10$, and in particular $\gamma = -2.8 \pm 0.1$ for the
CNM for $10 \la l \la 100$. This is  by far  the steepest power law
derived by us for the CNM, and we interpret the steep excess power
distribution $\mathfrak{E}(l)$ for the CNM as induced by phase
transitions. Thermal instabilities obviously also  affect  the power
spectra for WNM and LNM; they are no longer straight, but bent up
systematically with increasing power for high multipoles. The different
phases are correlated according to Eq. \ref{eq:crossPower}, but the power
spectrum for the total \hi\ remains straight.

In agreement with \citet{Kalberla2017} we find strong evidence that the
power distribution in the local ISM is significantly steepened by phase
transitions. The CNM is filamentary and anisotropic on small scales
(\citet{Kalberla2016b} and \citet{Kalberla2017}), but even the large CNM power ratios
$\mathfrak{R}(l)$ at low multipoles appear to be linked to large-scale
filaments, oriented along the magnetic field as observed by
\citet{Clark2014} and \citet{Kalberla2016}.

\section{Spectral indices and velocity channel widths}
\label{velwidth}

%========================================================================

\begin{figure}[th] %%  12
    \centering
    \includegraphics[width=9cm]{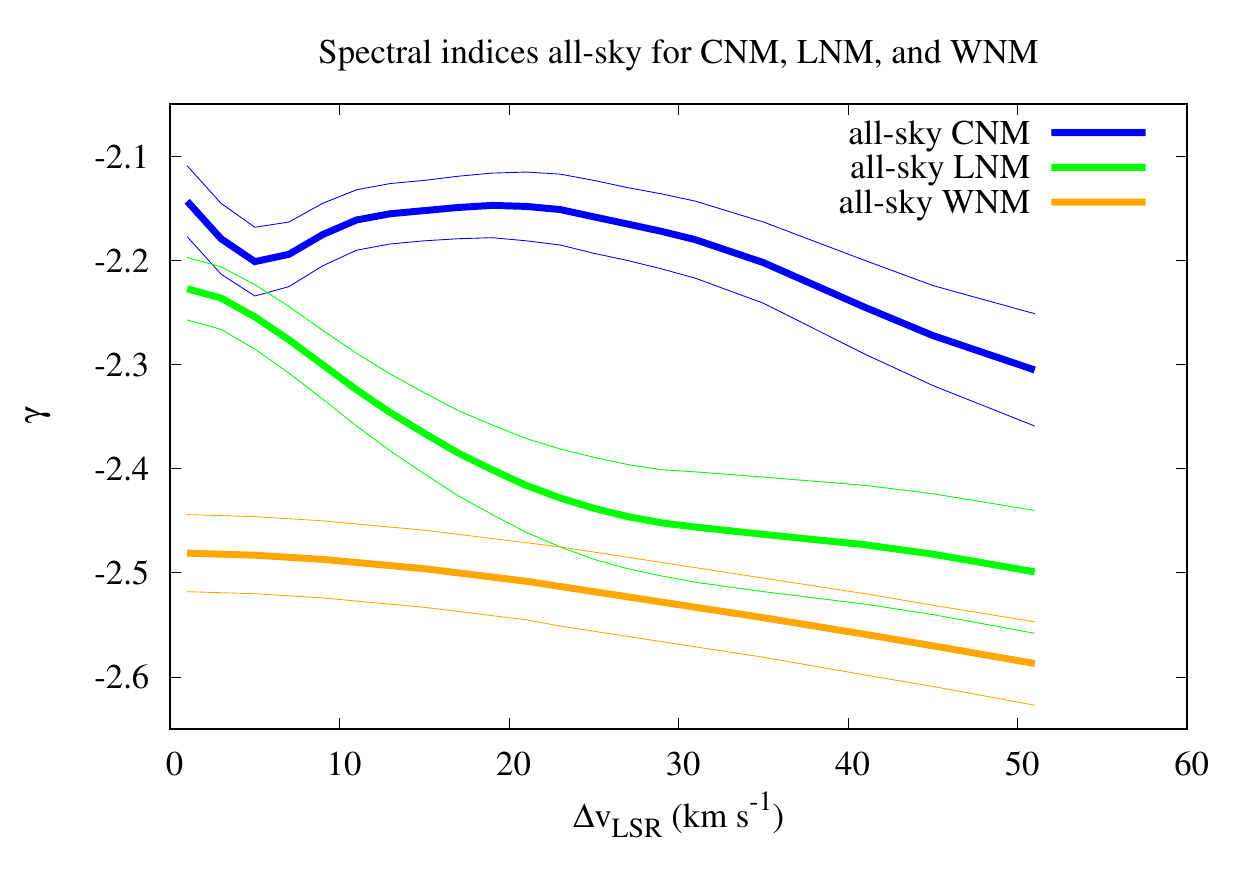}
    \caption{All-sky dependences of spectral indices for \hi\ phase
      fractions and velocity channel width.  The thin lines represent
      the scatter for one-sigma uncertainties. }
   \label{Fig_X_broad}
\end{figure}
%========================================================================

\begin{figure}[th] %%  13
    \centering
    \includegraphics[width=9cm]{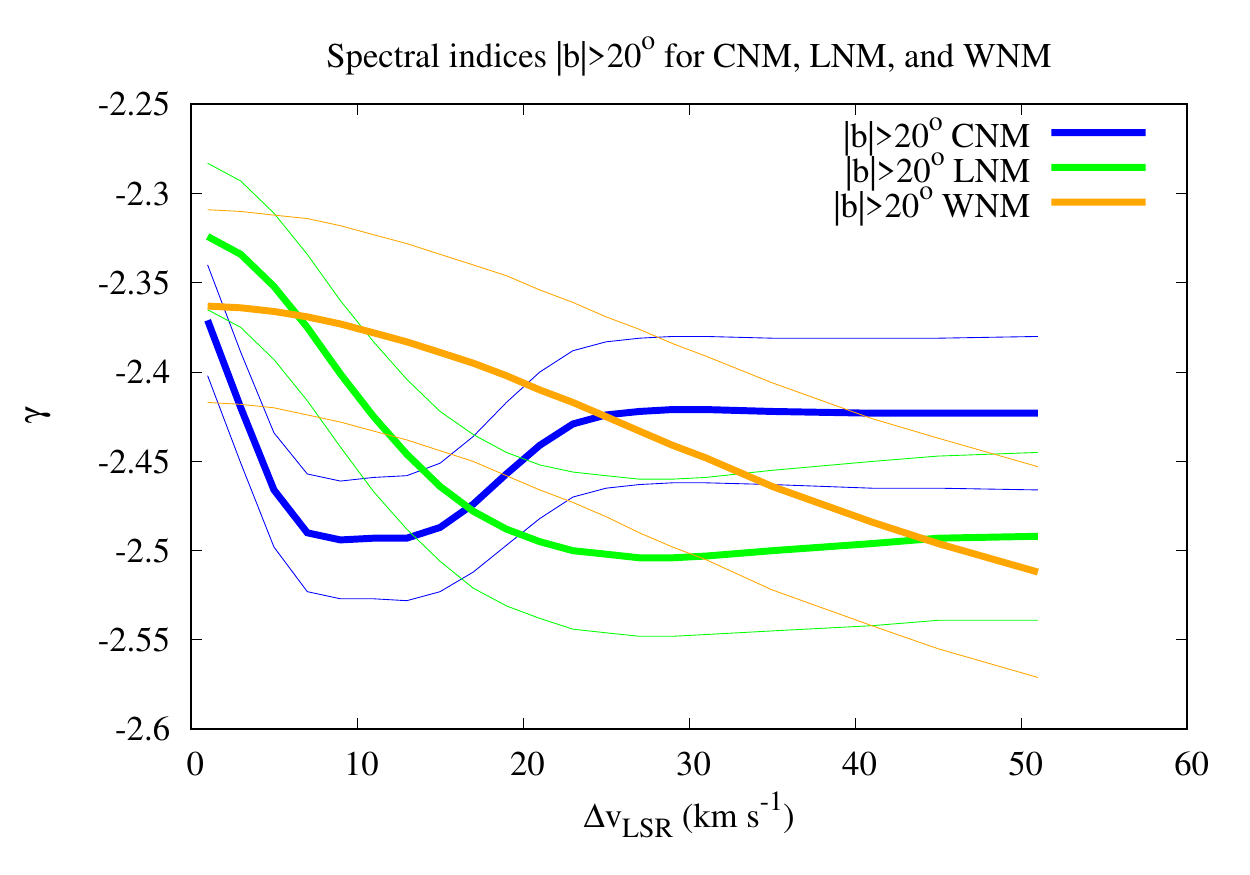}
    \caption{Dependences of spectral indices for \hi\ phase fractions
      and velocity channel width at $|b| > 20 \deg$.  The thin lines
      represent the scatter for one-sigma uncertainties. }
   \label{Fig_X_broad_high}
\end{figure}

%========================================================================

In this section we determine spectral indices for individual \hi\ phases
after integrating the column density distributions over several channels
with a total full width at half maximum (FWHM) of $\Delta
v_{\mathrm{LSR}}$. Figures
\ref{Fig_X_broad} and \ref{Fig_X_broad_high} show the results. From
the all-sky data we note that the spectral indices for WNM and LNM 
steepen continuously. The WNM has broad lines, resulting in a slow
change in spectral index. The characteristic LNM line widths are
narrower, corresponding to faster changes of the slope with $\Delta
v_{\mathrm{LSR}}$.  For the CNM the situation is more complex. In the case
 of the all-sky data we note a prominent minimum at $\Delta
v_{\mathrm{LSR}} = 5 $ \kms\ and a rise afterward. At high latitudes the
situation is similar, but the CNM spectral index now has  a broad minimum
around $\Delta v_{\mathrm{LSR}} \sim 10 $ \kms.

From the observed \hi\ column densities or the sum of all phases we
derive the spectral indices shown  in Fig. \ref{Fig_broad}. Spectral
indices decrease in a similar way for the all-sky data and for $|b| > 20
\deg$ until $\Delta v_{\mathrm{LSR}} = 16$ \kms. Afterward the indices
diverge; they  fall for all-sky data but rise for $|b| > 20
\deg$. This  result is unexpected, and  an explanation is needed.

\subsection{Velocity channel analysis}
\label{VCA}

%=========================================================================
\begin{figure}[th] %%  14
    \centering
    \includegraphics[width=9cm]{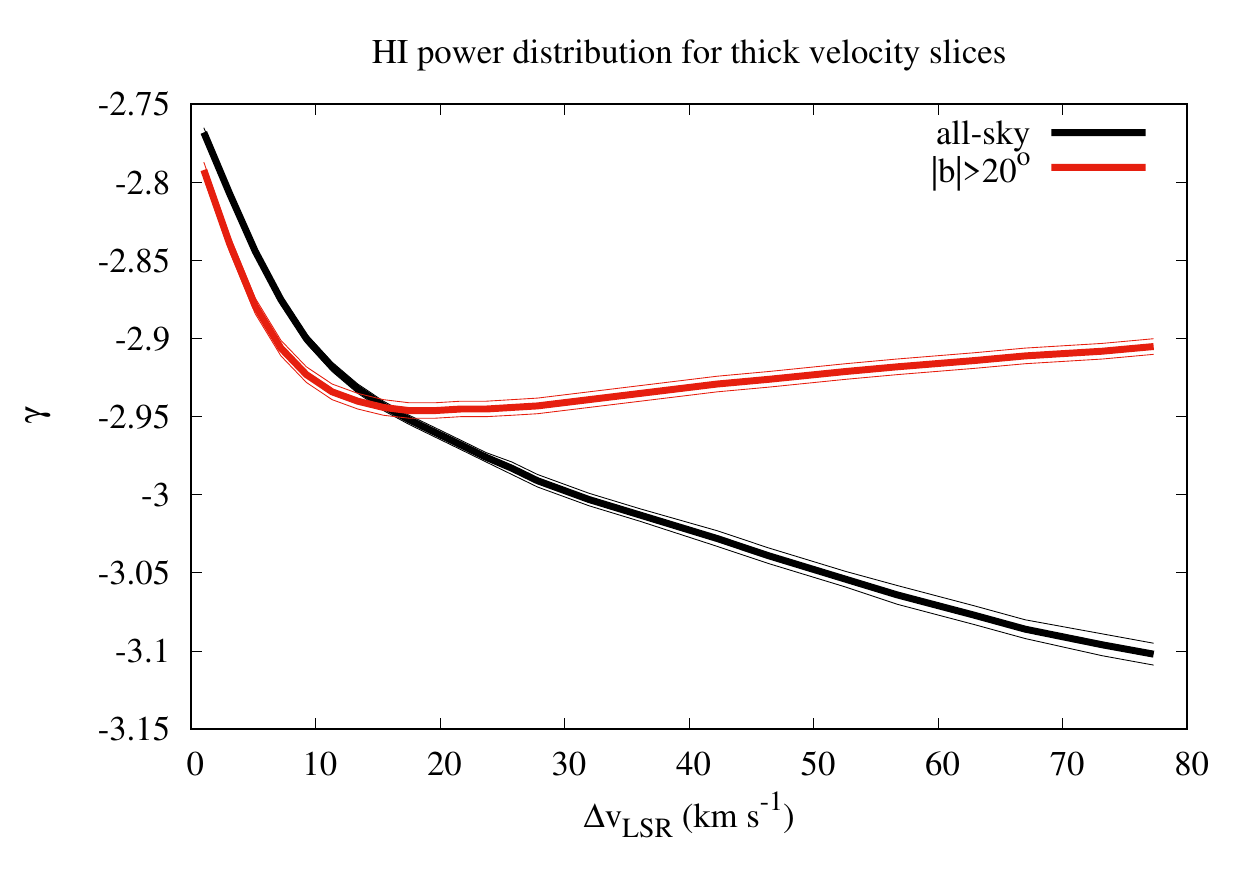}
    \caption{All-sky dependences of spectral indices for observed
      \hi\ column densities as a function of the velocity channel width.
      The thin lines represent the scatter for one-sigma
      uncertainties. }
   \label{Fig_broad}
\end{figure}

%=========================================================================

Power spectra extracted from narrowband data (thin velocity slices) are
affected by turbulence in two ways. Density fluctuations have an imprint
on the observed column densities, but  velocity fluctuations also need to
be taken into account. To separate effects caused by fluctuations
in density or velocity, the basic idea of VCA is to increase the width
$\Delta v_{\mathrm{LSR}}$ of the velocity slice. Velocity fluctuations
should average out, and when the velocity window $\Delta
v_{\mathrm{LSR}}$ is broad enough that all internal velocities are
covered (thick velocity slice) the observed emissivity should eventually
be dominated by density fluctuations \citep[][Eq 25]{Lazarian2000}. The
expected minimum velocity width is $\Delta v_{\mathrm{LSR}} \sim 17$
\kms\ \citep[][Sect. 4.3 ]{Lazarian2000}, the typical FWHM thermal line
width for the WNM.  Starting with a thin velocity slice and successively increasing
 the velocity width, a gradual steepening of the spectral
index should be observed until convergence. For our application
to \hi\ column densities we observe a steepening (Fig. \ref{Fig_broad}),
but in the case of high latitude data only until $\Delta v_{\mathrm{LSR}}
\sim 16$ \kms, which is the maximum linewidth for consistent
results. Considering the CNM in Figs. \ref{Fig_X_broad} and
\ref{Fig_X_broad_high} we find a steepening, but subsequently the
spectral index flattens again. The question arises whether VCA is
applicable to our data. We  take up the discussion later in
  Sect. \ref{VCArevisited}.

 Spectral indices close to $\gamma \sim -2.95 $ and within the
  uncertainties of $\Delta \gamma \la 0.1$ independent of velocity
  widths $0.82 < \Delta v_{\mathrm{LSR}} < 21.4$ \kms\ have been
  reported  by \citet[][Tables 3 and 4, and Fig. 29]{Khalil2006}. A
  constant spectral index is inconsistent with VCA;  these authors
  discuss in their Sect. 6 some other observations that do not show
  the steepening predicted by VCA. \citet{Yuen2019} note
  that the analysis of the data agrees well with VCA predictions and
  revise the \citet{Khalil2006} indices in their Table 3 to
  $\gamma_{\mathrm{thin}} = -2.6$ and $\gamma_{\mathrm{thick}} =
  -3.4$. No details are given how this result was obtained, but the
  authors refer to a forthcoming more detailed paper.

%=========================================================================

\begin{figure}[th] %%  15
    \centering
    \includegraphics[width=9cm]{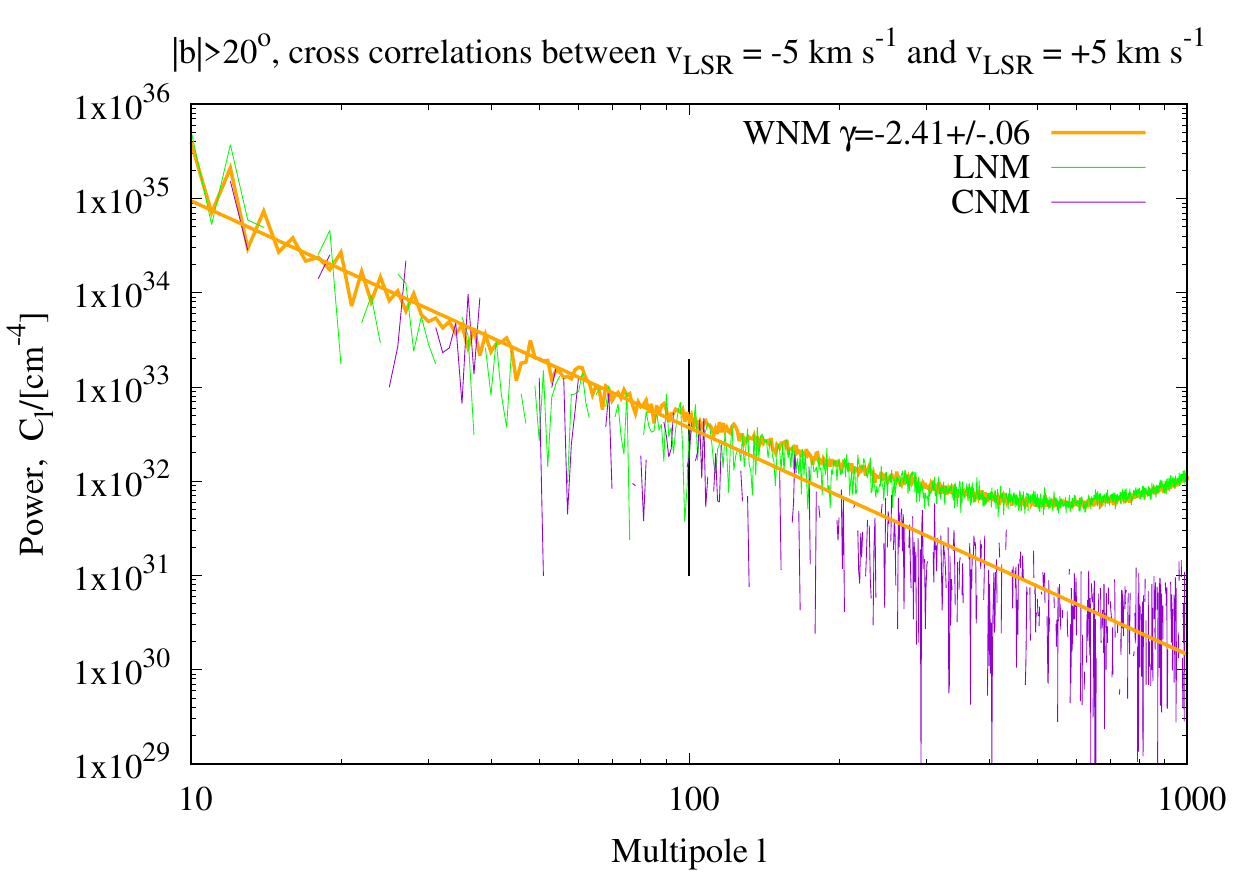}
    \caption{Cross-correlations at latitudes $|b| > 20 \deg$ between
      two channels at $v_{\mathrm{LSR}} = -5$ and $v_{\mathrm{LSR}} = +5$
      for the WNM, LNM, and CNM. }
   \label{Fig_cross_55a}
\end{figure}
%=========================================================================

\subsection{Velocity channel cross-correlations}
\label{velcross}

Trying to better understand the dependences between turbulent power
distribution and the velocity window width $\Delta v_{\mathrm{LSR}}$,
we calculate cross-correlations for single-channels in the wings of the
line signal. Hence we cross-correlate channels separated by a distance
$\Delta v_{\mathrm{LSR}}$ in velocity. This is a differential approach.
The cross power spectra that we use to study internal velocity
dependences of individual \hi\ phases are, according to the
Wiener-Khintchine theorem, the Fourier transforms of the cross-correlation function at particular characteristic lags in velocity.
In the case of a signal from a genuine turbulent medium, the cross
  power spectra should inform us on how well correlated the individual
  \hi\ phases remain throughout their velocity distribution. 

We consider high latitudes first; all-sky data is deferred
  until Sect. \ref{rotation}. Figure \ref{Fig_cross_55a} shows us the
cross-correlations between two channels at velocities $v_{\mathrm{LSR}}
= -5$ \kms\ and $v_{\mathrm{LSR}} = +5$ \kms. For the WNM we find a well-defined power spectrum with a slope of $\gamma = 2.41 \pm 0.06$ for
multipoles $10 < l < 100$ with enhanced power at high multipoles. For
the LNM no power law can be found for $10 < l < 100$ (only positive
values of $C_l$ can be plotted), but   for this phase we also obtain enhanced
power at high multipoles; the velocity cross power distributions for LNM
and WNM are  almost identical there. In the case of the CNM only a noisy
cross-correlation signal at high multipoles can be seen. In comparison
to the expected contribution of the uncertainties from the Gaussian
decomposition shown in Fig. \ref{Fig_Gauss_0} this signal ($\sim
10^{30}$ cm$^{-4}$) is barely significant. The local minimum of the
spectral index for the CNM phase in Fig. \ref{Fig_cross_55a} at $\Delta
v_{\mathrm{LSR}} = 10$ \kms\ indicates that the correlation of the CNM
is lost. This implies that on average any CNM structure seen in
  channels separated by $\Delta v_{\mathrm{LSR}} \ga 10$ \kms\ is
  unrelated. Homogeneous CNM clouds have an average FWHM line width of
  3.6 \kms\ \citep{Kalberla2018} and according to Sect. \ref{model} only
  3.2 \kms\ for velocities $|v_{\mathrm{LSR}}| \la 5 $ \kms. The
  internal velocity distribution of individual CNM clouds is too narrow
  to cause a limitation of this kind. The implication must be that CNM
  clumps in a turbulent flow decouple from each other if their center
  velocities differ by $\sqrt{10^2-3.6^2} \ga 9$ \kms. In the following
  we  investigate this remarkable result further;  in
  Sect. \ref{Constraints} an explanation is offered from constraints
  considered by \citet{Kolmogorov1941}.  Both the WNM and LNM maintain
  their correlation at a lag of 10 \kms. Their cross power spectra are
  identical for $l \ga 200$; in Fig.  \ref{Fig_X_broad_high} the LNM
  data overplot the WNM data.  Increasing the velocity separation for
  two channels further, we also observe  for the LNM a decorrelation of
  the cross power for $\Delta v_{\mathrm{LSR}} \ga 26$ \kms. This is
  discussed later in Sect. \ref{similar} and demonstrated in
  Fig. \ref{Fig_coherence_LW}. 

\subsection{Constraints on spectral coherence}
\label{coherence}

A different way to explore correlations in the velocity domain is by
defining the spectral coherence $S_l(v_1,v_2)$ between
two single channels at velocities $v_1$ and $v_2$
\begin{equation}
S_l(v_1,v_2)  =  C_l^2 (v_1,v_2) / [  C_l(v_1,v_1) \cdot  C_l(v_2,v_2)  ]
\label{EQ_coherence}
,\end{equation}
%magnitude-squared coherence
where $C_l(v_1,v_2)$ is the cross-correlation between individual PPV
channel maps at velocities $v_1$ and $v_2$ with corresponding autocorrelations $C_l(v_1,v_1)$ and $C_l(v_2,v_2)$. This relation has the
advantage that it is independent of beam smoothing as long as the
\hi\ emission is little affected by noise.  It is clear that $S_l = 1 $ for
$v_1 = v_2$ and the signal decorrelates for increasing channel
separation, but we require for a turbulent flow that some correlation
should be maintained if we want to define eddies as coherent
patterns of flow velocity, vorticity, and pressure.

\begin{figure}[th] %%  16
    \centering
    \includegraphics[width=9cm]{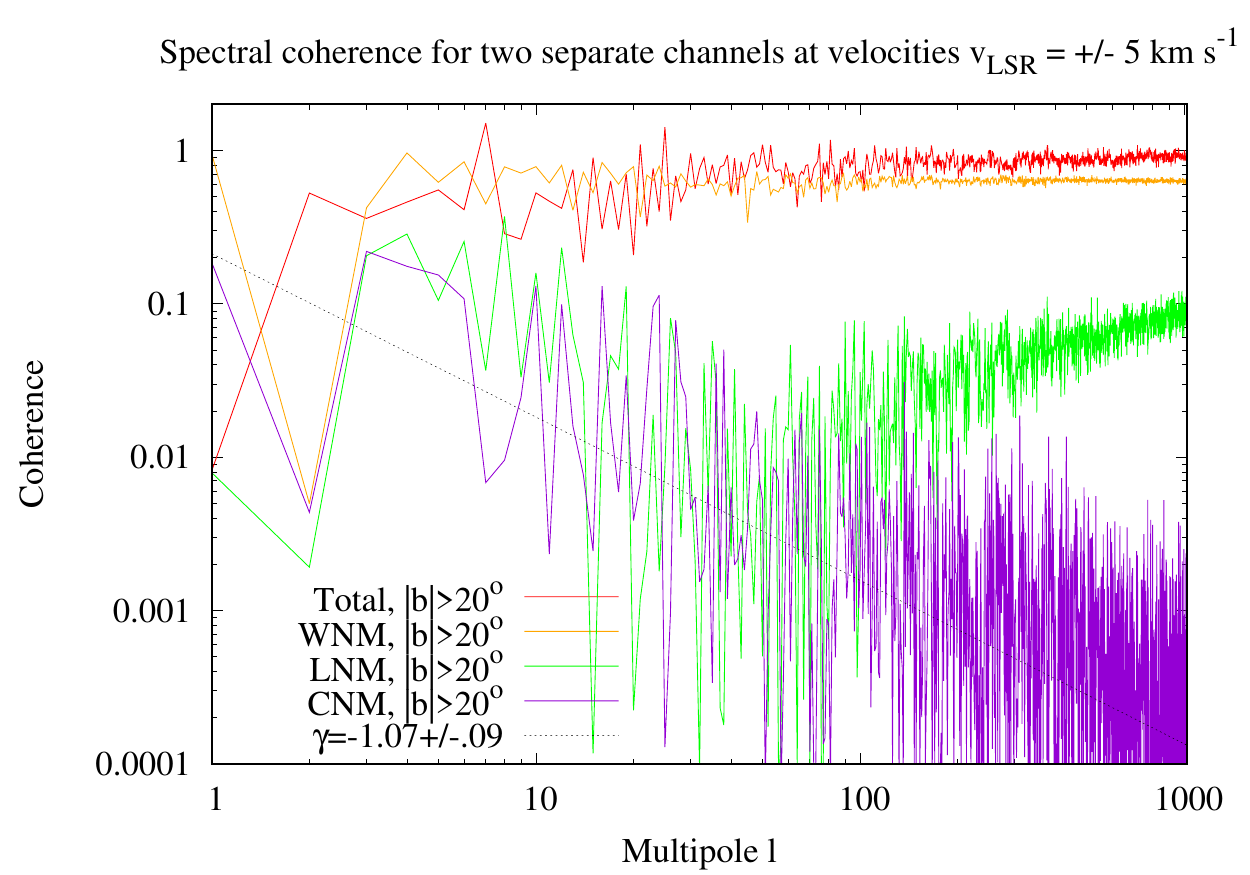}
    \caption{Spectral coherence for velocity slices at $v_{\mathrm{LSR}}
      = -5$ \kms\ and $v_{\mathrm{LSR}} = +5$ \kms\ for the total \hi,
      WNM, LNM, and CNM. }
   \label{Fig_coherence_10}
\end{figure}

We use high latitude data to calculate $S_l(v_1,v_2)$ for the WNM, LNM,
CNM, and the total \hi\ column density distributions at velocities $v_1
= -5$ \kms\ and $v_2 = +5$ \kms. The results are shown in
Fig. \ref{Fig_coherence_10}. The WNM distribution is coherent at all
multipoles $l$. This is expected since the velocity spread $v_2 - v_1 =
10$ \kms\ is small compared to the average FWHM width of 23.3 \kms\ of
the WNM emission \citep{Kalberla2018}. In the case of the LNM with an
average FWHM 9.6 \kms\ some spectral coherence remains at high
multipoles. This behavior is expected for an unstable phase that is
embedded in the WNM, but surrounds a clumpy CNM distribution (see
Sect. \ref{phases}). For the CNM with an average FWHM width of 3.6
\kms\ the fit $\gamma = -1.07 \pm 0.09$ at a channel separation of
$\Delta v_{\mathrm{LSR}} = 10$ \kms\ is within the uncertainties
consistent with random Gaussian noise ($\gamma = -1$). For comparison,
the spectral coherence is also shown for the total multiphase
\hi\ column density (Fig. \ref{Fig_coherence_10}, red). This is 
expected close to one, but interestingly the scatter in $S_l$ is large
compared to the WNM. Clearly the WNM derived from Gaussian components
provides a more consistent presentation of the large-scale
\hi\ distribution.

%=========================================================================

%=========================================================================

\begin{figure}[th] %%  17
    \centering
    \includegraphics[width=9cm]{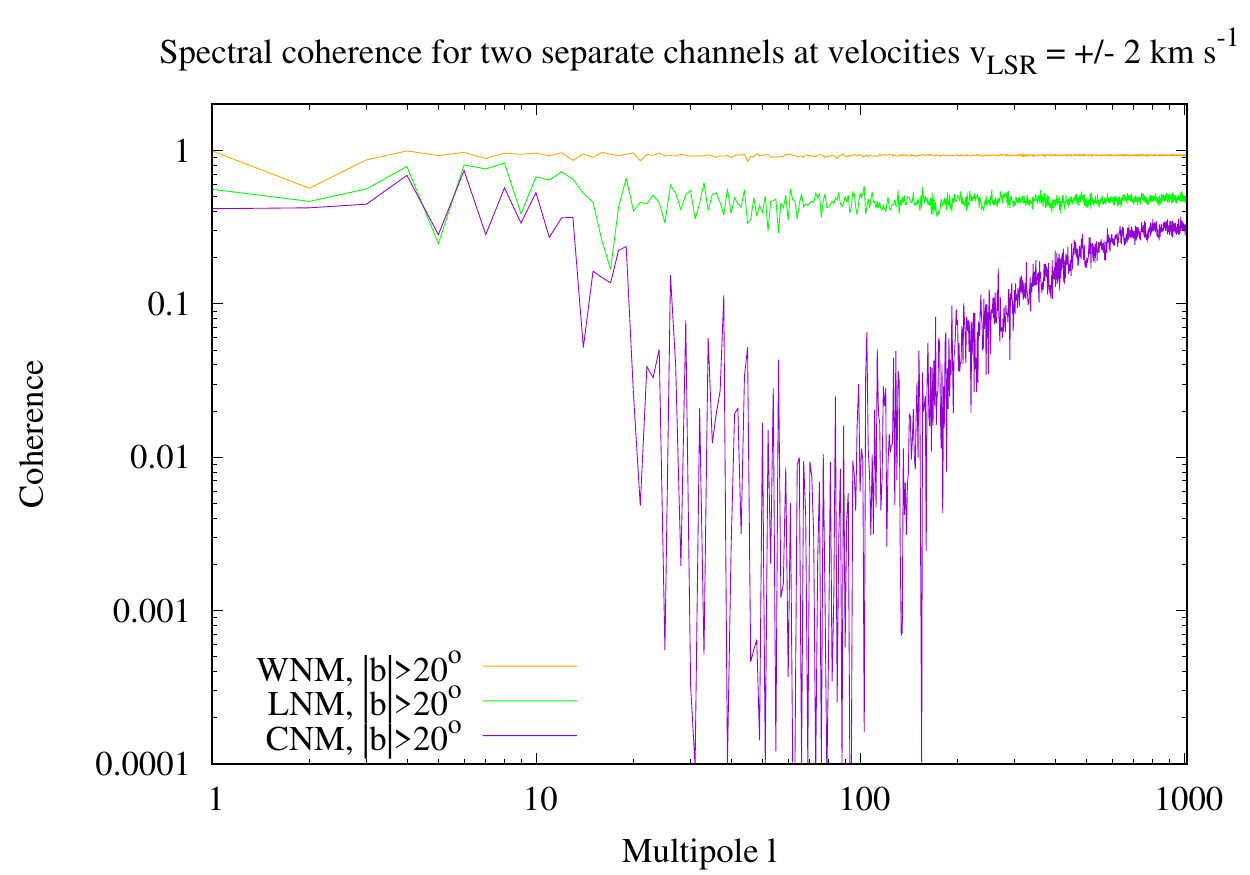}
    \caption{Spectral coherence for velocity slices at $v_{\mathrm{LSR}}
      = -2$ \kms\ and $v_{\mathrm{LSR}} = 2$ \kms\ for the WNM, LNM, and
      CNM. }
   \label{Fig_coherence_4}
\end{figure}
%=========================================================================

We repeat the calculations and determine $S_l$ for $v_2 - v_1 < 10$
\kms; coherence for the CNM is recovered. As an example we use
$S_l(v_1,v_2)$ for $v_1 = -2$ \kms\ and $v_2 = +2$ \kms. This velocity
lag corresponds roughly to the average FWHM width of 3.6 \kms\ for the
CNM. The result is shown in Fig. \ref{Fig_coherence_4}. For the WNM the
situation is similar to Fig. \ref{Fig_coherence_10}, but for the CNM we
find coherence for $l > 100$. This situation can be described as
follows: power in the CNM exists spatially predominantly on small scales,
and structures in velocity space can only be detected for small velocity
lags. Increasing the velocity lag leads to a gradual decorrelation; as
shown in Fig. \ref{Fig_coherence_10}, the spectral coherence is lost
completely for $\Delta v_{\mathrm{LSR}} = 10$ \kms; the CNM in the two
separate channels decouples if this value is exceeded.

For the LNM we find correlation on larger scales (up to smaller
multipoles) and at the same time for larger velocity lags. The LNM
surrounds the CNM, occupying a larger volume and at the same time a
larger velocity spread. Some evidence for this hierarchical structure
was given before by \citet{Kalberla2018}. Increasing the lag $v_2 - v_1$
we find a gradual decorrelation for the LNM, the LNM decouples for $\Delta
v_{\mathrm{LSR}} \ga 23$ \kms.

%=========================================================================
\begin{figure}[th] %%  18
    \centering
    \includegraphics[width=9cm]{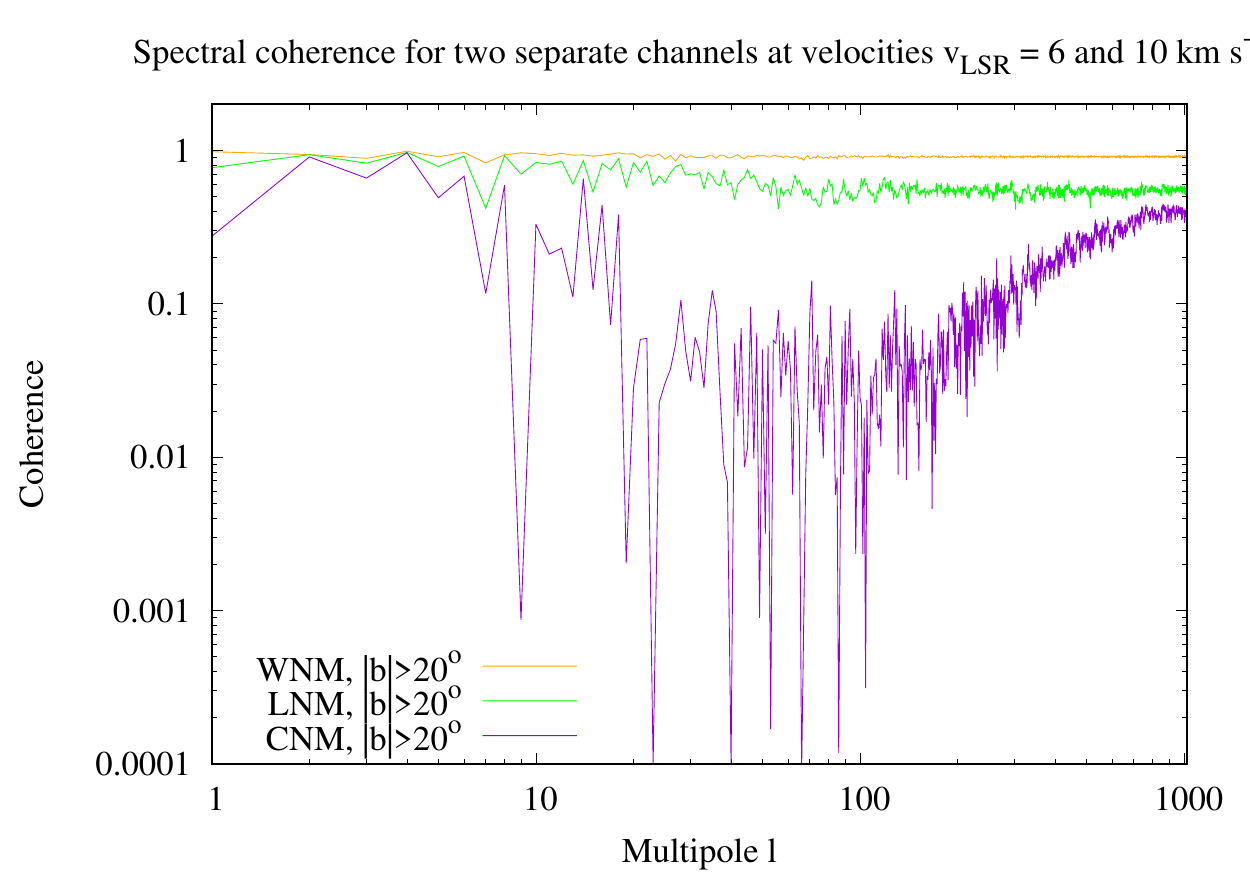}
    \caption{Spectral coherence for velocity slices at $v_{\mathrm{LSR}}
      = 6$ \kms\ and $v_{\mathrm{LSR}} = 10$ \kms\ for the WNM, LNM,
      and CNM. }
   \label{Fig_coherence_8}
\end{figure}
%=========================================================================

\begin{figure}[th] %%  19
    \centering
    \includegraphics[width=9cm]{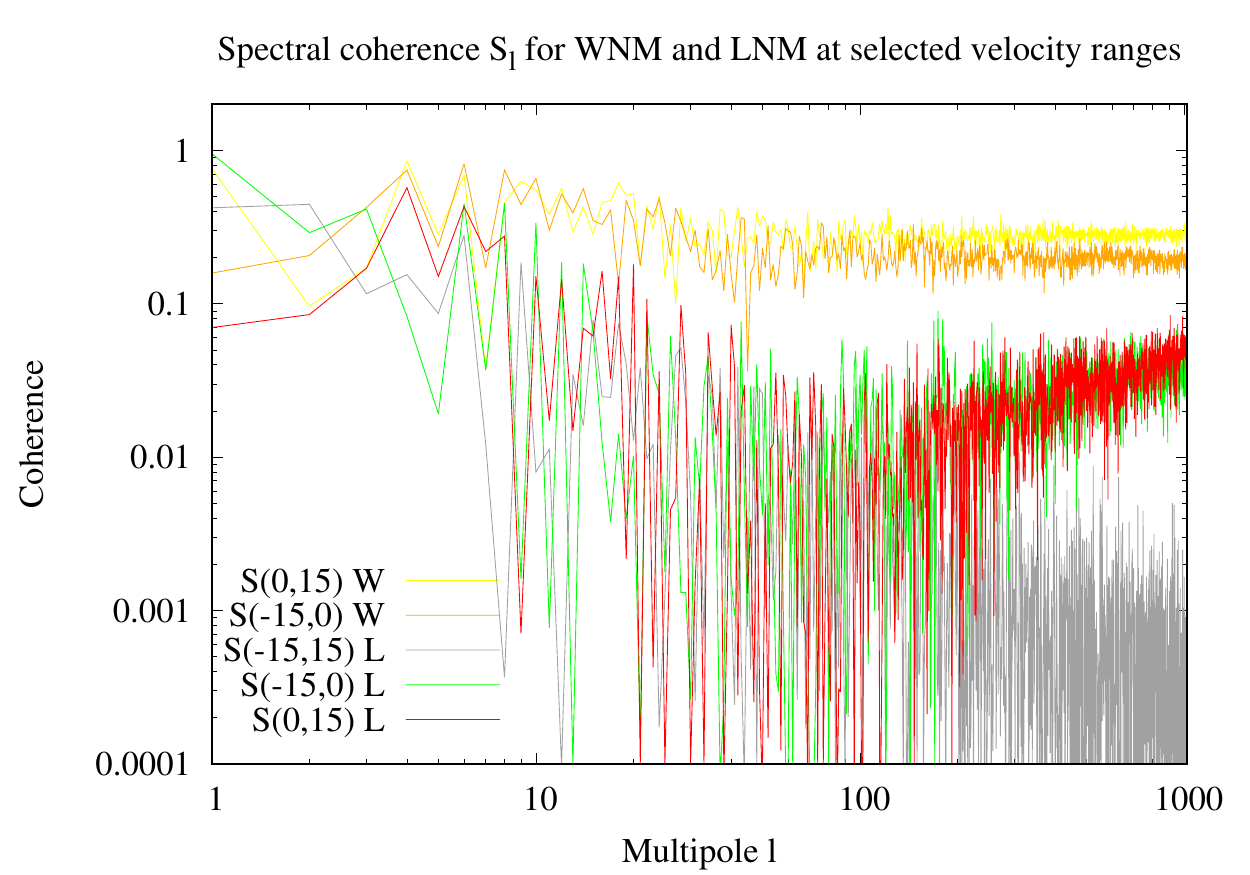}
    \caption{Spectral coherence $S_l(v_1,v_2)$ for WNM and LNM velocity
      slices.  The labels indicate the velocities $v_1$ and $v_2$ and
      the phases, WNM (W) or LNM (L). }
   \label{Fig_coherence_LW}
\end{figure}
%=========================================================================

\subsection{Self-similarities in velocity space}
\label{similar}

The missing correlation between CNM clouds with velocity lags $\delta
v_{\mathrm{LSR}} \ga 10$ \kms,\ and the good spectral coherence of the CNM
at lags $\delta v_{\mathrm{LSR}} \sim 4$ \kms\ shown in
Fig. \ref{Fig_coherence_4} do not imply that the CNM distribution is
restricted in velocity space. The CNM can exist independently at other
bulk velocities $(v_1+v_2)/2$ and other domains embedded in LNM.

To demonstrate this we repeat the calculations and determine
$S_l(v_1,v_2)$ for $v_1 = 6$ \kms\ and $v_2 = 10$ \kms. The result is
shown in Fig. \ref{Fig_coherence_8} and resembles  the CNM  in
Fig. \ref{Fig_coherence_4}. The local \hi\ is self-similar concerning
the properties of the phase distribution in velocity space even though  the column densities are significantly different at
$v_{\mathrm{LSR}} \sim 0$ and $ v_{\mathrm{LSR}} \sim 8$ \kms. Differences in phase fractions (Fig. \ref{Fig_VelPhase}) and spectral
indices (Fig. \ref{Fig_X_f_gamma_high} and \ref{Fig_X_gamma_high}) also
exist, but they have no significant imprint on the spectral
coherence of the CNM.  For the LNM we find in comparison to
Fig. \ref{Fig_coherence_4} an increase in spectral coherence at
multipoles $l \la 100 $, caused by the reduced contribution of  CNM  to
the \hi\ at velocities $\delta v_{\mathrm{LSR}} \sim 8$ \kms.

Self-similarities are also recognizable for the LNM though the velocity
limits are in this case less sharply defined than in the case of the CNM.
Figure \ref{Fig_coherence_LW} shows examples of lags between velocities
$v =$ -15, 0, and 15 \kms. The spectral coherence in the LNM is lost at
large velocity lags, but can be recovered by gradually decreasing the
width of the velocity lag, independent of the bulk velocity $(v_2 +
v_1)/2$. The example in Fig. \ref{Fig_coherence_LW} also shows  that the
spectral coherence values for the LNM at negative and positive bulk velocities
are comparable, while slight differences exist for the WNM.

\subsection{Spectral coherence at high multipoles } 
\label{model}

\begin{figure}[th] %%  20
    \centering
    \includegraphics[width=9cm]{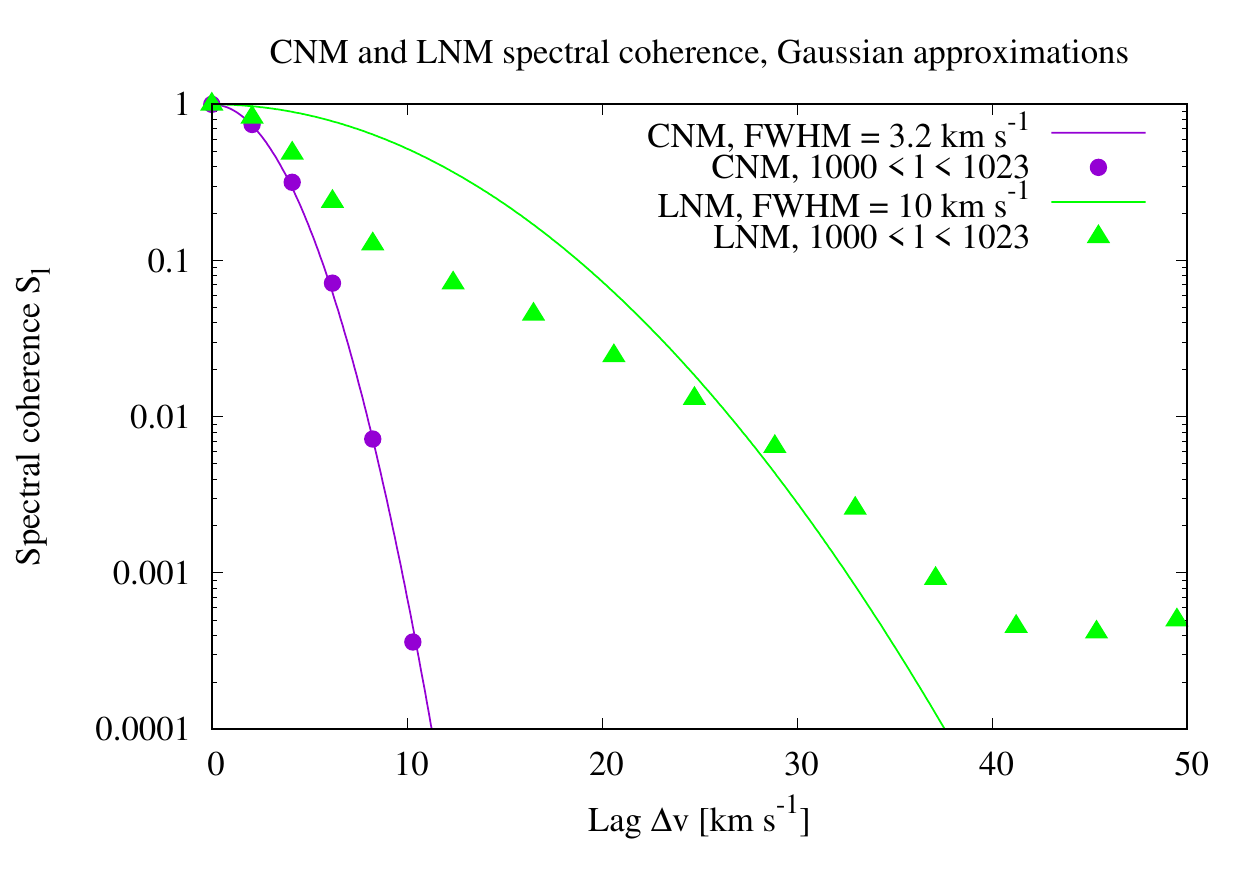}
    \caption{Dependences  of the spectral coherence $S_l(v_1,v_2)$ on
      the lag $\Delta v = v_2 - v_1$ for the CNM and LNM at $|b| > 20
      \deg$.  Averages for multipoles $l < 1000 $ are given (points) in
      comparison with Gaussian distributions with FWHM widths of 3.2
      \kms\ and 10 \kms\ (lines).  }
   \label{Fig_model_CNM}
\end{figure}

The dependence of $S_l(v_1,v_2)$ (Eq. \ref{EQ_coherence}) on the lag
$\Delta v = v_2 - v_1$ can be best studied at high multipoles. We
consider $v_{\mathrm{LSR}} = (v_2 + v_1)/2 = 0 $ \kms\ and use high
latitude data at $|b| > 20 \deg$. To suppress random fluctuations we
average $S_l(v_1,v_2)$ for $l \ga 1000$. The results are shown in
Fig. \ref{Fig_model_CNM} with symbols.  We try to model these $S_l$
distributions by assuming that they are caused by a sample of
\hi\ clumps with characteristic Doppler temperatures and corresponding
Gaussian line shapes. This works well for the CNM, but fails in the case of
the the LNM. For the CNM we fit a FWHM width of $ 3.20 \pm 0.01 $ \kms;
this Gaussian is shown in Fig. \ref{Fig_model_CNM}. For comparison we
plot for the LNM a Gaussian with FWHM width of 10 \kms.  It is
obvious that the LNM cannot be approximated by a Gaussian distribution
with a single characteristic line width. This implies that the CNM and
the LNM must have very different properties. The CNM can be described as
a phase that occupies a well-defined range in velocity width. The LNM as
an unstable phase may be ``best understood as a range of density and
temperature values'' \citep{Vazquez-Semadeni2012}. A large scatter in
Doppler temperatures (or line widths) observed by
\citet[][Fig. 7]{Kalberla2018} supports this proposal. The missing
stability of the LNM implies that \hi\ gas in this phase has to develop
either to CNM or to WNM; the line width distribution gets stretched out
in both directions.

%=========================================================================

%sigma           = 1.35839          +/- 0.00476      (0.3504%)
%FWHM = 3.20 +/- 0.01 
%=========================================================================

The fitted CNM line width of $ 3.20 \pm 0.01 $ \kms\ corresponds to a
Doppler temperature of $ T_{\mathrm D} = 223.7 \pm 2$ K. This
temperature was reported previously by \citet{Kalberla2016} for cold
filamentary \hi\ gas. The agreement is at first glance surprising
since filamentary structures, derived from unsharp masking (USM), have
typically low column densities. \citet{Kalberla2018} determined an
average CNM line width of $ 3.6 \pm 0.1$ \kms, corresponding to a
Doppler temperature of 283 K, from profiles with less than eight well-defined  Gaussian components. They assumed that the Gaussian analysis
would recover only a part of the weak components. The cold filamentary
USM structures, however, turn out to be characteristic of the small-scale structure of the CNM if velocities are restricted to the velocity
range $-5 < v_{\mathrm{LSR}} < 5$ \kms. 

Unsharp masking aims to extract \hi\ structures at the highest spatial
frequencies that can be observed by large single-dish radio
telescopes. The Doppler temperature was determined by
\citet{Kalberla2016} as the median and also by fitting a log-normal
probability density distribution\footnote[5]{For a turbulent medium the
  characteristic PDFs are log-normal \citep{Vazquez1994} as a result of
  the central limit theorem applied to self-similar random
  multiplicative perturbations. The geometric mean of a log-normal
  distribution is equal to its median. }. Here we  select coherent
structures at high multipoles and derive the column density weighted
geometrical mean Doppler temperature from the correlation functions.
The methods for both investigations are quite different, in each case
with uncertain assumptions, but the resulting Doppler temperatures agree
very well.

A characteristic temperature for the CNM can also be derived directly
from the sample of Gaussian components. From narrowband data at high
Galactic latitudes with $\delta v_{\mathrm{LSR}} = 1$ \kms\ we derive the
lowest median $ T_{\mathrm D} = 208 $ K at $ v_{\mathrm{LSR}} = 0$
\kms. For the velocity range $-5 < v_{\mathrm{LSR}} < 5 $ \kms\ we obtain
a median $ T_{\mathrm D} = 217 $ K and a mean $ T_{\mathrm D} = 229 $
K. These $ T_{\mathrm D}$ values are deconvolved for the instrumental
line broadening caused by the spectrometers and agree well with the
value $ T_{\mathrm D} = 223.7 \pm 2$ K, derived by fitting $S_l$. No
deconvolution is necessary for $S_l$ determined from
Eq. \ref{EQ_coherence} because bandwidth and beam effects cancel in this
case. Thus, the median Doppler temperature $ T_{\mathrm D} = 223 $ K for
small-scale CNM structures is well defined and confirmed by our current
investigations.

Extracting prominent cold filaments \citep{Kalberla2016} is 
equivalent to selecting \hi\ components with steep spectral indices (see
also Sect. \ref{veldependent}). A completely independent determination of
\hi\ small-scale structures with the Arecibo telescope by
\citet{Clark2014} leads, within the errors, to the same result; these
authors found a Doppler temperature of 220 K.  \citet{Clark2019} have
proven recently that anisotropic magnetized \hi\ small-scale structures
and narrow linewidths are dust-bearing density structures. They
found that anisotropic small-scale \hi\ channel map structures are
correlated in position and position angle with far infrared structures
at 857 GHz, implying that this emission originates from a colder, denser
phase of the ISM than the surrounding material.

Coherent CNM structures, described by $S_l$ according to
Eq. \ref{EQ_coherence}, can be interpreted as eddies (see \citealt[][Figs. 3 and 4,]{Clark2014} or \citealt[][Figs. 28 and
  29)]{Kalberla2016b}. These structures are caused by a well-defined
population of CNM clouds with characteristic log-normal distributions in
column density and Doppler temperature \citep[][Figs. 12 and
  13]{Kalberla2016b}. The properties of these CNM structures are well
defined. They are cold, dusty, magnetized, and aligned with the local
magnetic field (\citet{Clark2014} and \citet{Clark2019}). Typical temperatures of the
associated dust are $T_{\mathrm dust} \sim 18.5 $ K. These structures
are embedded in LNM with $T_{\mathrm dust} \sim 19 $ K
\citep{Kalberla2018}. Anisotropies associated with radio-polarimetric
filaments, explored by \citet{Kalberla2016b} and \citet{Kalberla2017},
suggest that these structures are shaped by  MHD 
turbulence in the presence of a magnetic field \citep{Goldreich1995}.
Filamentary features from these observations are probably 
mostly sheets with systematic velocity gradients perpendicular to the
field direction.  According to \citet{HeilesCrutcher2005}, ``edge-on
sheets should be edge-on shocks in which the field is parallel to the
sheet.'' Not all of the data can be explained this way. An alternative
interpretation of filaments as fibers, having an approximately
cylindrical geometry, was given by \citet{Clark2014}.

A characteristic Doppler temperature of 223 K implies for an average spin
temperature of 50 K \citep{Heiles2003}  a mean Mach number
of $ M \sim 3.8$ for the CNM, a value that agrees well with the estimate
of $ M \sim 3.7$ by \citet{Heiles2005}. Using a median spin temperature
of 61 to 79 K \citep[][Table 4]{Murray2018b} would result in $2.8 \la M
\la 3.3$.

According to \citet{Lazarian2000} small-scale structures in the \hi\ do
not reflect masses of real clumps, but are caustics produced by the
turbulent velocity field via projection; this process is also called
velocity mapping \citep[][Sect. 6]{Lazarian2000}. Observational
parameterized structures with properties as summarized in the last two
paragraphs were newly rejected as being real entities.
\citet{Lazarian2018} conclude that ``the filaments observed by
\citet{Clark2014} in thin channel maps can be identified with caustics
caused by velocity crowding.'' \citet{Clark2019} object and conjecture
instead that ``small-scale structures in narrow \hi\ channel maps are
preferentially cold neutral medium that is anisotropically distributed
and aligned with the local magnetic field.'' We also interpret these
structures as density enhancements, CNM eddies from a turbulent flow,
caused by phase transitions and observable at the resolution limit of
large single-dish telescopes. Phase transitions may be partly provoked
by turbulence \citep{Saury2014}, but cause at the same time
back-reactions on the power distribution (see also Sects. \ref{veldependent} and
\ref{noCaustic}).

\subsection{Linewidth regimes for different phases}
\label{linewidth}

The FWHM velocity window width for the steepest CNM power index
(Fig. \ref{Fig_X_broad_high}) is identical with the lag where spectral
coherence for the CNM is lost. This velocity window is also identical
with the mean LNM line width of 9.6 \kms\ determined by
\citet{Kalberla2018}. The model assumption of the LNM as a transition
between the stable clumpy CNM and diffuse WNM implies that the velocity
space covered by the LNM must be at least as large as the velocity
spread where we find coherence in the CNM. The LNM is unstable, and there is a high
probability of finding it associated with CNM, which 
causes a coupling between both phases in velocity space.

By generalizing such a hierarchical scheme in the velocity space, we 
assume that the spectral power for the LNM distribution, which is also
embedded in the WNM, should be correlated with the mean FWHM velocity
width of 23.3 \kms\ for the WNM \citep{Kalberla2018}. 
Figure \ref{Fig_X_broad_high} shows that this is the velocity width of the
steepest LNM spectral index. For broader velocity windows the LNM
spectral index is flattening since the correlation between channels
separated by $\Delta v_{\mathrm{LSR}} \ga 23$ \kms\ is lost.

We conclude that the mean FWHM velocity widths of the different phases
are not independent from each other; instead,  they reflect correlations between
the \hi\ phases. \citet{Saury2014} report from their simulations ``that
the turbulent motions inside clumps and the relative velocities between
them are related to the motions of the WNM from which they were
formed,'' which  means that ``a non-negligible fraction of the
measured velocity dispersion is caused by the relative motions of the
clumps along the line of sight, suggesting that the observed line
broadening is likely to be due to the relative clumps motions rather
than supersonic turbulence.''

%=========================================================================
\begin{figure}[th] %%  21
    \centering
    \includegraphics[width=9cm]{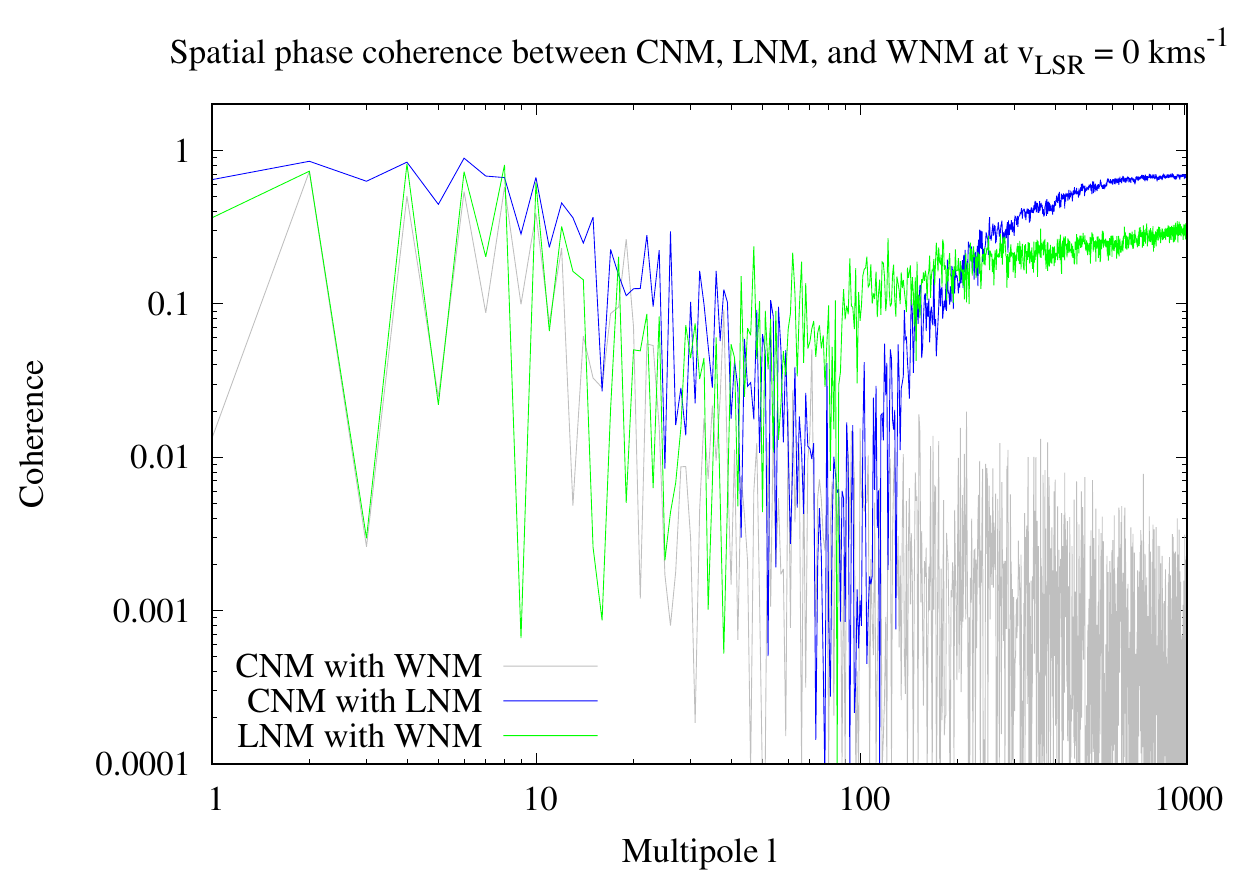}
    \caption{Single-channel spatial coherences $P_l$ at
      $v_{\mathrm{LSR}} = 0$ \kms\ for CNM, LNM, and WNM.  }
   \label{Fig_coherence_phase}
\end{figure}
%=========================================================================

\subsection{Spatial phase coherences }
\label{phases}

To study the spatial relationship between different phases we
define, in analogy to Eq. \ref{EQ_coherence}, the spatial phase coherence
\begin{equation}
P_l(\mathrm {A,B})  =  C_l^2 (\mathrm {AxB}) / [  C_l(\mathrm {AxA}) \cdot  C_l(\mathrm {BxB)}  ]
\label{EQ_phase_coherence}
\end{equation}
between phases A and B. Here $C_l(\mathrm {AxB)}$ is the cross-correlation between PPV channel maps for phases A and B, and $C_l(\mathrm
{AxA)}$ denotes the auto correlation for phase A, and $C_l(\mathrm
{BxB)}$ for phase B. A and B stand for CNM, LNM, and WNM; $P_l$ depends
in general not only on the \hi\ phases, but also on the velocity range
$v_1,v_2$.

Figure \ref{Fig_coherence_phase} shows $P_l$ for single channels at
$v_{\mathrm{LSR}} = 0$ \kms. Here $P_l(\mathrm {CNM,LNM})$ indicates that the
CNM is dominating high multipoles $l \ga 100$; $P_l(\mathrm {LNM,WNM})$
shows that the LNM is somewhat more extended with $l \ga
50$; $P_l(\mathrm {CNM,WNM})$ indicates that CNM and WNM are spatially
almost uncorrelated since the LNM is enveloping the CNM, see also
Fig. \ref{Fig_Cross_Power_high}. Within the noise a very weak
correlation $P_l(\mathrm {CNM,WNM})$ remains; this signal is noisy
because of the low volume filling factor of the CNM.  To find a relation
between CNM and WNM it is necessary to constrain the statistics by
selecting positions with a significant phase fraction for the CNM (see
\citealt{Kalberla2018}). 

The spatial phase coherence $P_l$ is self-similar  for changes in bulk
velocity $(v_1+v_2)/2$ and velocity spread $v_2-v_1$ as long as the
spectral coherence for the individual \hi\ phases is not violated. 

%=========================================================================
\begin{figure}[th] %%  22
    \centering
    \includegraphics[width=9cm]{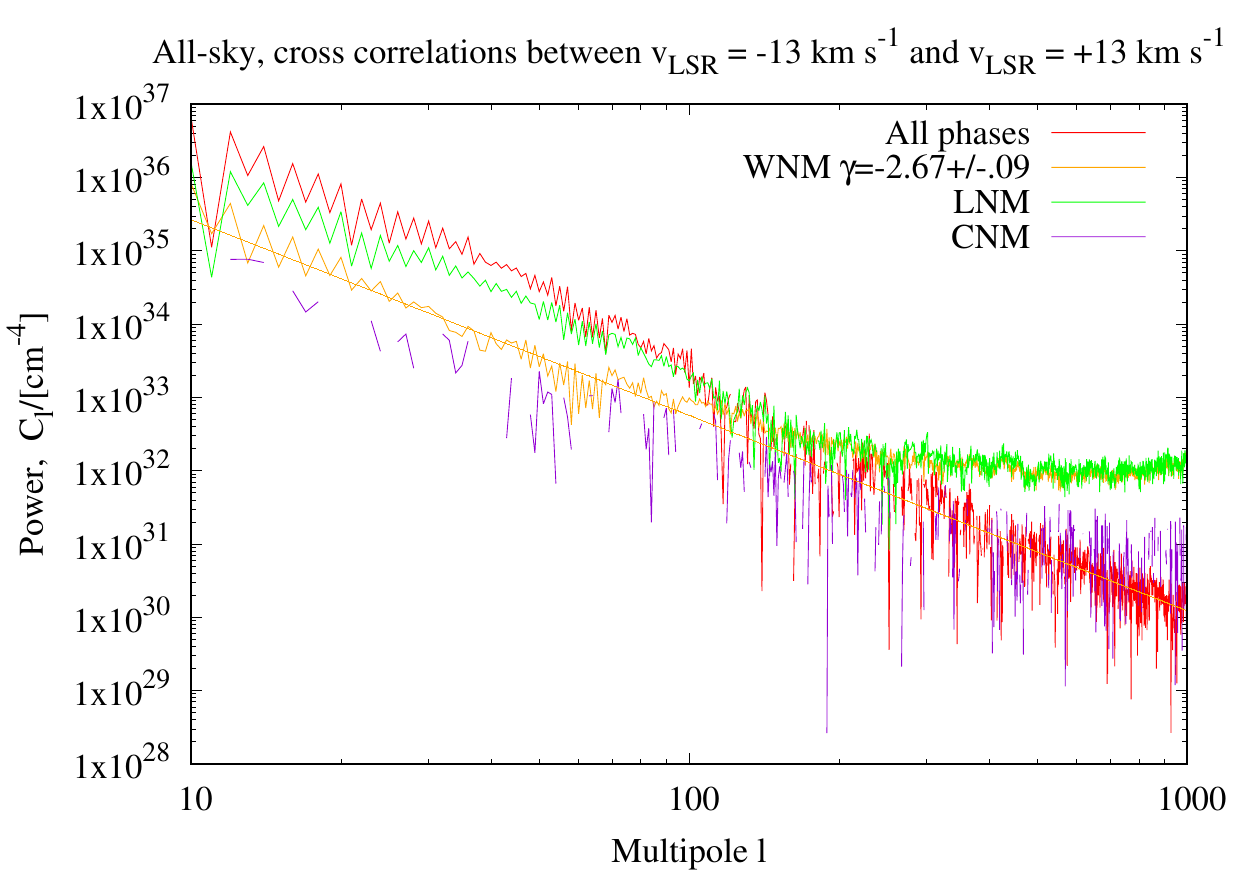}
    \caption{All-sky cross-correlations between $v_{\mathrm{LSR}} = -13$ and
      $v_{\mathrm{LSR}} = +13$ for the total \hi, WNM, LNM, and CNM. }
   \label{Fig_cross_1313}
\end{figure}

%=========================================================================

\subsection{Galactic plane data -- differential Galactic rotation}
\label{rotation}

Next we consider the all-sky data shown in Fig. \ref{Fig_broad}. Due to the
strong emission in the Galactic plane, the all-sky cross-correlations for
channels with a separation of $\Delta v_{\mathrm{LSR}} = 26$ \kms\ (see Fig. \ref{Fig_cross_1313}) are significant for WNM, LNM, and the sum
of all phases. However, what we observe is not a signal that we expect
for a genuine turbulent flow. The observed \hi\ column density
distribution in the Galactic plane is affected by differential Galactic
rotation, causing at Galactic longitudes $Glon$ a $\mathrm{sin} (2 \cdot
Glon)$ modulation  \citep[e.g., ][]{Mebold1972}. In particular, velocities
close to zero are affected by velocity crowding \citep{Burton1972},
leading to a systematic degradation of the signal for $l \la 100$; the
assumption of homogeneity and isotropy is no longer valid.  High column
densities from the Galactic plane dominate the statistics. Checking
cross-correlation spectra with smaller velocity separation we can trace
this problem to channel separations as low as $\Delta v_{\mathrm{LSR}} =
16$ \kms. HI4PI all-sky data, in short, are affected by confusion from the
Galactic plane, and are therefore  only of limited use to us.

%=========================================================================
\begin{figure*}[th] %%  23
   \centering
   \includegraphics[width=9cm]{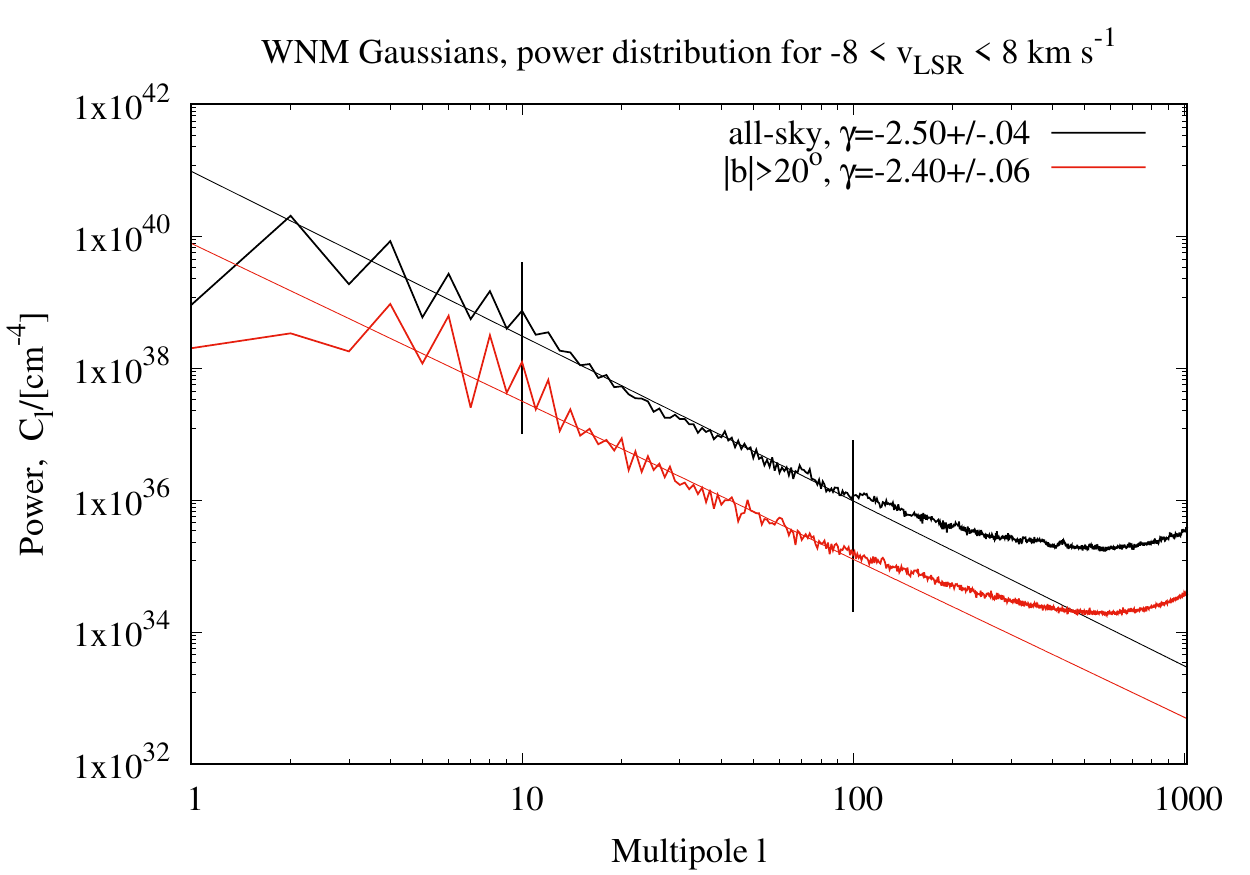}
   \includegraphics[width=9cm]{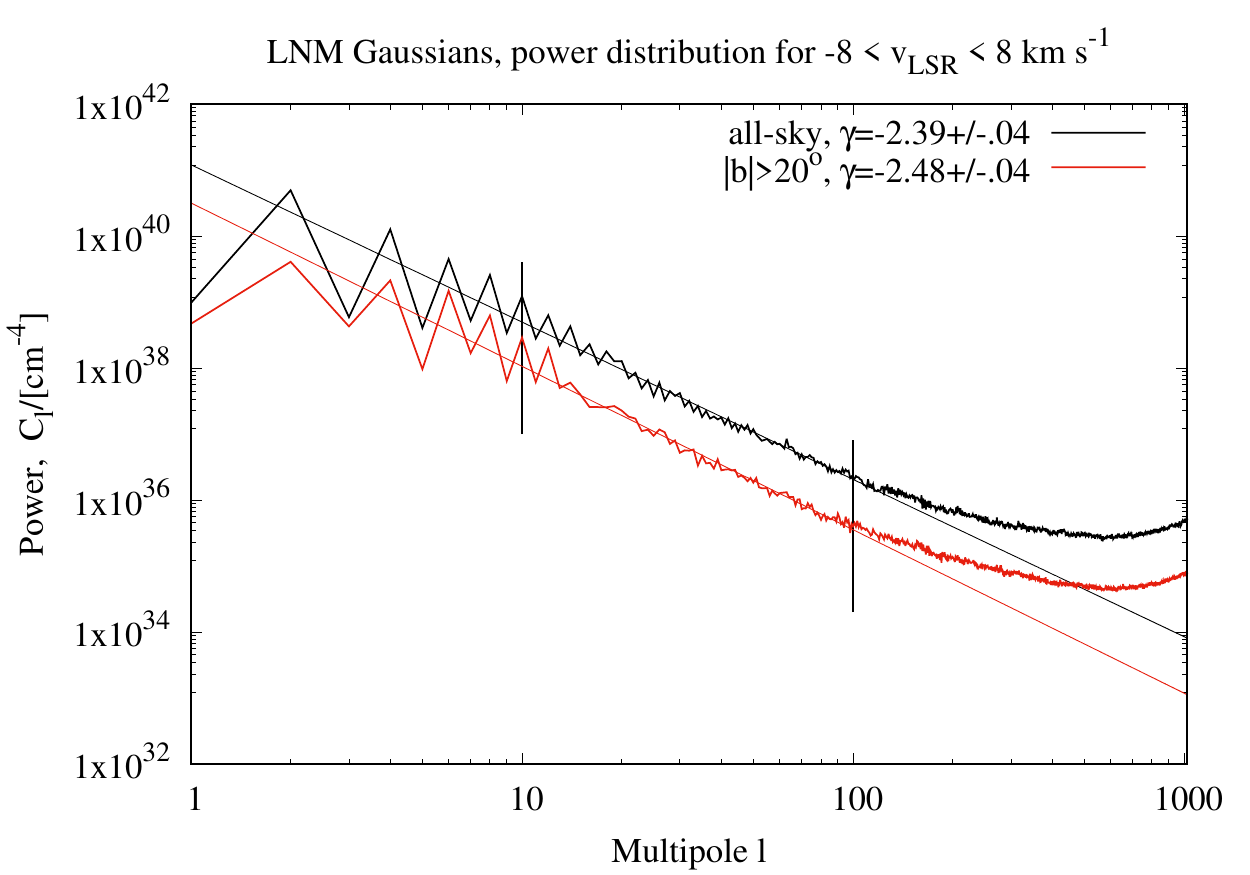}
   \includegraphics[width=9cm]{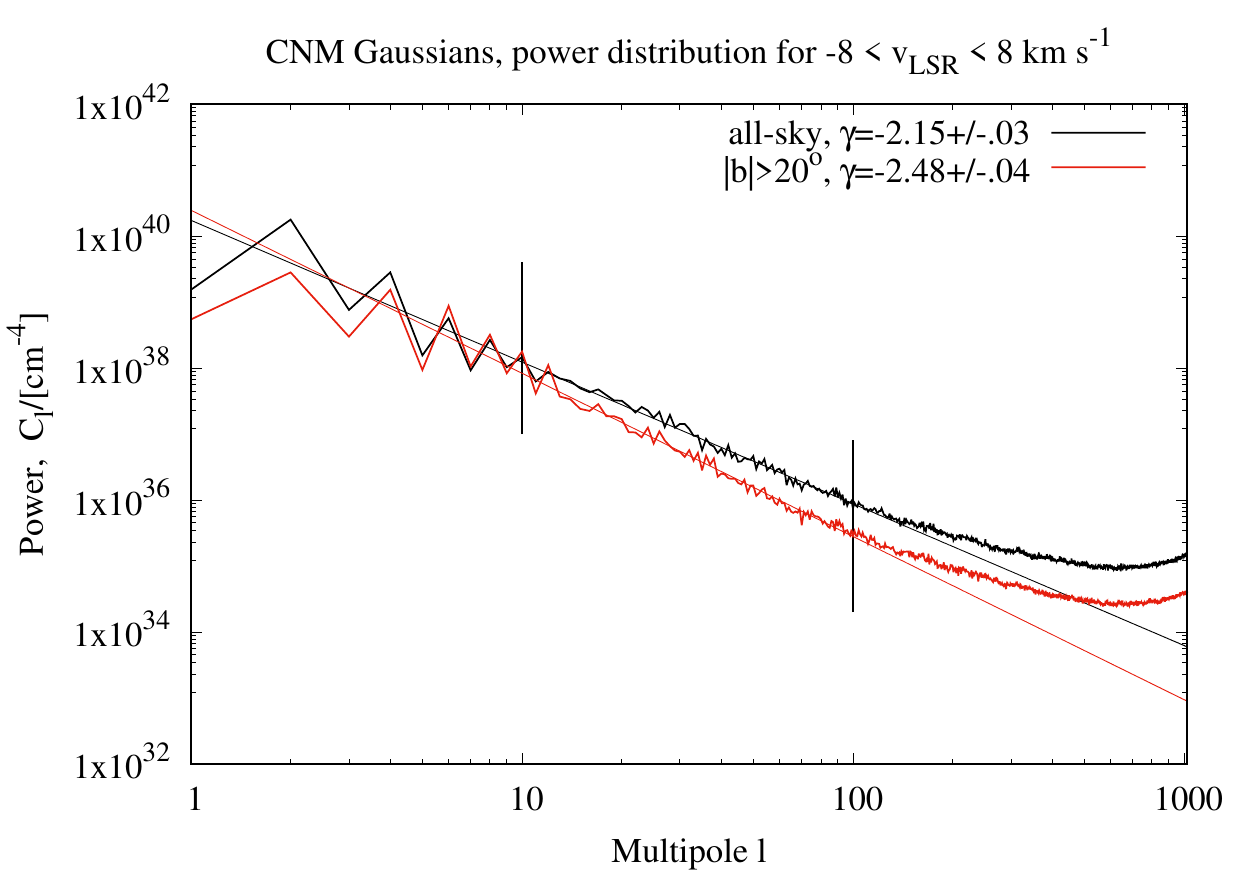}
   \includegraphics[width=9cm]{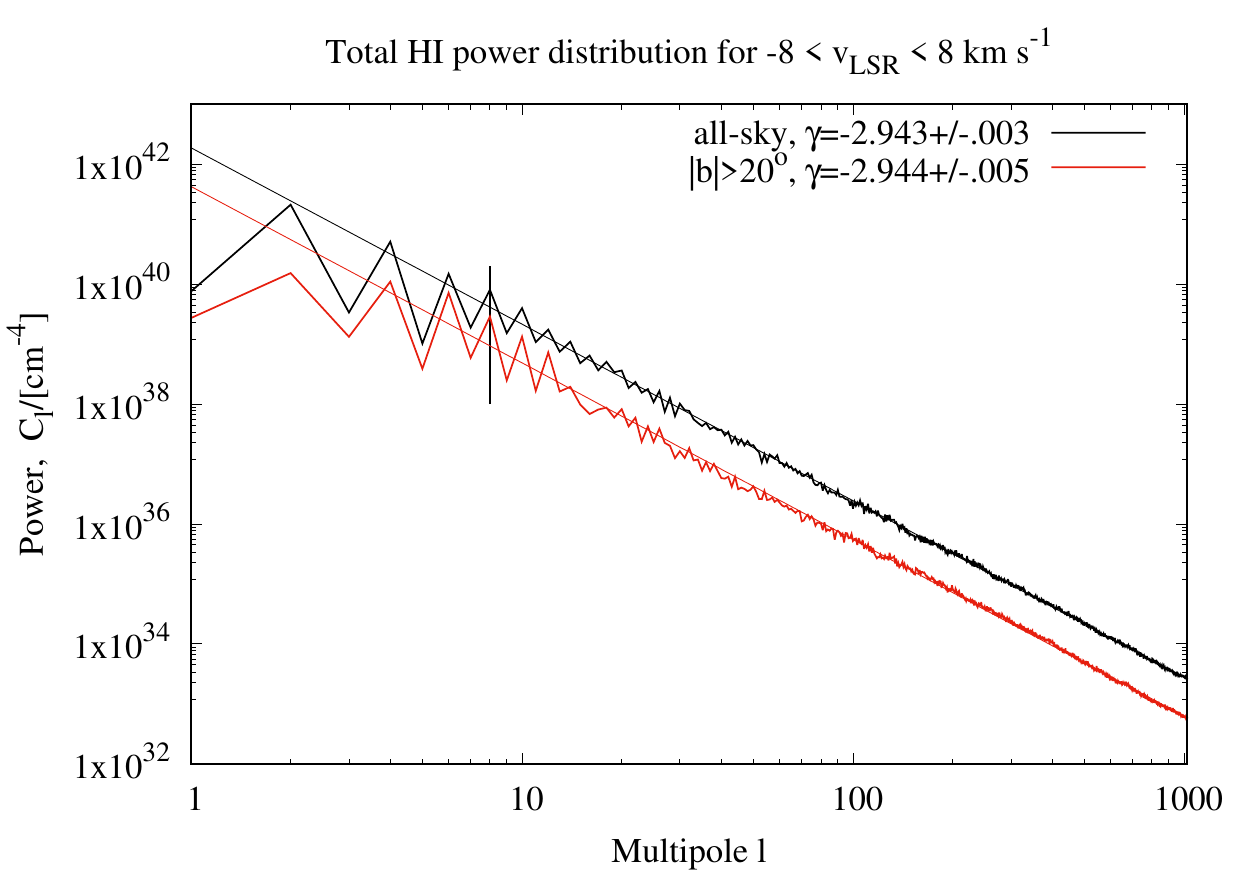}
   \caption{Power distributions for different \hi\ phases at $-8 < v_{
       \mathrm{LSR} } < 8 $ \kms.  Top left: WNM; top right: LNM; bottom
     left: CNM; and bottom right: Sum of all phases with uncertainties
     from the Gaussian decomposition (cyan and orange). Black lines show
     all-sky data; red lines are for $|b| > 20 \deg$. Spectral indices
     $\gamma$ for CNM, LNM, and CNM are derived at $10 < l < 100$ for
     the sum of all phases at $l > 8$, as indicated by the vertical
     lines. }
   \label{Fig_Gauss_88}
\end{figure*}
%=========================================================================

\section{Kolmogorov's local constraint }
\label{Constraints}

\citet{Kolmogorov1941} constrained his paper on ``turbulence in
incompressible viscous fluids for very large Reynolds numbers'' to
locally homogeneous and isotropic structures. The term ``locally''
was (except for stationarity in time) specified by Kolmogorov as
``restrictions are imposed only on the distribution laws of differences
of velocities and not on the velocities themselves.''

A remarkable result of our investigations in
  Sects. \ref{veldependent} and \ref{velwidth} 
is that the properties of the turbulent flow in the multiphase ISM are
limited concerning velocity differences (lags $(v_2 - v_1)$ used by
us). For the two phases, CNM and LNM, we find a decorrelation of the flow
in velocity (Fig. \ref{Fig_cross_55a}) and describe this
phenomenon as a loss in spectral coherence 
  (Fig. \ref{Fig_coherence_10}). The decorrelation depends on spatial
properties of the flow. The CNM dominates high multipoles $l \ga 100$
(Fig. \ref{Fig_coherence_4}), and decorrelation happens at lags $v_2
-v_1 \sim 10 $ \kms. The LNM exists at multipoles $l \ga 50$ 
  (Fig. \ref{Fig_coherence_phase}) and decorrelates at lags $v_2 -v_1
\sim 23 $ \kms. Thus, for both phases we have well-defined domains, and in
agreement with \citet{Kolmogorov1941} ``the hypothesis of local isotropy
is realized with good approximation in sufficiently small domains G of
the four-dimensional space.''    We find self-similarities, i.e., the properties of the CNM and LNM do not change
significantly when shifting the velocities (Figs.
  \ref{Fig_coherence_8} and \ref{Fig_coherence_LW}), but decorrelation
at large velocity differences remains, independent of the ``velocities
themselves'' that are investigated. Turbulence in the local ISM is
locally constrained but otherwise is  homogeneous and isotropic (for a comparison
between the power spectra from two hemispheres, see the lower right panel
of Fig. \ref{Fig_EBHIS_GASS}).

Restricting turbulence to be locally constrained, \citet{Kolmogorov1941}
probably also had  limitations from the experimental setup in mind; the domain
under investigation should not be ``lying near the boundary of the flow
or its other singularities.'' Except for the layered structure of the
\hi, affecting according to Sect. \ref{outer} multipoles $l \la 9$, and
apodization to avoid confusion from the Galactic plane, our
investigations are not affected by boundaries. Our global panoramic view
is based on a coordinate system that is comoving with the LSR in the
center of a locally turbulent flow and should not generate limitations
concerning ``velocity differences.''

\section{VCA revisited}
\label{VCArevisited}

With the results from Sect. \ref{velwidth} we  now apply a
modified velocity channel analysis \citep{Lazarian2000} to our data.
From the theoretical point of view VCA is quite appealing; however, from
the channel cross power analysis in the previous section we see that some
care is needed to avoid biases caused by a decorrelation of the
\hi\ database due to confusion that  is, unfortunately, not easily
recognizable from the data. From the multiphase composition of the ISM
we get additional constraints that are also not recognizable without a
decomposition in different phases.

\subsection{Characteristic broadband spectral indices}
\label{VCAchar}

We conclude from Fig. \ref{Fig_broad} that the best possible estimates
for multiphase broadband power spectra in our case should be derived
for $\Delta v_{\mathrm{LSR}} = 16$ \kms. This is rather narrow compared
to the typical FWHM width of 23.3 \kms\ for the WNM, but interestingly
this width is close to the FWHM width of $\Delta v_{\mathrm{LSR}} = 16.8$
\kms\ for the velocity distribution of filamentary
\hi\ structures. It has been suggested that these structures  indicate
processes that feed kinetic energy to the local ISM \citep[][Sect. 5.13
  and 5.14, Fig. 22]{Kalberla2016}.  We show the broadband spectral
distributions resulting for $\Delta v_{\mathrm{LSR}} = 16$ \kms\ in
Fig. \ref{Fig_Gauss_88}, and obtain remarkably well-defined results: 
$\gamma = -2.943 \pm 0.003$ for all-sky and $\gamma = -2.944 \pm 0.005$
for high latitudes. The derived uncertainties are formal one-sigma
errors, but systematic fluctuations on the order
of $\Delta \gamma \sim 0.05$ may be more characteristic.

Decomposing broadband distributions for individual phases leads,
according to Fig. \ref{Fig_Gauss_88} for multipoles $10 < l < 100$, to
all-sky spectral indices $-2.15 > \gamma > -2.5$, comparable to the
single-channel results at $v_{\mathrm{LSR}} = 0 $ \kms, shown in
Fig. \ref{Fig_Gauss_0}. These results are questionable since they may be
affected by spurious effects from differential Galactic rotation.  At
high latitudes we obtain $\gamma = -2.40$ for the WNM and $\gamma =
-2.48$ for the other phases. These values differ only slightly from the
single-channel result $\gamma \sim -2.36$ at $v_{\mathrm{LSR}} = 0 $
\kms. The cross power spectral index for the WNM at high latitudes with
a channel separation of $v_{\mathrm{LSR}} = 10 $
\kms\ (Fig. \ref{Fig_cross_55a}) is, within the errors, identical to the
auto power WNM spectral index $\gamma = -2.40$ for $\Delta
v_{\mathrm{LSR}} = 16 $ \kms\ (Fig. \ref{Fig_Gauss_88}).

At high Galactic latitudes the spectral indices for broadband power
spectra may also be determined from Fig. \ref{Fig_X_broad_high},
assuming that for single \hi\ phases the minima of the derived
distributions are more characteristic. We obtain in this case for CNM and
LNM $\gamma \sim -2.5$;  the value for the WNM also appears to be
consistent with this result, though a minimum is not yet reached at
$\Delta v_{\mathrm{LSR}} = 50 $ \kms. All-sky results according to
Fig. \ref{Fig_X_broad} cannot be used this way; these data are
obviously affected by confusion.

\subsection{Modified VCA spectral index for the density field}
\label{VCArho}

The observational determined 3D spectral index $\gamma$, according to
\citet{Lazarian2000} or \citet[][Table 3]{Lazarian2009}, can be used to
estimate the spectral index of the density correlation function
$\Gamma_{\rho}$ and  for the velocity correlation function
$\Gamma_{v}$, respectively. For data that are averaged over velocity the observed
fluctuations in column density must be due to density fluctuations. In
the case of a shallow 3D density distribution with $\Gamma_{\rho} > 0$ the
observed thick slice power index corresponds to $\gamma = -3 +
\Gamma_{\rho} > -3$; this transformation is essentially a reduction in
dimensionality.

Not foreseen in the framework of VCA are limitations due to a
decorrelation of the velocity field. We use for CNM and LNM the minima
of the spectral index distribution from Fig. \ref{Fig_X_broad_high},
both close to $\gamma = -2.5$. This value also appears  reasonable for
the WNM. We obtain $\Gamma_{\rho} \sim 0.5 $. Using indices from
Fig. \ref{Fig_Gauss_88} would result in $\Gamma_{\rho} \sim 0.6 $ for
the WNM and $\Gamma_{\rho} \sim 0.52 $ for LNM and CNM. We disregard
all-sky data since they are most probably biased by confusion from the
Galactic plane. Using observed \hi\ column densities with a 3D spectral
index $\gamma \sim -2.94$ (Fig. \ref{Fig_Gauss_88} bottom right) would
imply $\Gamma_{\rho} \sim 0.06 $. Given all the constraints discussed
previously we consider this value (or $\gamma \sim -2.94$ for the 3D
density field) as characteristic of the total \hi\ gas phase, but
unrelated to the multiphase composition.

\subsection{Modified VCA spectral index for the velocity field}
\label{VCAvelo}

For thin velocity slices in the case of a shallow 3D density distribution
with $\gamma > -3$, the observed power index corresponds 
to $\gamma = -3 + \Gamma_{\rho} +
\Gamma_{v}/2 $ \citep[][Table 3]{Lazarian2009}. Here $\Gamma_{v}$ is the spectral index of the velocity
correlation function. For this relation it is assumed that intensity
fluctuations in thin velocity slices are significantly affected by
velocity effects. Line emission that is shifted in velocity can mimic
density fluctuations, so-called caustics. On Galactic scales this effect
is known as velocity crowding \citep{Burton1972}. For shallow density
spectra (i.e., $\gamma > -3$) and independent random velocity and
density distributions, the pure velocity effect should  dominate the density fluctuations on large scales (\citealt{Lazarian2006}),
where fluctuations are supposed to be large compared to the mean
density. \citet{Lazarian2018} conclude that on scales larger than 3
  pc, corresponding to multipoles $l \la 100$ at an assumed distance of
  100 pc, fluctuations in channel maps are dominated by velocity
  fluctuations. Thus, velocity caustics should mimic real physical
  entities such as filaments.  \citet{Clark2019} reject this
  interpretation and find that the \hi\ intensity features are real
  density structures in a multiphase medium and not velocity caustics.
  Cold CNM filamentary structures exist predominantly at small
  scales, or $l \ga 100$, and should (at least for shallow
  \hi\ power spectra) be largely unaffected by velocity fluctuations. 

Thus, according to \citet{Lazarian2000} and \citet{Lazarian2009} $\Gamma_{v}/2$ can
be determined from changes between thin and thick velocity slice power
distributions. We use results at high latitudes shown in
Fig. \ref{Fig_X_broad_high} and find changes in the power law index in
the range 0.12 to 0.16. Accordingly, we estimate $\Gamma_{v} \sim 0.28$
for the individual \hi\ phases. In the case of total \hi\ column densities
we obtain from Fig. \ref{Fig_broad}, $\Gamma_{v} \sim 0.3$. These results
are low in comparison to the \citet{Kolmogorov1941} index $\Gamma_{v} =
2/3$, but intermittency and supersonic motions may modify this index
\citep{Kolmogorov1962}. For a compressible ISM the spectral indices may
depend on the sonic Mach number \citep[e.g.,][]{Burkhart2010}. A 3D
spectral index of $\gamma \sim -2.5$ for $10 \la l \la 100$ implies
accordingly Mach numbers near eight, though from the Gaussian
decomposition lower values are expected: $M_\mathrm{CNM} = 4.4$,
$M_\mathrm{LNM} \la 3.6$, and $M_\mathrm{WNM} \sim 1.4$
\citep{Kalberla2018}.

Our estimates of $\Gamma_{v}$ according to a modified VCA are
lower limits only. The HI4PI survey data are limited in velocity
resolution to $\delta_v = 1$ \kms\ for GASS and $\delta_v = 1.5$
\kms\ for EBHIS, the asymptotical thin slice limit is not reached. At
$\Delta v_{\mathrm{LSR}} = 0 $ \kms\ we should see in
Figs. \ref{Fig_X_broad} and \ref{Fig_X_broad_high} a horizontally
tangential approach.  Our resolution is insufficient to derive a
significant thin slice velocity limit. The demanding condition $\delta_v
\ll 1$ \kms\ for a narrow line width is quite a surprise since the
estimate by \citet[][Sect. 4.3]{Lazarian2000} was that a thin slice of
$\delta_v \sim 2.6$ \kms\ should be a sufficient velocity slice width
for the \hi. Very thin slices in velocity may cause serious
observational problems; the instrumental noise would be significantly
increased and the noise term $N_l$ then needs   to be explicitly taken
into account.

\subsection{Phase transitions versus velocity caustics }
\label{noCaustic}

In the previous subsection we derived VCA predicted spectral index
changes with constraints on spectral coherence from
Sect. \ref{velwidth}. Here we   explore how the VCA solution
depends on multipole $l$. In particular we want to derive how much the
CNM power distribution $ P_{\mathrm {CNM}}(l) $ is modified by bandwidth
changes $\Delta v_{\mathrm{LSR}}$ in velocity. We define the ratio
\begin{equation}
  \mathfrak{R_{\rm VCA}}(l) = P_{\rm broad}(l) / P_{\rm narrow}(l),
\label{eq:vca}
\end{equation}
for the narrowband (single-channel) power $P_{\rm narrow}(l)$ and broadband power $P_{\rm broad}(l)$, integrated over the CNM coherence width
$\Delta v_{\mathrm{LSR}} = 10$ \kms, both at identical center
velocities. 

\begin{figure}[th] %%  24
%    \centering \includegraphics[width=9cm]{ratio_plot_0.eps}
    \centering \includegraphics[width=9cm]{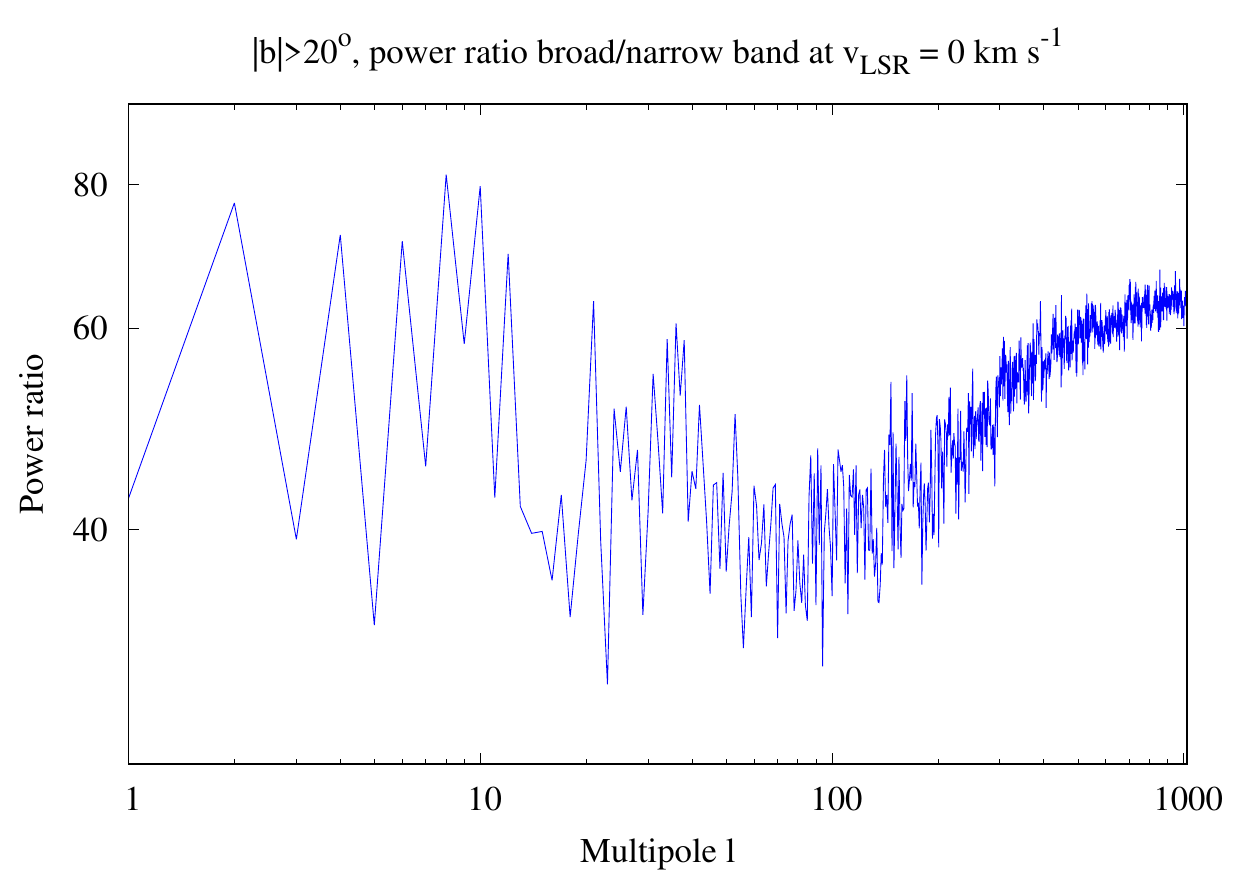}
    \caption{Power ratio $\mathfrak{R}(l)$ for broad- and narrowband
      power distributions at $v_{\mathrm{LSR}} = 0$ \kms. }
   \label{Fig_ratio0}
\end{figure}

\begin{figure}[th] %%  25
    \centering
    \includegraphics[width=9cm]{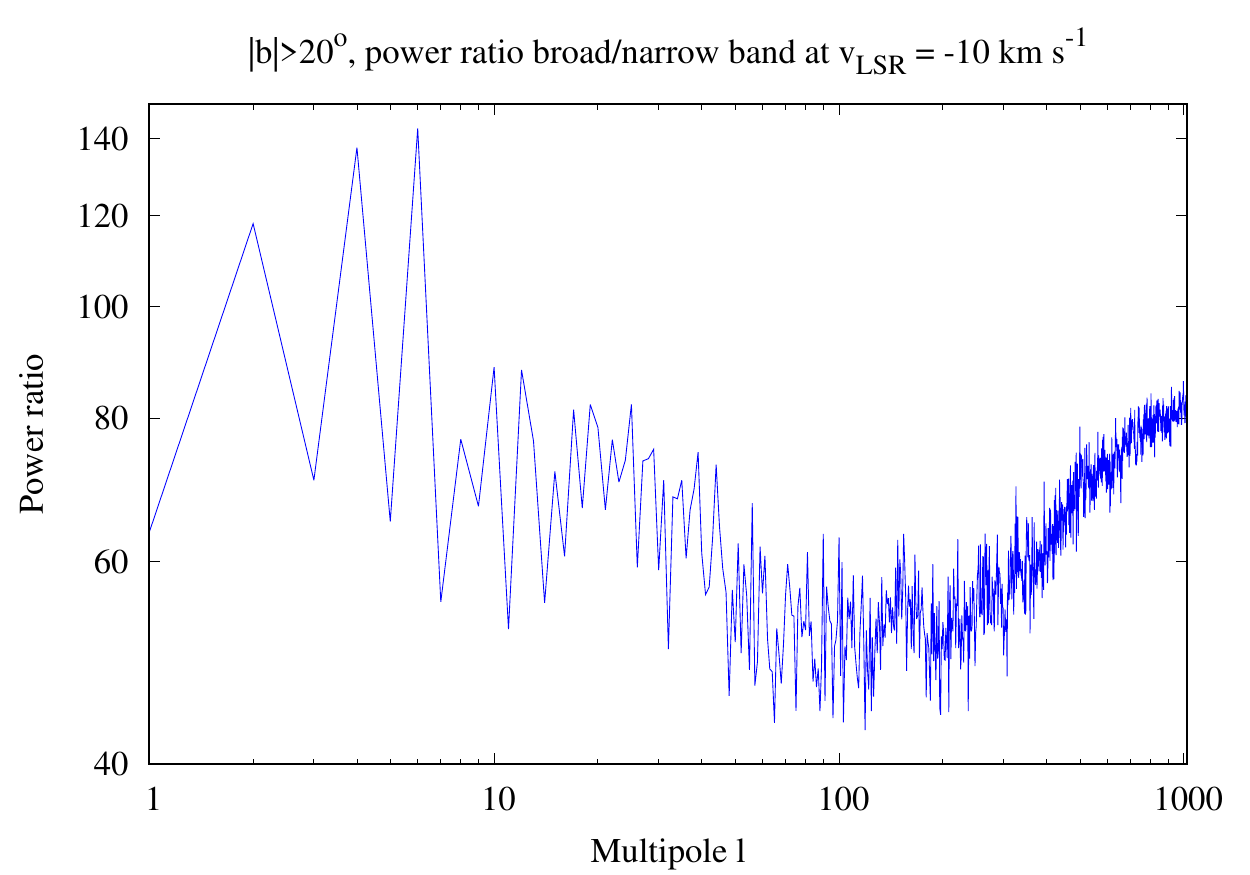}
    \caption{Power ratio $\mathfrak{R}(l)$ for broad- and narrowband
      power distributions at $v_{\mathrm{LSR}} = -10$ \kms. }
   \label{Fig_ratio10}
\end{figure}

We  chose two velocity settings. In the first case, for
$v_{\mathrm{LSR}} = 0$ \kms, there are significant fluctuations in the
CNM spectral index. The second velocity, $v_{\mathrm{LSR}} = -10$, was
chosen such that the CNM spectral index has little fluctuations (see
Fig. \ref{Fig_X_gamma_high}).  We show $\mathfrak{R}_{\rm VCA}(l)$ for
$v_{\mathrm{LSR}} = 0$ \kms\ in Fig. \ref{Fig_ratio0} and for
$v_{\mathrm{LSR}} = -10$ in Fig. \ref{Fig_ratio10}. In
each case we observe more power than expected from the increased
bandwidth, $\mathfrak{R}_{\rm VCA}(l) > 10$. There are in addition two
multipole ranges with different slopes in log-log presentation.

%=========================================================================
%=========================================================================
%=========================================================================

For $l \la 100$ we observe a steepening of the spectral index with
$\delta \gamma = -0.12 \pm 0.04$ at $v_{\mathrm{LSR}} = 0$
\kms\ (Fig. \ref{Fig_ratio0}) and $\delta \gamma = -0.21 \pm 0.04$ at
$v_{\mathrm{LSR}} = -10$ \kms\ (Fig. \ref{Fig_ratio10}). This steepening
is expected from the velocity correlation function
(\citet{Lazarian2000} and \citet{Lazarian2009}), and we derive accordingly $\Gamma_{v}
\sim 0.24$ at $v_{\mathrm{LSR}} = 0$ \kms\ and $\Gamma_{v} \sim 0.4$ at
$v_{\mathrm{LSR}} = -10$ \kms, consistent with the results from the
previous subsection.

For $l \ga 100$ we observe the opposite trend, i.e., the power increases for
high multipoles. The sign for $\delta \gamma$ is positive, but the
distribution cannot be approximated with a constant spectral
index. Could this multipole range be biased? According to
\citet{Lazarian2000} the velocity slice thickness must be increased to
steepen the power law distribution; however, this approach leads  to
contradictions. For an increase in slice thickness beyond $\Delta
v_{\mathrm{LSR}} \sim 10$ \kms\ we observe for low mutilpoles a
flattening of the spectral index (see Sect. \ref{velwidth},
Fig. \ref{Fig_X_broad_high}). The power at high multipoles increases
further.  The power for $l \ga 100$ cannot be explained as being caused by
velocity caustics. It must originate from phase transitions, causing
extra power at high multipoles as discussed in Sect. \ref{model}.

The question arises whether velocity caustics or phase transitions are
dominating spectral index changes. The results from Sect. \ref{veldependent}
indicate that spectral index changes from phase transitions can amount
to $\delta \gamma \sim -0.4 $, twice as large as changes determined in
this section and by a VCA analysis in velocity width in the previous
section. Such changes are usually attributed to velocity caustics.  CNM
structures at $l \sim 1000$, discussed in Sect. \ref{model}, need to be
interpreted as magnetized dust-bearing density structures
\citep{Clark2019}. These entities are
organized in larger filamentary structures, up to scales of tens of
degrees (\citet{Clark2014} and \citet{Kalberla2016}); the most prominent structure is
Loop I. Contemporary observations of such objects with large single-dish
telescopes (Arecibo, GBT, Effelsberg, Parkes), also with the DRAO
interferometer, interpret these structures consistently as clouds or
filamentary cloud complexes, real density structures with a well-defined
range of physical parameters,
(e.g., \citet{Clark2014}, \citet{Martin2015}, \citet{Kalberla2016},
\citet{Blagrave2017}, and \citet{Clark2019}).
An interpretation that most of the structures should be due to
velocity caustics \citep{Lazarian2018} is not supported by
observations.

\section{Intermediate and high velocity clouds}
\label{IVC_HVC}

\begin{figure}[th] %%  26
    \centering
    \includegraphics[width=9cm]{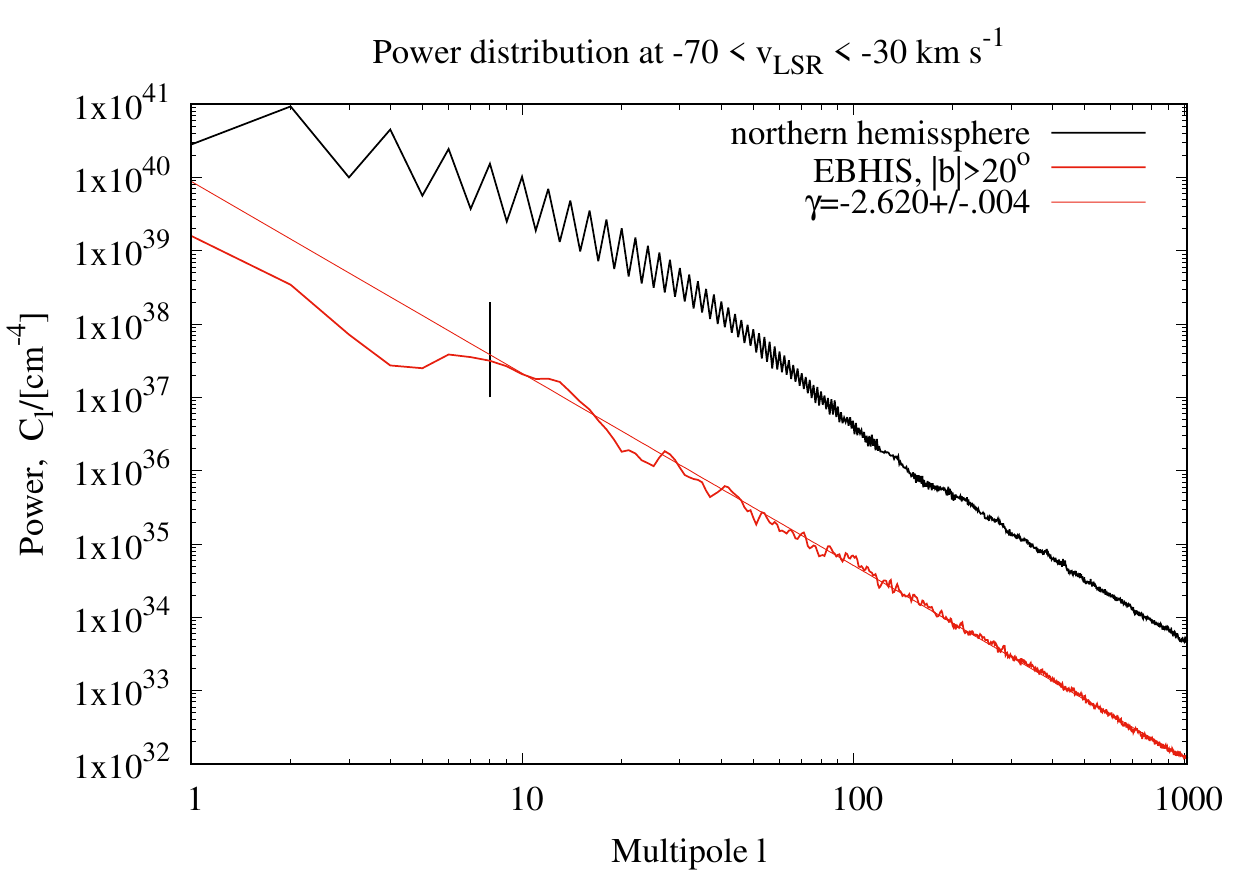}
    \caption{EBHIS power spectra at intermediate velocities $-70 < v_{
        \mathrm{LSR} } < -30 $ \kms\ and fit for $l > 8$. }
   \label{Fig_IVC}
\end{figure}

Power spectra for thick velocity slices, discussed in the previous
section, may be affected by emission from IVCs for large $\Delta
v_{\mathrm{LSR}}$. We have chosen the most prominent IVC emission in the
velocity range $-70 < v_{\mathrm{LSR} } < -30 $ \kms\ and calculated the
power distribution. The IVC emission is dominant on the northern
hemisphere. To avoid any possible instrumental biases from a telescope
mix we extracted EBHIS data, using the apodization schemes discussed in
Sect. \ref{Beam}.  Our result is shown in Fig. \ref{Fig_IVC}. The
all-sky data are obviously affected by spurious effects from
differential Galactic rotation and do not represent genuine turbulent
features. High latitude data show a straight power spectrum with an
index $\gamma = -2.620 \pm 0.004$, within possible systematical
uncertainties $\Delta \gamma \sim 0.05$ in good  agreement with
$\gamma = -2.68 \pm 0.04$ obtained by \citet{Martin2015}, and $-2.60 \pm
0.04 \la \gamma \la -2.48 \pm 0.06$ by\citet{Blagrave2017}.  In comparison
to a thick slice index of $\gamma = -2.94$ for the local \hi,\ this is
significantly flatter. No decomposition of the ICV emission in
components from different \hi\ phases is applied since such a separation
for sources with unknown distances and beam smoothing effects appears to
be ambiguous.

%=========================================================================
%=========================================================================

\begin{figure}[th] %%  27
    \centering
    \includegraphics[width=9cm]{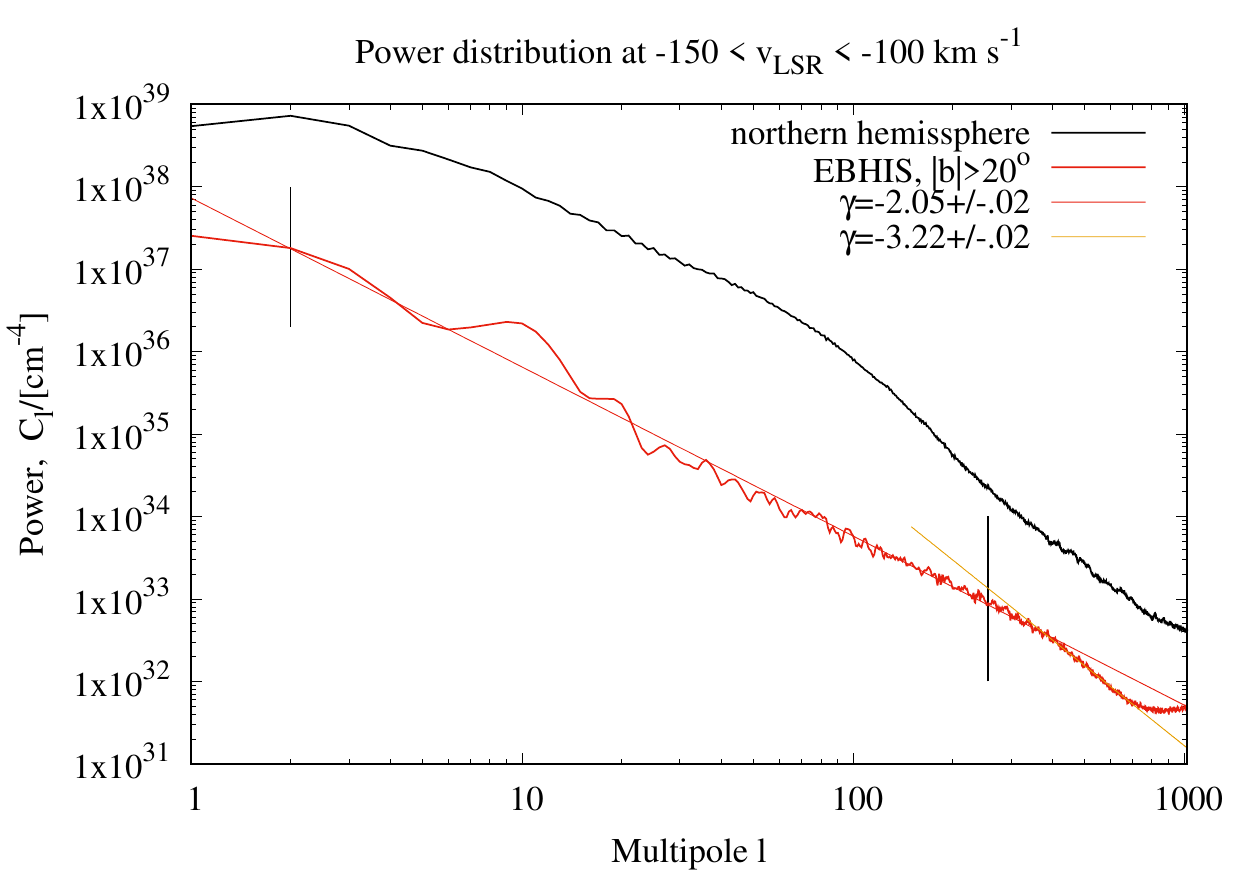}
    \caption{EBHIS power spectra at intermediate velocities $-150 < v_{
        \mathrm{LSR} } < -100 $ \kms\ and fits for $2 < l< 256$ and $350
      < l < 700$. }
   \label{Fig_HVC}
\end{figure}
%=========================================================================

For the sake of completeness we also calculated  power spectra for HVCs
in the velocity range $-150 < v_{\mathrm{LSR} } < -100 $ \kms. In
this case as well we only used  data from the northern hemisphere. The resulting
power spectrum in Fig. \ref{Fig_HVC} is useless in the all-sky case, but
at high latitudes it is straight for $2 < l < 256$ with $\gamma =
-2.0$. This is even flatter than the index $\gamma = -2.59 \pm 0.07$
observed by \citet{Martin2015} and $\gamma = -2.85 \pm 0.07$ by
\citet{Blagrave2017}. Our power spectrum shows for multipoles $ l > 350$
an S-shaped feature that cannot be explained by instrumental
biases. For $350 \la l \la 700$ there is a steepening to $\gamma \sim
-3.2$ with an abrupt flattening at high multipoles. This S-shaped
feature comes from high latitude HVCs and is even more pronounced if we
apodize data for $|b| > 30\deg$. An interpretation without further
detailed investigations, which are beyond the scope of the current paper,
is difficult. Perhaps this S-feature is caused by fragmentation and
dissipation of HVCs on scales near 1\deg. The observed differences in
spectral indices between our data and those of
\citet{Martin2015} and \citet{Blagrave2017} may
imply changes between different HVC complexes.

\section{Low multipoles -- outer scale}
\label{outer}

It is broadly believed that the outer scale $L$, first mentioned in
Sect. \ref{GaussPower} and most likely defined by energy injection from
old supernova remnant shock waves, must be close to $L \sim 100$ pc
\citep{Haverkorn2008}. According to
\citet{Chepurnov1998}, \citet{Cho2002}, and \citet{Mertsch2013} the critical multipole
$l_{\mathrm {crit}}$ for a broken power law depends on $L$ and the scale
height $H$ of the turbulent medium, $l_{\mathrm {crit}} \sim 2 \pi H/L$.
The half width at half maximum scale height for the \hi\ layer was
determined to $ 115 \la H \la 140$ pc (\citealt[][Fig. 10]{Dickey1990} and
\citealt[][Fig. 14)]{Kalberla2007}, resulting in $7 \la l_{\mathrm {crit}}
\la 9 $. This is in good agreement with our finding that the power
spectra tend, within the noise, to be shallow at $ l \la 9$. Strong
multipole disparities are also mostly restricted to $ l \la 9$.
(e.g., Figs. \ref{Fig_Gauss_0} and \ref{Fig_scatter}).

\begin{figure}[th] %%  28
   \centering
   \includegraphics[width=9cm]{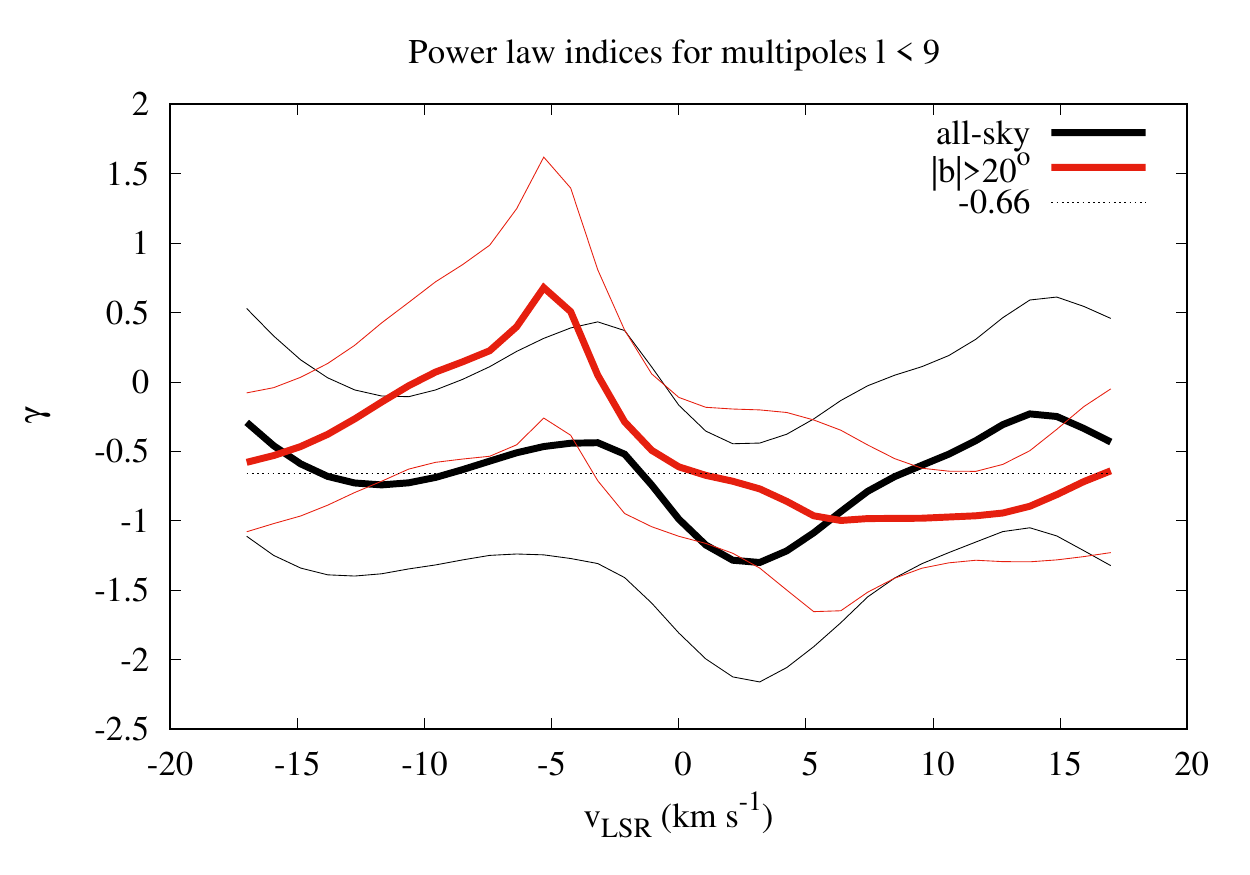}
   \caption{Power law indices at low multipoles $l < 9$ for all-sky
     (black) and high latitudes (red). Thin lines indicate one-sigma
     upper and lower limits. The average $\gamma = -0.66 $ is indicated
     by a dotted line. }
   \label{Fig_flat}
\end{figure}

Low multipoles $l \le 8$ were in all cases fit separately from the rest
of the power distribution, using in this case constant weights. Figure
\ref{Fig_flat} shows the derived velocity dependent power law indices. As
an average over all fits for $ -16 < v_{\mathrm{LSR}} < 16 $ \kms\ we
determine $\gamma = -0.66 \pm 0.13$ for the all-sky data and $\gamma =
-0.65 \pm 0.09$ for $|b| > 20 \deg$.

Our estimate $ l_{\mathrm {crit}} \sim 8$ is consistent with
\citet{Regis2011} who determined $l_{\mathrm {crit}} = 10 \pm 3 $ for
the all-sky Galactic continuum emission from five radio maps and
$l_{\mathrm {crit}} = 5^{ +10}_{-4} $ at high latitudes. Our power law
index $\gamma = -0.65 \pm 0.09$ for $l \la l_{\mathrm {crit}}$ appears
to be in conflict with the proposed value $\gamma = -1$. However, we need
to consider that the index, derived for $ l_{\mathrm {crit}} \sim 8$, is
probably influenced by the planar distribution of the Galactic \hi. On
large scales the column density distribution is dominated by cosecant
latitude effects. Removing this effect leads to a change in the power
distribution at low multipoles (see Sect. \ref{Gauss_f}, and  
Fig. \ref{Fig_f_Gauss_0} in comparison to Fig. \ref{Fig_Gauss_0}).

We use all-sky data, but our main conclusions are drawn from high
latitudes, $|b| > 20 \deg$. This region covers 67\% of the sky with
typical \hi\ column densities below $10^{21} \mathrm{cm}^{-2}$.  We
compare this region with that used by \citet{Planck2016b}. To study the
angular power spectrum of polarized dust emission at intermediate and
high Galactic latitudes, sky fractions between 30\% and 80\% were
investigated by these authors to derive best-fit power law
exponents. For multiploes $l > 40$ there are no significant changes in
the best-fit power law exponents, in particular no changes for sky
fractions of 70\% and 80\%. From these results, since gas and dust at
high Galactic latitudes are well correlated with each other
\citep{Planck2011}, we expect no significant cosecant latitude effects
on our analysis of \hi\ column densities for multipoles $l \ga 10$.

%=========================================================================
%=========================================================================

\section{Summary and discussion}
\label{Summary}

We use high resolution 21 cm line data from the HI4PI survey
\citep{Winkel2016c} to determine spatial power spectra for the \hi\ in
the local ISM. These panoramic observations enable us to
derive spatial power spectra for multipoles $9 \la l \la 1023$. The
power distribution for multipoles $ l < 9$ is consistent with the
assumption that turbulence has an outer scale close to 100 pc. Most of
the observed multiphase \hi\ power spectra are exceptionally straight in
log-log presentation and can be fit well with constant spectral indices,
$\gamma \sim -2.94$ for the local gas, $\gamma \sim -2.6$ for IVCs, and
$\gamma \sim -2$ for HVCs. HVCs deviate from this rule and show some
steepening at high multipoles. HI4PI power spectra for the emission
  from the complete Galactic plane can be contaminated seriously by
confusion from differential Galactic rotation.  \hi\ data for
  several low latitude subfields within a range $ 98\deg \la l \la
  145\deg $ and $-20.4 < v_{\mathrm{LSR}} < 0.2 $ \kms\ were analyzed
  previously by \citet{Khalil2006}. They derived $\gamma = -2.95 \pm
  0.05$, in good agreement with our result for the local gas at high
  latitudes. 

The extraordinarily straight power distribution for the total \hi\ at
high latitudes with its well-defined broadband index of $\gamma \sim
-2.944 \pm 0.005$ (with possible systematic fluctuations on the order of
$\Delta \gamma \sim 0.05$; see Sect. \ref{Beam}) is in excellent
agreement with dust spectral index $\gamma_{\mathrm {dust}} = -2.9 \pm
0.1$ determined by \citet{Miville2016} who also conclude  from the
straight power distribution that there is no sign of energy dissipation
at the 0.01 pc scale. The power spectrum for the dust has to be
interpreted as a spectrum of turbulent density fluctuations. Gas and
dust are well mixed. The value $\gamma \sim -2.94$ for the total (phase
independent) \hi\ distribution must therefore be interpreted as
characteristic of the total \hi\ density distribution.

Phase transitions lead to modified distribution laws for individual
phases \citep{Saury2014}. The observed filamentary CNM structures are
well aligned with the magnetic field (\citet{Clark2014} and \citet{Kalberla2016}),
resulting in asymmetries (\citet{Kalberla2016b} and \citet{Kalberla2017}). The 3D
space is no longer populated isotropically \citep{Goldreich1995}. This
probably implies  a reduction in dimensionality, suggesting a flattening
of the 3D power spectra for those parts of the \hi\ that share the
magnetic field orientation, the CNM, and the LNM.

We derive single-phase power spectra for CNM, LNM, and WNM from a
Gaussian decomposition with characteristic linewidth regimes. These
individual phases show power spectra that deviate significantly from the
multiphase power spectra. For all of the phases we get straight power
spectra for $ 10 < l < 100$ with indices $\gamma \sim -2.5$. In addition
for $ l \ga 100$ a strong increase in the power up to high multipoles is
observed.  The single-phase power spectra are, according to
Eq. \ref{eq:crossPower}, highly correlated; the auto power relations for
all phases must be considered together with all cross power terms
between these phases.  This correlation leads to counterintuitive
consequences and contradictions if we try to describe the \hi\ with
several independent phases.

We find distinct correlations in velocity space. Coherence in the CNM is
limited to a velocity spread of $\sim 10$ \kms. Coherent features at
high multipoles are identified as cold clumps with median Doppler
temperatures of 223 K. Our analysis recovers structures that were
previously described by \citet{Clark2014} and \citet{Kalberla2016}. Trying to
interpret these CNM structures as velocity caustics leads to
contradictions. In particular, according to VCA the spectral index
  should steepen; however, we observe it to be undefined for velocity
  widths beyond 10 \kms. These CNM features must be real entities in
accordance with the findings of \citet{Clark2019}. \hi\ filaments on
large scales are made up of aligned small-scale CNM structures. These CNM
features show  velocity fluctuations on larger scales, but  these
large-scale structures cannot be explained as being caused by velocity caustics.

Phase transition must be
considered as local events with limitations in the cloud-cloud velocity
dispersion for the resulting CNM clumps.  For different bulk velocities
we obtain coherence for different samples of CNM objects. The spectral
coherence of the CNM is self-similar and statistically invariant under
translation in velocity. The correlation between CNM and LNM implies
that the LNM exists in connection with the CNM as a more extended phase,
enveloping the CNM spatially but also covering  a wider velocity spread
of $\sim 23$ \kms\ as a transition to the more extended WNM. Self-similarity in velocity also applies to the LNM. Hence, associated with
the hierarchical wave number scaling for phases with different linewidth
domains we obtain a remarkable homogeneity in velocity space. Comparing
power spectra for different hemispheres we find a high degree of
isotropy.

The remarkable loss of correlation at large velocity lags is not in
contradiction to the theoretical setup for the analysis of turbulent
flows as considered by \citet{Kolmogorov1941}. The treatment of a
turbulent flow needs to be restricted to a sufficiently small domain in
four-dimensional space.  Homogeneity and isotropy was explicitly defined
by Kolmogorov as ``local'' only. These constraints, as described by
Kolmogorov, correspond to the coherence conditions derived by us.

Restrictions in the velocity domain imply constraints for a velocity
analysis, including the use of velocity centroids and explain 
discrepancies discovered previously by \citet[][Appendix C]{Kalberla2017}. Our
multiphase 3D spectral index $\gamma = -2.94 $, derived with velocity
constraints, is in conflict with steeper power law indices derived by
\citet{Dickey2001} and \citet{Stanimirovic2001}; however, these authors
considered very different conditions. \citet{Dickey2001} got thick slice
3D spectral indices up to $-4.0$ in the Galactic plane for velocities
$-20 < v_{ \mathrm{LSR} } < -80 $ \kms, covering distances $1.5 \la d
\la 13$ kpc in the inner Galaxy. \citet{Stanimirovic2001} found a thick
slice 3D spectral index of $-3.4$ for the Small Magellanic Cloud,
analyzing structures between 30 pc and 4 kpc perpendicular to the line
of sight, integrating in velocity over $ \Delta v_{ \mathrm{LSR} } = 90
$ \kms.

We use VCA \citep{Lazarian2000} modified with restrictions in
  velocity space to disentangle density and velocity effects on
observed power spectra.  Taking velocity decorrelations into account, we
derive single-phase estimates of $\Gamma_\rho \sim 0.5$ for the spectral
index of the density correlation function and $\Gamma_v \sim 0.3$ for
the spectral index of the velocity correlation function for $10 \la l
\la 100$. For the total observed \hi\ column density distribution we
determine $\Gamma_\rho \sim 0.06$ and $\Gamma_v \sim 0.3$ for all
multipoles $ l \ga 9$. The significance of the derived $\Gamma_v \sim
0.3$ must be questioned. A steepening of the power spectra as expected
for velocity caustics can also be caused by phase transitions. For
$v_{\mathrm{LSR}} = 3 $ \kms\ we determine a CNM spectral index change
$\delta \gamma \sim -0.4$ from phase transitions, larger than the change
$\delta \gamma \sim -0.14$, indicated by VCA.

Spectral indices for single-channel power spectra are steepest at
velocities close to zero \kms\ where CNM phase fractions are high and
WNM phase fractions are low. We interpret this as an indication of
local phase transitions. Phase transitions do not change the total
column densities significantly, but decrease locally the line
widths. Thus, thermal instabilities produce enhanced power in the line
centers at the expense of the power in the wings of the line. At the
same time phase transitions cause fragmentation, small-scale structure
is generated, giving rise to enhanced power at high multipoles ($l \ga
100$). This process is necessarily accompanied by a decrease in power on
scales that are characteristic of the regions that were affected by
phase transitions and fragmentation. Power is reduced on
intermediate scales and the power spectrum steepens for scales that are
little affected by fragmentation ($l \la 100$).

Recent high resolution simulations of thermal instabilities and
collapsing cold clumps by \citet{Wareing2019} show that the density
power spectrum can rapidly rise and steepen as the structures grow in
the simulations. They find sheets and filaments on typical scales of 0.1
to 0.3 pc, consistent with high power for the CNM at large
  multipoles and also consistent with small-scale structures observed
by \citet{Clark2014} and \citet{Kalberla2016}.  Previously
\citet{Audit2005} and \citet{Federrath2016} have derived a preferred scale of 0.1 pc
from simulations. Our results appears to be in conflict with
\citet{Saury2014} who find for subsonic turbulence that phase
transitions lead to shallow power spectra with $\gamma \sim -2.4$. The
discrepancy would be resolved if their analysis were sensitive to the
single-phase CNM for which we observe $\gamma \sim -2.37$ at high
latitudes.

Steep broadband 3D power law indices $\gamma \sim -3.6$ that were
previously considered to be characteristic for \hi\ in emission deviate
according to \citet[][Fig 10]{Hennebelle2012} significantly from power
law indices for molecular lines and other data.  The single-phase power
indices $\gamma \sim -2.5 $, derived by us, are in much better agreement
with the other indices. This also applies  to the comparison by
\citet[][Fig. 11]{Ghosh2017}. Cold filamentary \hi\ structures were
found to be correlated with {\it Planck } 353 GHz polarized dust
emission \citep{Kalberla2016}. Our broadband CNM power index of $\gamma
= -2.48 \pm 0.04 $ fits well to the mean {\it Planck } intermediate
latitude sky exponent of $\gamma = -2.42$ at 353 GHz for the EE and BB
polarized dust power spectra at multipoles $40 < l < 370$
\citep{Planck2016b}.  \citet{Ghosh2017} use in their dust model for the
south Galactic pole region the CNM as a tracer of the dust polarization
angle, and find for EE, BB, and TE modes spectral indices close to
$-2.4$, a value that they consider  to be characteristic of the
turbulent magnetic field as well. This is consistent with the results from
\citet{Vansyngel2017} and more recently with \citet{Planck2018},
who obtain an index of $-2.42 $ for the EE, $-2.54 $ for the BB, and
$-2.49$ for the TE mode.  \citet{Adak2019} obtain for their dust
model for the north Galactic pole region an index of $-2.4 \pm 0.1$ for EE, BB,
and TE modes and the turbulent magnetic field, in excellent
agreement with our value of $\gamma = -2.48$ for CNM and LNM at high
latitudes.

%=========================================================================
%=========================================================================
%=========================================================================
%=========================================================================

%=========================================================================
\begin{acknowledgements}
  We acknowledge the second anonymous referee for the constructive
  criticism. P.K. thanks J{\"u}rgen Kerp for the inspiring discussions and
  B{\"a}rbel Koribalski for the help with the LVHIS data.
  U.H. acknowledges the support from the Estonian Research Council grant
  IUT26-2, and from the European Regional Development Fund (TK133).  This
  research has made use of NASA's Astrophysics Data System.  EBHIS is
  based on observations with the 100m telescope of the 
  Max-Planck-Institut f\"ur Radioastronomie (MPIfR) at Effelsberg. The Parkes
  Radio Telescope is part of the Australia Telescope, which is funded by
  the Commonwealth of Australia for operation as a National Facility
  managed by CSIRO. Some of the results in this paper have been derived
  using the HEALPix package.
   \end{acknowledgements}

\begin{appendix} 

\section{Observations and data reduction} 
\label{Obs}

\subsection{HI4PI 21-cm line survey data}
\label{HI4PI}

We use HI4PI data \citep{Winkel2016c}, combining 21 cm line data from
Effelsbeg-Bonn \hi\ Survey (EBHIS), observed with the 100m Effelsberg
radio telescope (\citet{Winkel2016a} and \citet{Winkel2016b}) and from the Galactic
All Sky Survey (GASS), observed with the 64m Parkes telescope
(\citet{Naomi2009}, \citet{Kalberla2010}, and \citet{Kalberla2015}).   EBHIS data from the
first data release \citep{Winkel2016b} and  GASS data from the final
data release \citep{Kalberla2015} were calibrated to a common intensity
scale.

The multi-beam receivers  scanned the sky, dumping spectra with
short integration times (0.5 sec for EBHIS and 5 sec for GASS). Each of the individual dumps was corrected for stray
radiation.  The telescope data were gridded using Gaussian kernels of
5\farcm4 for EBHIS and 7\farcm5 for GASS, resulting in an
effective FWHM beam size of 10\farcm8 for EBHIS, and 14\farcm5 for
GASS. An nside = 1024 HEALPix database was chosen, appropriate for an
angular resolution of $\Theta_\mathrm{pix} = 3\farcm44 $
\citep{Gorski2005}. This implies that our database is oversampled;
neighboring positions are not independent from each other. Bandwidth
limitations need to be taken into account, but a deconvolution (discussed
below) for nside = 1024 is rather simple and well defined. For both
telescopes a large number of dumps from the multi-beam receiver
were averaged to generate profiles at the HEALPix grid positions,
leading in each case to well-defined Gaussian beam shapes.

Combining data from different hemispheres we use a border line at $
\delta = -2\deg$. To avoid discontinuities caused by different
beam sizes, we take a linear interpolation between the two surveys for $
-4\deg < \delta < 0\deg$.

The 21 cm line surveys were observed with different spectrometers. For
EBHIS the spectral resolution is $\delta v_{\mathrm{LSR}} = 1.29$ \kms;
for GASS it is $\delta v_{\mathrm{LSR}} = 0.82$ \kms. The average
  noise for a single channel outside the \hi\ line is 90 mK for the
EBHIS and 55 mK for the GASS. We  chose to interpolate the data to
a common intermediate velocity grid with $\delta v_{\mathrm{LSR}} = 1.03$
\kms, which is the velocity resolution of the LAB \citep{Kalberla2005};
the same database has been used previously  \citep{Kalberla2016}. 
  This interpolation causes for the EBHIS a slight increase in the
  average noise to 100 mK.

%=========================================================================
%=========================================================================
%=========================================================================
%=========================================================================

\subsection{Instrumental effects }
\label{processing}

The observed intensities $\widetilde{I_{\mathrm{obs}}}$ can for
individual telescope dumps be described by the convolution between the
true distribution $\widetilde{I_{\mathrm{sky}}}$ on the sky and the beam
function $\widetilde{B_{\mathrm{tel}}}$ of the telescope. In addition, we
need to take into
account the instrumental noise $\widetilde{N_{\mathrm{dump}}}$,  which   affects individual dumps:
\begin{equation} % Eq. 3
\widetilde{I_{\mathrm{obs Dump}}} =  \widetilde{B_{\mathrm{tel}}} \ast \widetilde{I_{\mathrm{dump}}}  + \widetilde{N_{\mathrm{dump}}} .
\label{eq:TB_smo1}
\end{equation}
To get the intensity distribution on a HEALPix grid, the dump database
is interpolated using a Gaussian weighting with an smoothing kernel
$\widetilde{B_{\mathrm{grid}}}$. This implies an additional convolution
and we obtain
\begin{equation} % Eq. 3
\widetilde{I_{\mathrm{obs HEAL}}} = \widetilde{B_{\mathrm{grid}}} \ast (
\widetilde{B_{\mathrm{tel}}} \ast \widetilde{I_{\mathrm{dump}}} +
\widetilde{N_{\mathrm{dump}}} ) = \widetilde{B_{\mathrm{data}}} \ast
\widetilde{I_{\mathrm{sky}}} + \widetilde{B_{\mathrm{grid}}}  \ast
\widetilde{N_{\mathrm{dump}}} 
\label{eq:TB_smo2}
.\end{equation}
For convenience, we define here the smoothing function 
$\widetilde{B_{\mathrm{data}}} = \widetilde{B_{\mathrm{grid}}} \ast
\widetilde{B_{\mathrm{tel}}} $ that was applied to the data.

To calculate the power distribution we need to autocorrelate and perform a Fourier
transform on Eq. \ref{eq:TB_smo2}, leading to 
\begin{equation} 
P_{\mathrm{obs}} = B^2_{\mathrm{data}}\ P_{\mathrm{sky}} +
B^2_{\mathrm{grid}}\ P_{\mathrm{Noise}}  + 
  2 P_{\mathrm{cross}} , 
\label{eq:TB_power2}
\end{equation}
with the cross power $P_{\mathrm{cross}}$ from $\widetilde{B_{\mathrm{data}}} \ast
\widetilde{I_{\mathrm{sky}}}$ and $\widetilde{B_{\mathrm{grid}}} \ast
\widetilde{N_{\mathrm{dump}}}$
\begin{equation} 
P_{\mathrm{sky}} = P_{\mathrm{obs}} /B^2_{\mathrm{data}} -
 P_{\mathrm{Noise}}\ B^2_{\mathrm{grid}} /B^2_{\mathrm{data}}  -
  2 P_{\mathrm{cross}} /B^2_{\mathrm{data}} 
\label{eq:TB_power3}
\end{equation}
or with $B_{\mathrm{data}} = B_{\mathrm{grid}}\ B_{\mathrm{tel}} $
simplified to 
\begin{equation} 
P = P_{\mathrm{obs}} /B^2_{\mathrm{data}} -
 P_{\mathrm{Noise}} /B^2_{\mathrm{tel}}  
-  2 P_{\mathrm{cross}} /B^2_{\mathrm{data}} 
\label{eq:TB_power4}
\end{equation}
This result makes it necessary to correct the beam smoothing with
different kernel functions for observed data and the noise term, and  also to
consider  the question of whether the observations are affected by a
cross term between noise and data. Smoothing and noise effects are
discussed below in separate subsections. We also need to  consider  that
EBHIS and GASS have different beam functions and different noise
contributions. In data processing we have to  process the two different hemispheres separately.

%=========================================================================

\begin{figure}[th] %%  A1
   \centering
   \includegraphics[width=9cm]{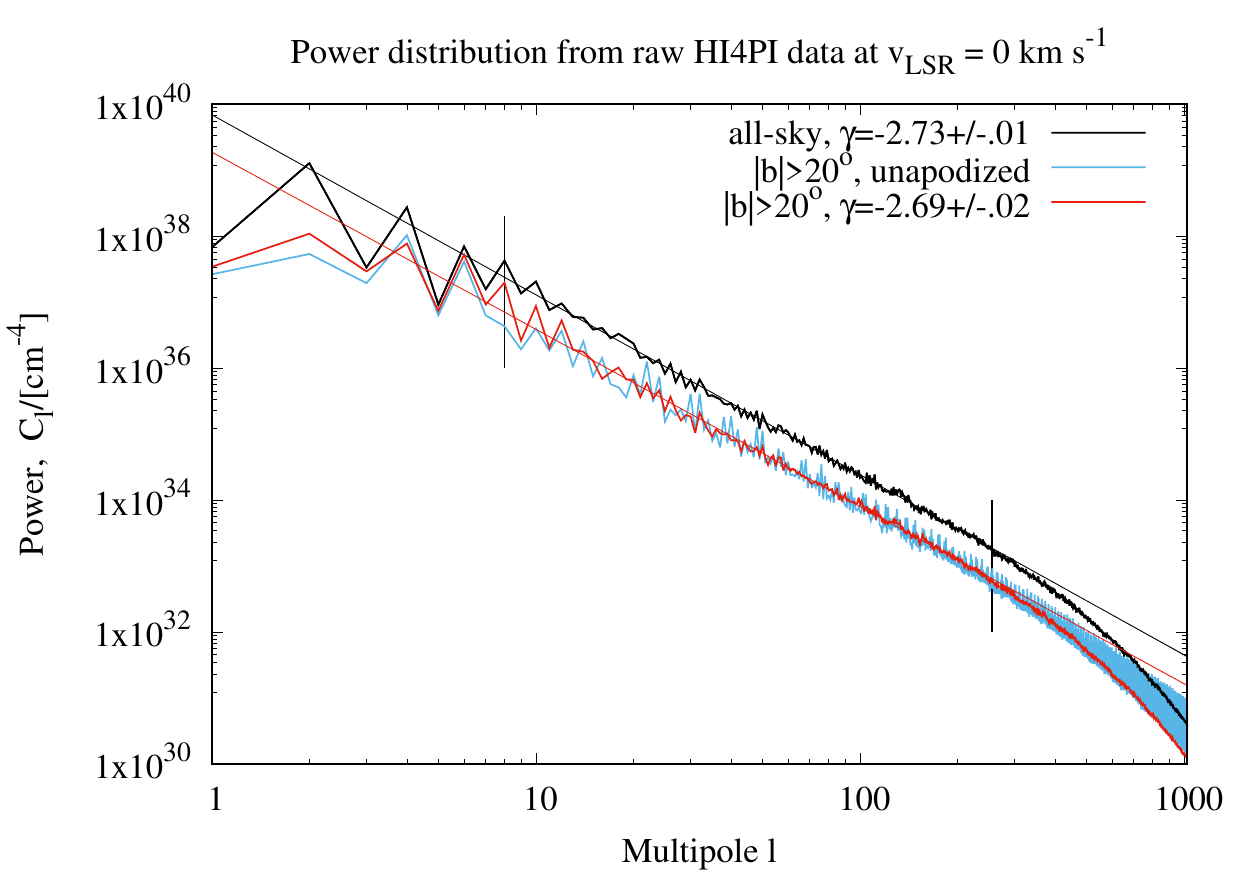}
   \caption{Raw single-channel spatial power distribution for
       observed \hi\ column densities at $ v_{\mathrm{LSR}} = 0 $ \kms\
      without corrections for observational artifacts. To fit 
     the spectral indices $\gamma$ we only used multipoles $ 8 < l < 256 $     . These limits are indicated by the vertical lines.}
   \label{Fig_Power_obs}
\end{figure}

%=========================================================================

\subsubsection{Apodization }
\label{Apodize}

If we want to analyze only a part of the sphere, in our case high
latitudes with $|b| > 20\deg$, we need to take the window function into
account that masks the observations. Since this window is applied to the
brightness temperature distribution on the plane of the sky, it implies a
smoothing to the observed power distribution $P_{\mathrm  {obs}}(l)$.
Sharp edges of the window function may cause aliasing or
power leakage between different multipoles $l$. To avoid such aliasing
it is necessary to smooth the sharp edges of the window function by
apodizing \citep{Planck2016}.

Figure \ref{Fig_Power_obs} shows the raw power spectrum for the observed
column densities in a single channel at $v_{\mathrm{LSR}} = 0 $ \kms. We
use this case here to demonstrate  our data processing. The
upper black profile presents the all-sky power spectrum; below we
show two versions of the power spectrum at high latitudes, $|b| >
20\deg$. The cyan spectrum in Fig. \ref{Fig_Power_obs} shows
$P_{\mathrm {obs}}(l)$ after setting in ANAFAST the parameter
theta\_cut\_deg (or gal\_cut in healpy) to 20\deg. We observe strong
aliasing caused by an observational window with sharp cutoff at
20\deg;  an apodization is needed to overcome spurious
effects.

%=========================================================================
\begin{figure*}[th] %%  A2
   \centering
   \includegraphics[width=9cm]{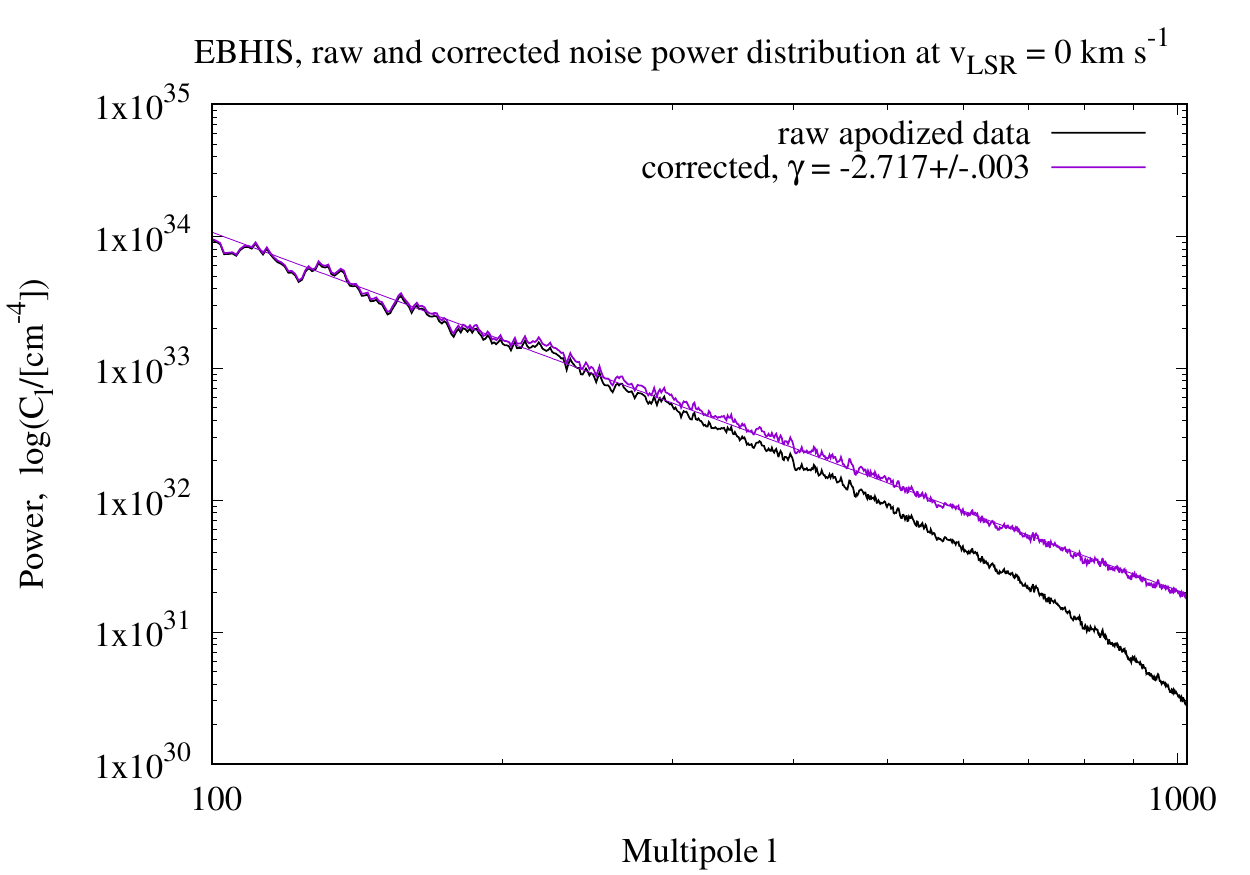}
   \includegraphics[width=9cm]{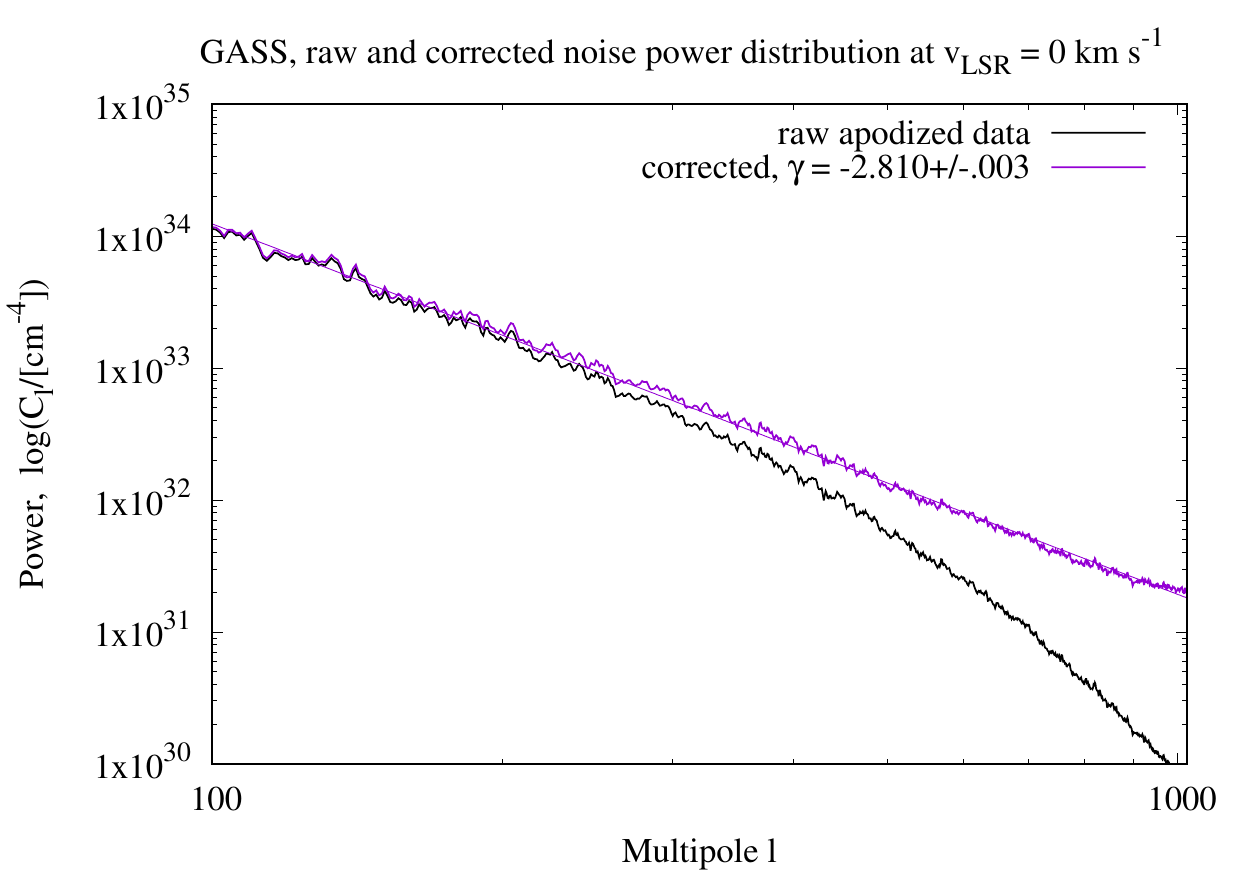}
   \includegraphics[width=9cm]{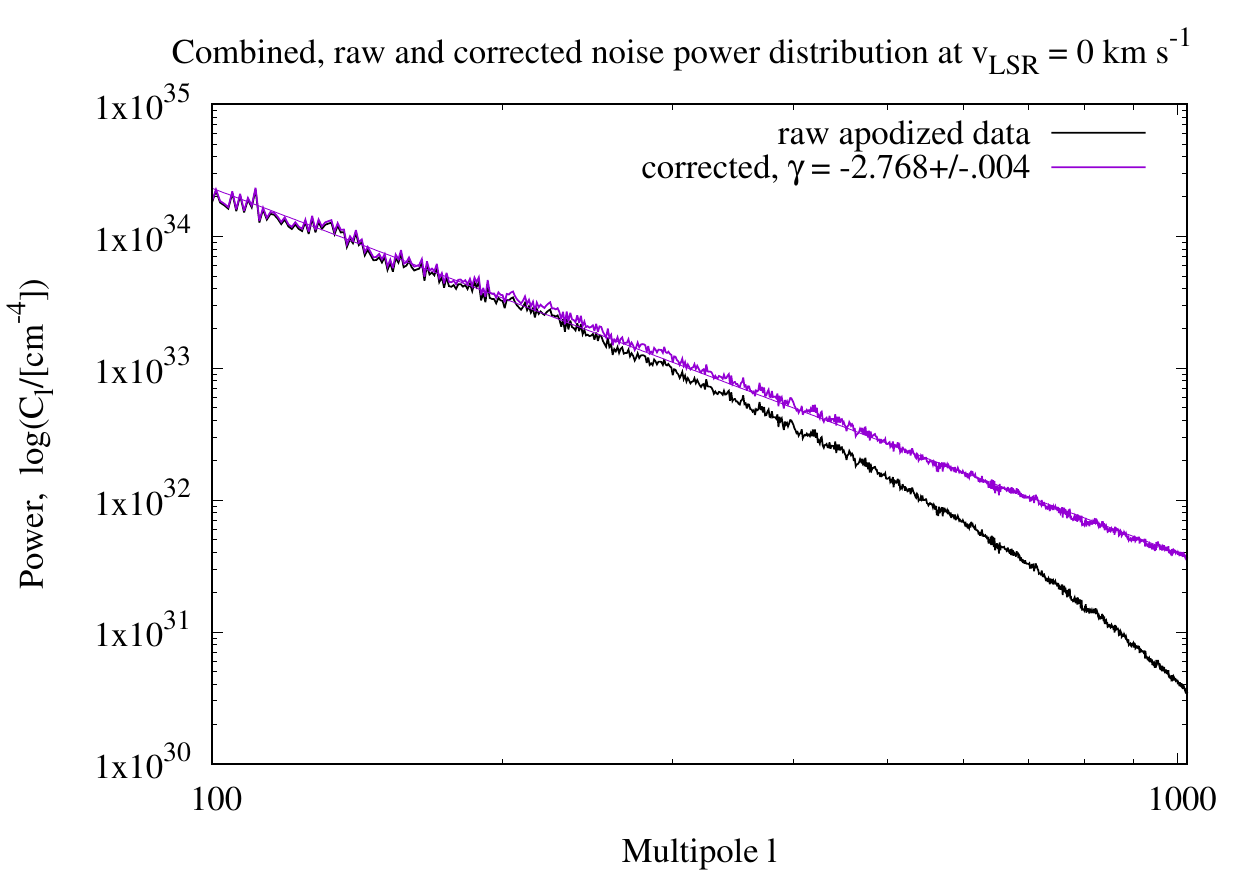}
   \includegraphics[width=9cm]{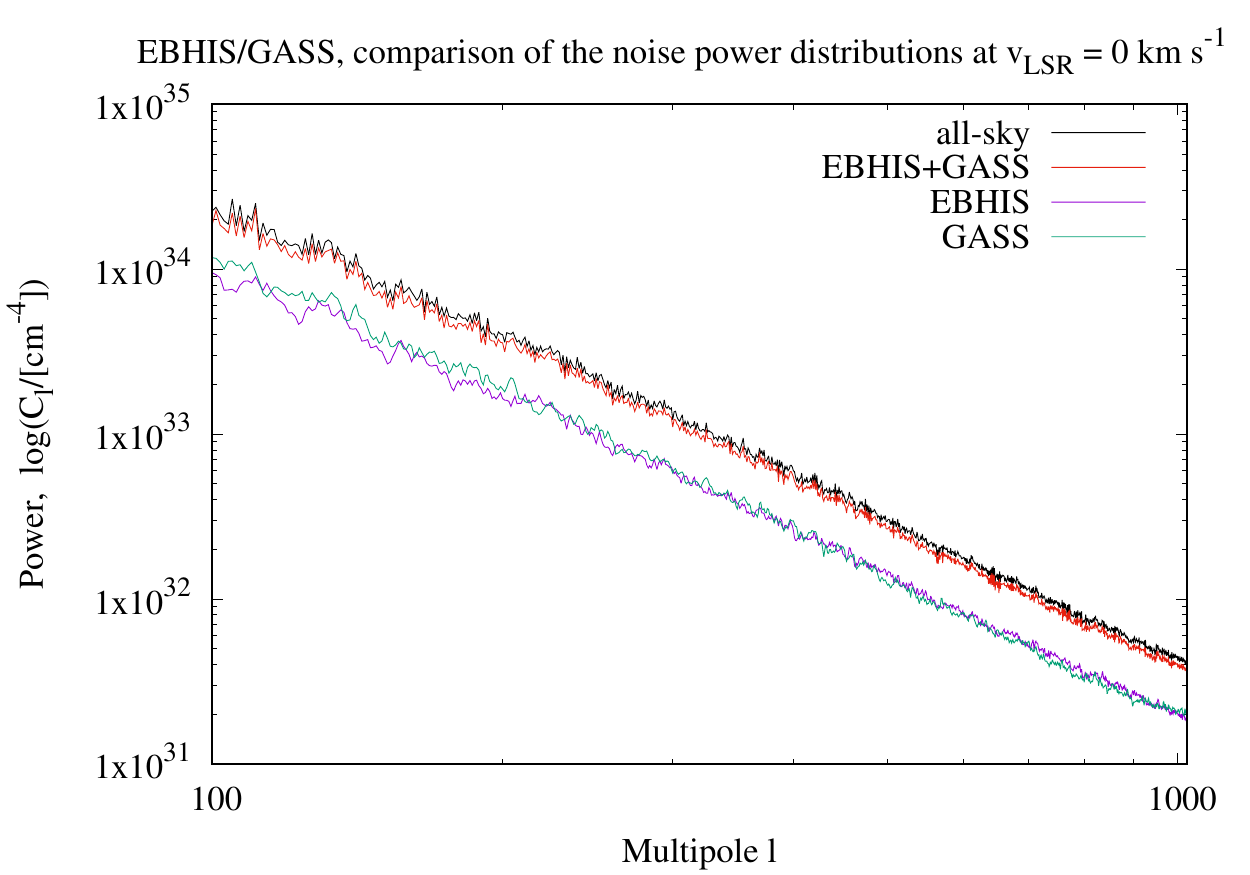}
   \caption{Power distributions at $v_{\mathrm{LSR}} = 0 $ \kms, comparing  
     single-channel \hi\ column densities for EBHIS (top left) and GASS
     (top right) with the all-sky combination of both surveys (bottom left). 
     The black lines indicate power spectra derived from raw
     observations, the magenta lines are beam corrected power spectra
     with fits. On the lower panel to the right we compare the
     beam corrected power spectra.   }
   \label{Fig_EBHIS_GASS}
\end{figure*}

%=========================================================================

Similar to \citet{Kalberla2016b} we use a cosine taper \citep[a Tukey
  window,][]{Harris1978} for apodization in latitude $b$. After a few
tests we  chose a taper width with a half-period of 15\deg,
weighting smoothly from one at $|b| = 20 \deg$ to zero at $|b| = 5
\deg$. The resulting power spectrum is shown in
Fig. \ref{Fig_Power_obs} in red, overplotted on the unapodized
case. Testing a narrower cosine taper, for example with a half-period of
10\deg, we found still significant unwanted aliasing effects. A broader
cosine taper for $|b| = 20 \deg$ is also not appropriate since it leads
to a leakage of Galactic plane emission from $|b| < 5\deg$. Any
apodization window leads to unwanted side effects \citep{Harris1978}, but
this apodization-ringing decreases as the width of the window
increases. Remaining problems (Fig. \ref{Fig_Power_obs}, red curve) are
acceptable for our analysis with minor errors, as demonstrated  in
Fig. \ref{Fig_f_Gauss_0}.  The optimal width of the Tukey window that we applied
 is slightly larger then the $\sigma = 5\deg$ Gaussian taper used
for the apodization map generated by the cosmological parameters team
\citep{Planck2016}.

\subsubsection{Beam smoothing }
\label{Beam}

Observed brightness temperatures are smoothed by the beam function. In
addition we need to take into account that for each cell of
the HEALPix grid the data are interpolated. The resulting
effective beam, has a FWHM beam width of 10\farcm8 for EBHIS
\citep{Winkel2016a}. For the GASS we needed to redetermine the FWHM
beamwidth since we use the third data release \citep{Kalberla2015} with
some improvements over the second release \citet{Kalberla2010}. We
searched the Local Volume \hi\ Survey \citep{Koribalski2018} for
appropriate point-like narrowband \hi\ sources and used the barred
Magellanic irregular galaxy IC~4662 \citep{vanEymeren2010} which has at
$v_{ \mathrm{LSR} } = 324 $ \kms\ a bright compact {2\arcmin} core. In
addition, the beam width was determined from the ultra-compact
high velocity cloud HVC289+33+251 (\citet{Putman2002} and \citet{Bruens2004}) with a
FWHM size of 4\farcm4. Fitting the sizes of these sources we derive a
FWHM beamwidth of $ 14\farcm5 \pm 0\farcm2$ which compares well with the
beamwidth of 14\farcm4 determined by \citet{Calabretta2014}. This $
14\farcm5$ beam is a considerable improvement over the second GASS data
release with an effective beam width of 15\farcm9 in the presence of
residual instrumental errors.

We consider once more a single channel at $v_{\mathrm{LSR}} = 0 $ \kms\
 (see Fig. \ref{Fig_Power_obs}). The upper all-sky power spectrum
(black) and the lower one (red and apodized for $|b| = 20 \deg$),
both suffer for $l \ga 256$ from beam smoothing. We observe a steepening of
the spectral index at high multipoles, which without beam correction
may be mistaken  as a real effect, for example as steepening caused by a
change in slice thickness \citep{Lazarian2000}. For the raw data shown
in Fig. \ref{Fig_Power_obs} essentially only a limited range, $l < 256$,
as indicated by a vertical bar, can be used to fit    the
spectral index $\gamma$.  It is essential to correct for beam smoothing
if we want to avoid instrumental biases.

To evaluate beam smoothing effects we separate the emission observed in
different hemispheres. We use a 15\deg\ cosine apodization scheme,
extracting data for EBHIS at declinations $\delta > 13\deg$ and for GASS
at $\delta < -17\deg$. Figure \ref{Fig_EBHIS_GASS} (top) shows the
correlation coefficients for EBHIS (left panel) and GASS (right
panel). The black lines show significant differences in beam smoothing
caused by different beam widths.  We correct in both cases the observed
power spectra by applying corrections based on the effective FWHM beam
widths, $P = P_{\mathrm{obs}} /B^2_{\mathrm{data}} $ according to the
first term in Eq. \ref{eq:TB_power3}. $B_{\mathrm {data}}(l)$ is
telescope dependent and includes beam smoothing and smoothing caused by
gridding.  The power spectra, including their fits for $8 < l < 1023$
are shown in magenta. A comparison between observed and beam corrected
power spectra shows that differences can be seen for $l \ga 256$,
corresponding to angular scales of $\theta \sim 180\deg/l \sim
42\arcmin$ or about three to four FWHM beam widths. In the case of noisy
data, the beam smoothing effects may not be recognizable until scales
corresponding to one beam width. This does not imply that beam
corrections are not necessary.

Merging disjunct data from EBHIS and GASS makes it necessary to correct
at once for beam effects from both telescopes. For an isotropic power
distribution this is fortunately very simple; at any multipole $l$ the
power from the two telescopes is summed; we can correct for beam effects by
using a weighted average beam function, taking the borderline at $\delta
= -2\deg$ into account.  The lower left plot in
Fig. \ref{Fig_EBHIS_GASS} can serve as a proof. Finally, we compare the
corrected power spectra in the lower panel to the right  side of
Fig. \ref{Fig_EBHIS_GASS}. The spectra from EBHIS and GASS agree well;
the sum of the two power distributions is shown in red. This sum contains
apodized regions; therefore, we also plot  the unapodized all-sky
power spectrum. This comparison shows a nearly isotropic power
distribution between the two disjunct hemispheres, but there are systematic
changes in the spectral index. We obtain $ -2.717 > \gamma > -2.810 $
with an average of $\gamma = -2.768$, hence a deviation of up to $\Delta
\gamma \sim 0.05$ from the all-sky index. This is more than a factor of
ten larger than the formal one-sigma uncertainty for an individual fit
in the range for $8 < l < 1023$. The change of the spectral index
reflects changes in the source distribution and is predominantly caused
by differences at low multipoles.  

\subsubsection{System noise }
\label{Noise}

%=========================================================================

\begin{figure}[th] %%  A3
   \centering
   \includegraphics[width=9cm]{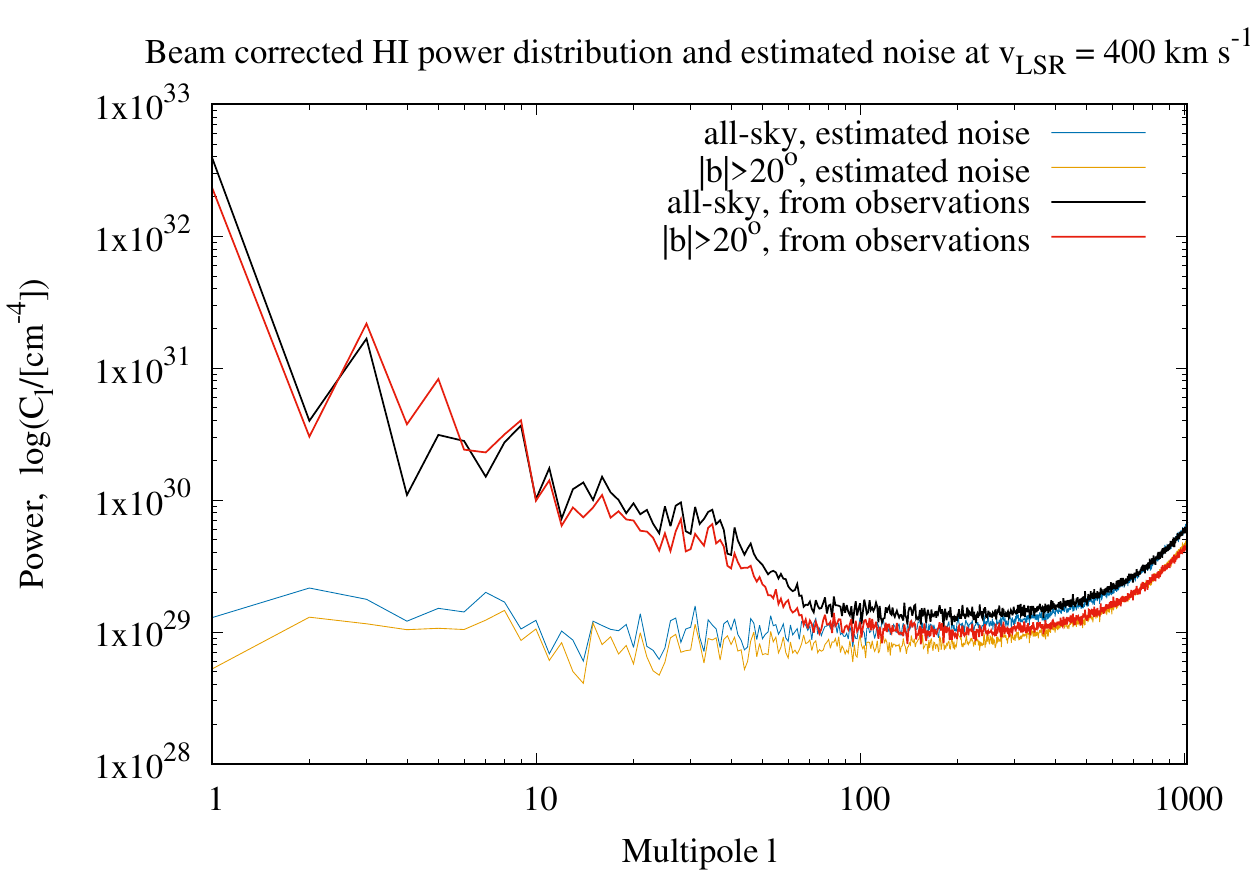}
   \caption{Spatial power distribution $P_{\mathrm {obs}}(l) /
     B^2_{\mathrm{data}}(l)$ for a single channel at $v_{\mathrm{LSR}}
     = 400 $ \kms\ after beam deconvolution. The estimated noise
     distribution $ P_{\mathrm{Noise}} /B^2_{\mathrm{tel}} $ (cyan and
     orange) is overlayed with the observed distribution (black and
     red).}
   \label{Fig_noise_obs}
\end{figure}

%=========================================================================

It has been demonstrated by \citet{Kalberla2016b} that a beam correction
according to Eq. \ref{eq:Power_obs_1} leads in general to an
amplification of instrumental uncertainties at high multipoles.  However, for
$l \la 1023$ we do not observe an obvious increase in
Fig. \ref{Fig_EBHIS_GASS}.

To check whether our analysis might be degraded by unrecognized system
noise, we generate all-sky random noise maps as expected from the system
performance of the two telescopes.  The system noise $T_{ \mathrm{sys} }$
contains several contributions. Most important is the thermal noise from
the receiver system and the elevation dependent ground radiation
including spill-over.  Next there is the position dependent continuum
sky background. All these components are variable, but we use here an
average thermal contribution $T_{ \mathrm{sys} } = 30$ K. The line
signal $T_{ \mathrm{B}} (v_{\mathrm{LSR}}) $ adds to the frequency
independent part and the noise contribution $T_{ \mathrm{Noise}}
(v_{\mathrm{LSR}}) $ to the \hi\ data can then be approximated as
\citep{Haud2000}
\begin{equation} % Eq. 4
  T_{ \mathrm{Noise}} (v_{\mathrm{LSR}}) = \sigma_{ \mathrm{av}} [T_{
      \mathrm{sys}}  + T_{ \mathrm{B}} (v_{\mathrm{LSR}}) ] / T_{\mathrm{sys}},
\label{eq:noise}
\end{equation}
where $\sigma_{ \mathrm{av}}$ is the average noise level, determined at
velocities without \hi\ line emission.

%=========================================================================
\begin{figure}[th] %%  A4
   \centering
   \includegraphics[width=9cm]{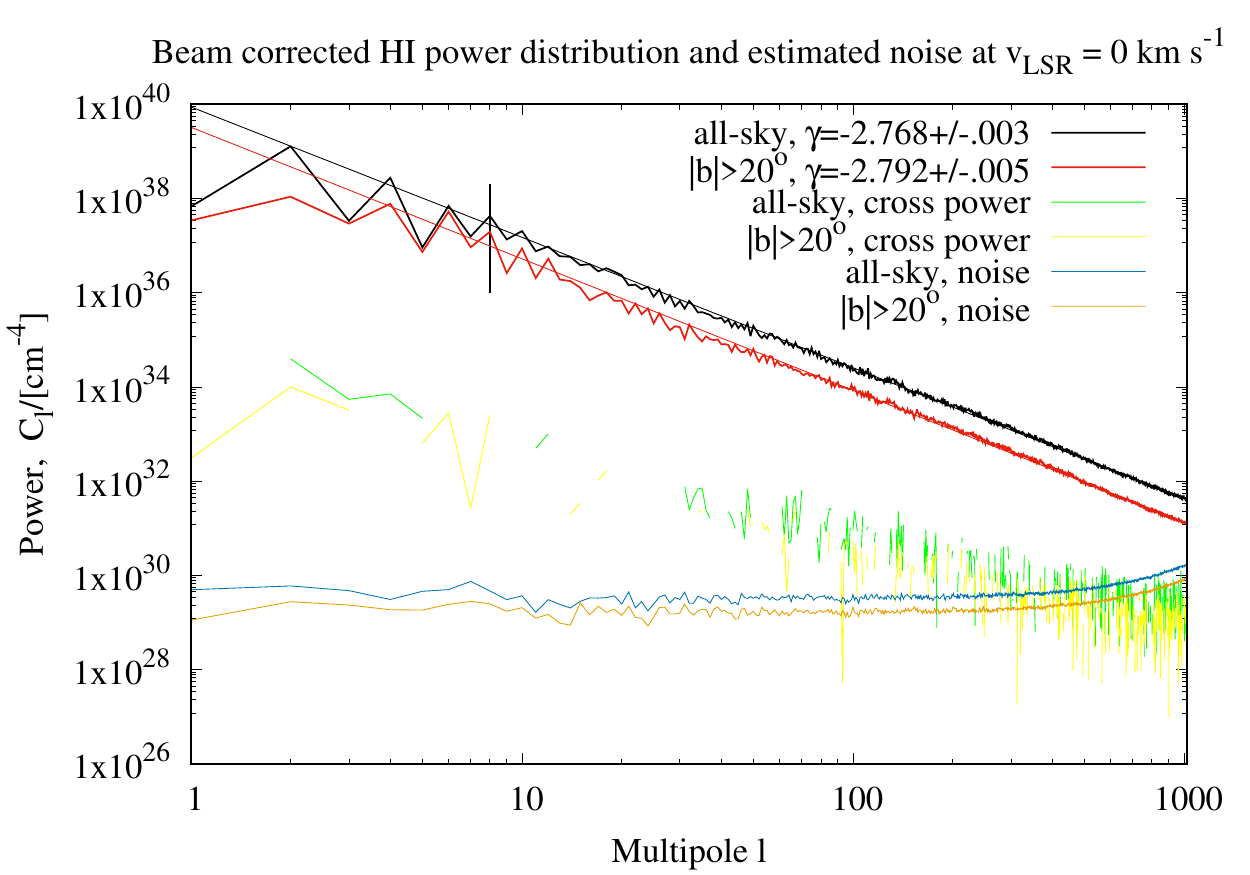}
   \caption{Beam corrected power distribution $P_{\mathrm obs}(l) /
     B^2_{\mathrm Data}(l)$ for single-channel column densities at $
     v_{\mathrm{LSR}} = 0$ \kms, all-sky (black) and $|b| > 20 \deg$
     (red). The estimated noise power $P_{\mathrm{Noise}}
     /B^2_{\mathrm{tel}}$ according to Eq. \ref{eq:noise} is shown in
     cyan and orange. The noise cross power term $2
     P_{\mathrm{cross}}(l)/B^2_{\mathrm{data}} $ is shown in green and
     yellow; only positive values are shown.}
   \label{Fig_cor_noise}
\end{figure}
%=========================================================================

Figure \ref{Fig_noise_obs} shows a noise power spectrum for such a
simulation of the observational conditions for $T_{ \mathrm{B}}
(v_{\mathrm{LSR}}) \sim 0$ K after deconvolution for the telescope beams
at velocities without \hi\ line emission. We overlay this estimate with
the power distribution $P_{\mathrm {obs}}(l)/B^2_{\mathrm {data}}(l)$
derived for a single channel observed at the velocity $v_{\mathrm{LSR}}
= 400$ \kms, without significant \hi\ emission. The cross term $2
P_{\mathrm{cross}}$ in Eq. \ref{eq:TB_power4} is therefore
zero. Comparing the observed power distribution with the expected distribution,  we
find agreement for $l \ga 300$ but significant deviations at low
multipoles, typically for single-dish telescopes (e.g., \citealt[][Figs. 6
  and 9)]{Martin2015}. Assuming a constant $\sigma_{ \mathrm{av}}$ is only
a poor approximation of the observed noise power (see
\citealt[][Fig. 9]{Kalberla2010} and \citealt[][Fig. 14]{Winkel2016a}). The
observational setup causes unavoidable deviations from a random noise,
in particular blocky structures, which in turn lead to enhanced power at
low multipoles.  To improve the noise template we need  to replace the
averages $\sigma_{\mathrm{av}}$, $T_{\mathrm{sys} }$, with position
dependent values in Eq. \ref{eq:noise}. For each of the telescope dumps
and HEALPix positions the rms fluctuations in the baseline regions are
known and can also be used. This more sophisticated scheme was used to
control the Gaussian decomposition
(\citet{Haud2000}, \citet{Kalberla2015}, and \citet{Kalberla2018}, but here we found that it  was not 
necessary to elaborate Eq. \ref{eq:noise} further.

We apply our simple noise model to the single channel \hi\ data at
$v_{\mathrm{LSR}} = 0$ \kms\ shown in Fig. \ref{Fig_Power_obs}. In
comparison to Fig. \ref{Fig_noise_obs} the line emission term $T_{
  \mathrm{B}} (v_{\mathrm{LSR}})$ in Eq. \ref{eq:noise} is significant,
leading to an enhancement of the noise power. In addition,  $T_{ \mathrm{B}}
(v_{\mathrm{LSR}}) \neq 0 $ K implies  that the noise depends on
the intensity distribution on the sky, and $2
P_{\mathrm{cross}}/B^2_{\mathrm{data}}$ from Eq. \ref{eq:TB_power4}
needs to be taken into account.  In Fig. \ref{Fig_cor_noise} we compare
the estimated noise contribution and the cross power with the beam
corrected power spectrum. The noise contribution $P_{\mathrm{Noise}}
/B^2_{\mathrm{tel}}$ is critical at high multipoles, but due to the good
signal-to-noise ratio of the HI4PI data this noise power is, in the
worst case, at $l = 1023$  more than an order of magnitude below the
\hi\ signal $P_{\mathrm {obs}}(l) / B^2_{\mathrm {data}}(l)$. The noise
cross power $2 P_{\mathrm{cross}}/B^2_{\mathrm{data}}$ scales with
$P_{\mathrm {obs}}(l) / B^2_{\mathrm {data}}(l)$, but is about three
orders of magnitude below the relevant \hi\ power signal. Altogether,
this is a very comfortable result, reflecting   that the HI4PI
survey has a good signal-to-noise ratio.

For \hi\ in the profile wings considered by us, for example at
$|v_{\mathrm{LSR}}| = 16$ \kms, the noise power is still more than a
factor of five below the line signal. Our noise template according to
Eq. \ref{eq:noise} can be applied to different velocity widths and we
obtain in this case a simple scaling relation.  Averaging $n$ channels
decreases the noise power contribution by a factor of $n$, but at the
same time the \hi\ power increases. For a total bandwidth of $\Delta
v_{\mathrm{LSR}} = 16$ \kms\ the expected noise contribution
is, in the worst case,   more than four orders of magnitude below the
observed signal. Noise limitations, including other remaining
instrumental uncertainties, are not important for HI4PI data as used by
us. For most of our analysis we can safely  disregard the noise term
$N_{\mathrm{oise}}(l)$ in Eq. \ref{eq:Power_obs_1}.

\subsection{Intrinsic uncertainties in parameter fitting}

\begin{figure}[th] %%  A5
   \centering
   \includegraphics[width=9cm]{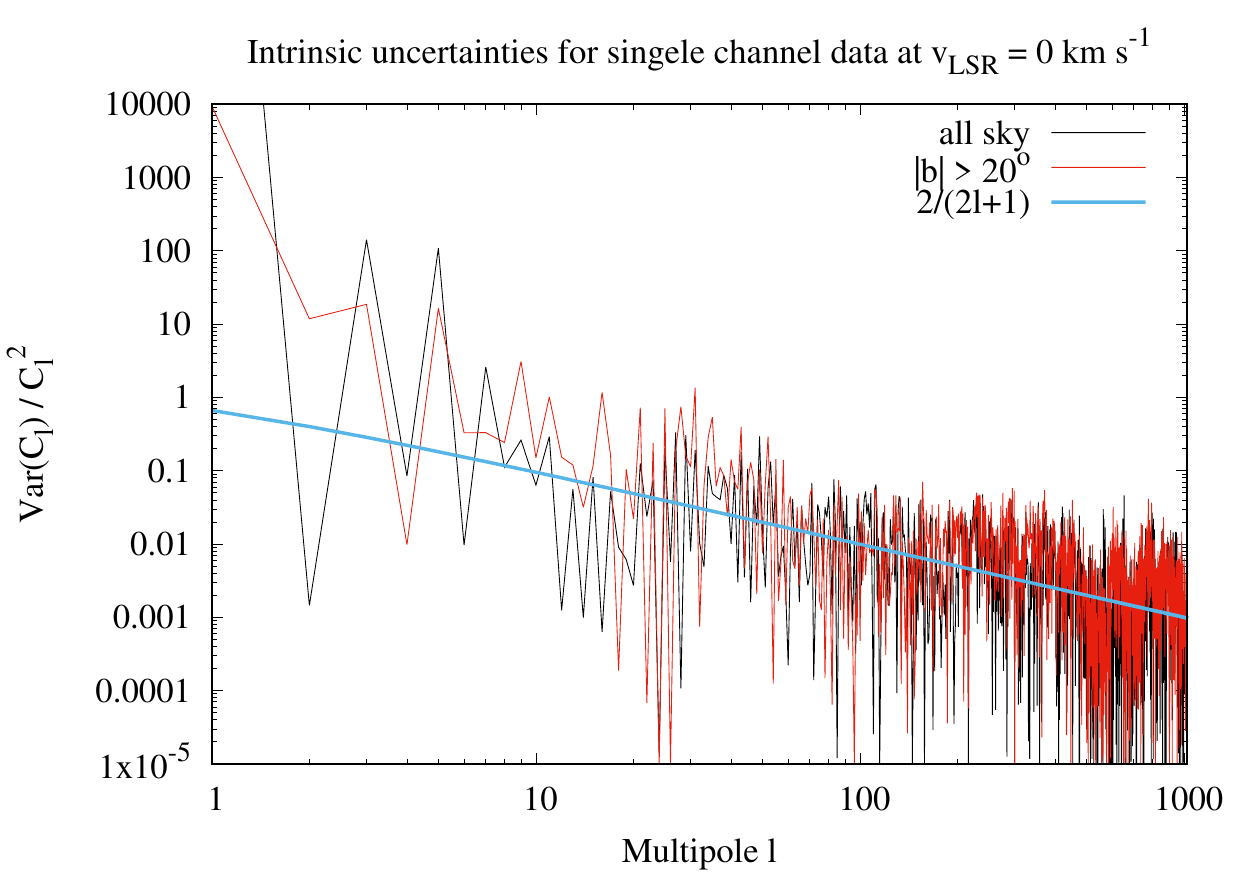}
   \caption{Intrinsic uncertainties $Var(C_l) / C_l^2 $  for single-channel data at $ v_{\mathrm{LSR}} = 0 $ \kms\ in comparison to
     the expected relation $2/(2l+1)$.}
   \label{Fig_scatter}
\end{figure}

We learn from the noise simulations that the scatter of $C_{\mathrm
  l}(l) $ is for white noise increasing at low multipoles $l$ (see
Fig. \ref{Fig_noise_obs}). This is a general property of angular power
spectra with finite counts \citep[][Eq. 5]{Campbell2015} and we expect
for the estimator $\hat{C_l}$
\begin{equation} % Eq. 5
  Var(\hat{C_l}) / [ N_l + C_l ]^2 = 2/(2l+1) .
\label{eq:scatter}
\end{equation}
$Var(\hat{C_l})$ is usually called cosmic variance. In
Fig. \ref{Fig_scatter} we show an example of single-channel data at $
v_{\mathrm{LSR}} = 0 $ \kms. When fitting power spectra we take
dispersions $\propto \sqrt{ C_l^2 \cdot 2/(2l+1)}$ into account. Fitting
is done iteratively with constant weights for an initial estimate of
$\hat{C_l}$, next using derived weights to improve the error estimate
for the final fit with $\chi^2/N_{\mathrm {DOF}} \la 1 $. We quote for
the power spectral indices formal one-sigma uncertainties from the
fit. Systematic fluctuations between different hemispheres amount to
$\Delta \gamma \sim 0.05$ (Sect. \ref{Beam}) and may be more
characteristic for the overall uncertainties. Differences between
the initial estimate and the final fit are typically less than or about one sigma.
In general we find a large scatter for $C_l$ at low multipoles $l \la 8$
with  tendencies for even-odd oscillations. Large-scale symmetries
of the Galactic \hi\ emission cause an enhanced disparity between even
and odd modes \citep{Mertsch2013}. These disparities disappear mostly
for higher multipoles, and therefore we exclude in general multipoles $l
\le 8$ when fitting spectral indices for the \hi\ distribution.

%=========================================================================

%=========================================================================

\end{appendix}

\end{document}